\documentclass[reprint,amsmath,amssymb,aps,pra,notitlepage,twocolumn]{revtex4-1}
\usepackage{graphicx}
\usepackage{dcolumn}
\usepackage{bm}

\begin{document}


\title{Properties of atomic pairs produced in the collision of Bose-Einstein condensates}



\author{Pawe{\l} Zi\'n}
\email{Pawel.Zin@ncbj.gov.pl}
\affiliation{National Center for Nuclear Research, ul. Ho\.za 69, 00-681 Warsaw, Poland}

\author{Tomasz Wasak}
\affiliation{Faculty of Physics, University of Warsaw, ul. Pasteura 5, 02-093 Warsaw, Poland}




\begin{abstract}
Collisions of Bose-Einstein condensates can be used as a mean to generate correlated pairs of atoms. The scattered massive particles, in analogy to photon pairs in
quantum optics, might be used in the violation of Bell's inequalities, demonstration of Einstein-Podolsky-Rosen correlations, or sub-shot noise atomic
interferometry. Usually, a theoretical description of the collision relies either on stochastic numerical methods or on analytical treatments involving various
approximations.  Here, we investigate elastic scattering of atoms from colliding elongated Bose-Einstein condensates within Bogoliubov method, carefully controlling
performed approximations at every stage of the analysis.  We derive expressions for the one and two particle correlation functions. The obtained formulas, which relate the
correlation functions with condensate wavefunction, are convenient for numerical calculations. We employ variational approach for condensate wavefunctions to obtain
analytical expressions for the correlation functions, which properties we analyze in details. We also present a useful semiclassical model of the process, and compare its
results with the quantum one. The results are relevant for recent experiments with excited helium atoms, as well as for planned experiments aimed at investigating the
nonclassicality of the system.
\end{abstract}

\maketitle

\newcommand{\x}{{\bf r}}
\newcommand{\K}{{\bf k}}
\newcommand{\dk}{  \Delta {\bf k}}
\newcommand{\DK}{\Delta {\bf K}}
\newcommand{\KK}{{\bf K}}
\newcommand{\X}{{\bf R}}

\newcommand{\B}[1]{\mathbf{#1}} 
\newcommand{\f}[1]{\textrm{#1}} 

\newcommand{\half}{{\frac{1}{2}}}

\newcommand{\vv}{{\bf v}}
\newcommand{\p}{{\bf p}}

\section{Introduction}\label{Int}

The production of correlated pairs of particles is an important requirement for probing the foundations of quantum mechanics. For example, in quantum optics correlated
pairs of photons were used to demonstrate violation of Bell's inequalities \cite{bellAspect} or the Hong-Ou-Mandel effect \cite{oum}. In atomic physics, pairs of atoms
were produced in the process of the four-wave mixing in optical lattice \cite{tunableParis}, collision-induced deexcitation of the cloud \cite{viennaTB}, and collision of
Bose-Einstein condensates \cite{keterle1,paryz0,paryz1}.  These many-body systems can be used to demonstrate sub-Poissonian relative atom number statistics \cite{paryz2},
the violation of the Cauchy-Schwarz inequality for matter waves \cite{paryz3, wasakCSI}, atomic Hong-Ou-Mandel effect \cite{atomicHOMtheo,atomicHOM}, or ghost-imaging
\cite{ghost}.  The nonclassicality of the states of the system can be potentially employed in the violation of Bell's inequality for atoms~\cite{atomBell} or atomic
interferometry~\cite{njpWasak}.

In order to have a high degree of control of  the fragile states of the pairs of atoms, it is necessary to posess an accurate model describing the system as well as the 
processes underlying generation of the correlated pairs. In this paper we analyze elastic scattering of atoms from a pair of colliding Bose-Einstein condensates (BECs).
Such condensate collisions 
were investigated in theoretical papers \cite{bach,zin1,zin2,zin3,zin4,deuar1,deuar2,gardiner1,gardiner2,paryz10,karen0,karen1, wasakdeuar}.
As a result of binary collisions between the particles that constitute the counter-propagating clouds, atomic pair scatter out from the condensates with opposite velocities.
In the spontaneous regime, where bosonic enhancement does not influence single collision event, the direction of velocity of outgoing particles
is random. As a result of superposition principle, the quantum state of single atomic pair is entangled in momentum directions.
Such systems, with non-classical correlations between massive particles entangled in external degrees of freedom, are ideal for fundamental experimental 
tests of quantum mechanics such as Einstein-Podolsky-Rosen gedanken experiment \cite{Wieden}.

In order to prove the presence of non-classical correlations between scattered atoms, or to utilize the entanglement in quantum-enhanced atomic interferometry, 
properties of the state of the system need to be well known. 
In particular, an important quantities are the widths of the correlation volume, which describes the extent of the correlation between the two particles.
To find those widths one has to analyze properties of the two particle correlation function.
This function was recently measured experimentally \cite{paryz0,paryz3}, and also was the subject of theoretical studies based on perturbation 
theory \cite{bach,zin2,zin3,karen1} or stochastic calculations \cite{deuar1,deuar2,gardiner1,gardiner2,karen0}.

In the case of perturbation theory, a number of uncontrolled approximations were introduced in order to obtain
analytical results. The main approximations were to neglect the interaction between atoms within BECs during the collision process, 
and omission of the influence of the mother-condensates on the scattered atoms.
Therefore, it is important to know how the interactions between atoms change the state of the system, and how they modify the non-classicality of the correlations
between the particles. The results are of importance for the planned experiments that might test EPR-correlations, non-locality, 
or utilize such systems in quantum-enhanced metrology.

Here, we derive, within the perturbation theory, approximate analytical expressions for the two particle correlation function taking 
into account the contribution from the interaction between atoms. 
We show that there are two types of correlations:
``back to back'' -- related to the fact that particle are scattered in pairs,
and  ``local'' -- which describe bosonic bunching effect. 
The obtained formulas connect correlations functions with colliding condensate wavefunctions
being a solution of time-dependent Gross-Pitaevskii (GP) equation.
The derived expressions have the form convenient for numerical analysis and as such can be useful for interpretation of the experimental data.
Additionally, we derive approximate analytical form of the colliding condensates wavefunctions using time-dependent
variational approximation. Using these wavefunctions we obtain analytical expressions for the two-body correlation functions. 
We analyze properties of the correlation functions among which the most important are the correlation
widths for which we present explicit expressions. Finally, we present the properties of the correlation volumes.

The paper is constructed as follows. 
In Section~\ref{model} we present the system and the methods to describe it based on the Bogoliubov theory.
In this Section we derive general expressions for a two particle correlation function which structure can be divided into the two types of correlations. 
Also, we present the results of variational approximation for static and colliding Bose-Einstein condensates.
In Section~\ref{Sbb} we analyze the ``back to back'' correlations, and present approximate expressions for the pair correlation function.
Then using the approximate condensate wave functions given by variational approach, we derive explicit formulas for the pair correlation function and
analyze its properties in important cases. We unify the obtained results in explicit expressions, and present the properties of the correlation volume.
Additionally, we introduce 
a semiclassical model and compare its results with the results of the quantum one. 
In Section~\ref{lc} we analyze the ``local'' pair correlations, which are directly related to single particle correlation function $G^{(1)}$. 
We introduce a classically motivated function describing properties of the source of atoms and relate it with the single particle correlation function. 
Further, we present explicit formula for this function obtained from a quantum model of the process.
We calculate $G^{(1)}$ using the approximate condensate wave functions given by variational approach and analyze its properties in certain cases. 
Also in this section, we unify the obtained results in explicit expressions, and present the properties of the correlation volume.
The details of the derivations of most of the formulas as well as the conditions presented in the main body of the paper can be found in the Appendices.

\section{Theoretical model}\label{model}

In the limit of low energies the collision of bosonic atoms can be described by a single parameter called the scattering length~\cite{zderzenia}. 
In our considerations we assume that the interaction potential is effectively described by this parameter. 
Additionally, we  assume that the atomic gas is in the dilute limit, i.e., $na^3 \ll 1$, where $n$ 
is a maximal density of the colliding clouds and $a$ is the scattering length. 
Furthermore, we restrict our considerations to the so-called collisionless regime, in which the probability of a secondary collision of the scattered atom with the atoms 
from the condensate is much smaller than unity. This condition requires that the size $\sigma$ of the cloud in the direction
of the atomic velocity is much smaller than the mean free path of the scattered particles $\ell_\mathrm{mfp} = 1 / 8\pi a^2 n$, that is  $\sigma/\ell_\mathrm{mfp} \ll 1$.
Another assumption we use is that the total number of scattered atoms is much smaller than the number of atoms in the moving clouds.

We remark one fact that determines our interests in specific observables.
After the collision the system consists of the two condensates and the halo of scattered atoms.
Some of the scattered atoms are located on the condensates, because their mean velocity points in the direction of the velocities of the condensates, 
and both of these vectors are of the same magnitude. 
However, the analysis of those scattered atoms is difficult experimentally, because the density of the halo is small compared to the density of the condensates. 
For this reason, in all the considerations below, we restrict our study to the atoms that are scattered away 
from the collision direction of two condensates (we take it to be $z$ axis).
Specifically, we assume that the part $\K_r = k_x {\bf e}_x + k_y {\bf e}_y$ of the wavevector $\K$ 
of the scattered atom perpendicular to the long axis of the condensate satisfies the following conditions: 
$|\K_r|/k > 1/2$. 


\subsection{Bogoliubov method}

The system that obeys all of the above restrictions is well described by the Bogoliubov method~\cite{przeglad}. 
In this method moving condensates are described by single particle wave-function $\psi(\x,t)$
which satisfy Gross Pitaevskii equation
\begin{equation} \label{Gp}
 i \hbar \partial_t \psi (\x,t) = \left( - \frac{\hbar^2}{2m} \triangle + g|\psi(\x,t) |^2  \right) \psi(\x,t),
\end{equation}
where $g= \frac{4\pi \hbar^2 a}{m}$ parametrizes the interaction strength between atoms.
The properties of scattered atoms are described by the field operator  $\hat \delta (\x ,t)$ 
which undergoes time evolution given by
\begin{eqnarray}\label{glowne}
i \hbar \partial_t \hat \delta (\x ,t) =  H_0 (\x ,t) \hat \delta (\x ,t) +  B (\x ,t) \hat \delta^\dagger (\x ,t)
\end{eqnarray}
with:
\begin{eqnarray}\label{H0}
H_0(\x ,t)  &=& - \frac{\hbar^2}{2m} \triangle + 2 g |\psi (\x,t)|^2,
\\ \nonumber
\\ \label{B0}
B(\x,t) &=& g \psi^2 (\x,t).
\end{eqnarray}
We assume that the initial state of noncondensed particles is vacuum~\cite{dopiska}, i.e.,
\begin{equation}\label{stan}
 \hat \delta(\x,0)|0 \rangle = 0.
\end{equation}
In the calculation of mean values of the product of $\hat\delta$ operators, taken for various positions, the Wick's theorem can be applied. This is a consequence of the lineary of the
equation of motion, Eq.~\eqref{glowne}, and the fact that the initial state is vacuum.
Therefore, all the correlation functions of arbitrary order decompose into products of anomalous density
\begin{equation}\label{Mdef}
M\left( \x_1,\x_2,T \right) =  \langle \hat \delta (\x_1 ,T)
\hat \delta (\x_2 ,T)\rangle,  
\end{equation}
and single particle correlation function 
\begin{equation}\label{G1}
G^{(1)} \left( \x_1,\x_2,T \right) = \langle \hat \delta^\dagger (\x_1 ,T)
\hat \delta (\x_2 ,T)\rangle.    
\end{equation}
For example, the two particle correlation function is of the following form:
\begin{eqnarray} \nonumber
&& G^{(2)}\left( \x_1,\x_2,T \right) = \langle \hat \delta^\dagger (\x_1 ,T) \hat \delta^\dagger (\x_2 ,T)
\hat \delta (\x_2 ,T) \hat \delta (\x_1 ,T)\rangle
\\ \nonumber
&& = G^{(1)} \left( \x_1,\x_1,T \right)G^{(1)} \left( \x_2,\x_2,T \right)
+  \left|G^{(1)} \left( \x_1,\x_2,T \right) \right|^2 +
\\ \label{G2}
&&\quad + \left| M \left( \x_1,\x_2,T \right)  \right|^2.
\end{eqnarray}
In this equation, the first term, given by $G^{(1)} \left( \x_1,\x_1,T \right)G^{(1)} \left( \x_2,\x_2,T \right)$, is a product of single particle densities, 
and so it represents uncorrelated particles.
The presence of second and third terms are responsible for nontrivial corelations between particles.


In the next sections, we show that the terms $\left| M \left( \x_1,\x_2,T \right)\right|^2$ and $\left|G^{(1)} \left( \x_1,\x_2,T \right) \right|^2$ 
represent correlations of particles with opposite and collinear velocities, respectively. 
The appearance of correlation of particles with opposite velocities, called ``back to back'' or ``cross'' correlation, 
is a consequence of the fact that atoms are scattered in pairs of opposite momenta.
The correlation of particles with collinear velocities, called ``local'' correlation, is related to the bosonic bunching effect.

In the considered system a bosonic enhancement effect can take place.
This effect was both predicted theoretically~\cite{keterle1,zin1,zin2} and observed experimentally~\cite{keterle1}.
In this paper, we restrict to the regime where the effects of the bosonic enhancement are negligible. 
Then, the Heisenberg equation of motion, Eq.~\eqref{glowne}, can be approximately solved with help of perturbation theory.
In this case, the formula for the anomalous density reads
\begin{eqnarray}\nonumber
M(\x_1,\x_2,T)   
&=&  \frac{1}{i \hbar} \int_0^T \mbox{d} t \int \mbox{d}\x \, K(\x_1 ,T; \x,t) 
\\ \label{M}
& & \ \ \ \ \ \ \ \ \ \ \ \ \ \ \ \ \times K(\x_2 ,T; \x,t)   B (\x ,t),
\end{eqnarray}
where $K(\x_1,t_1;\x_2,t_2)$ is a single body propagator of Hamiltonian given in Eq.~(\ref{H0}).
Furthermore, on the grounds of perturbative approach, the following relation between one body correlation function and anomalous density can be established: 
\begin{equation}\label{pertG1}
G^{(1)} \left( \x_1,\x_2,T \right) =\int \mbox{d} \x \, M^*\left( \x_1,\x,T \right)M\left( \x,\x_2,T \right).
\end{equation}
The details of the derivation of the above formulas are presented in Appendix~\ref{AppBog}. 
The conditions for the validity of the first order perturbation calculus 
are derived and discussed in Appendix~\ref{VP}.

\subsection{Properties of the condensates}\label{mod}

Let us now further specify the properties of the considered system.
We initially deal with a single condensate described by a wavefunction $\psi(\x)$ satisfying stationary GP equation
\begin{equation}\label{sGP}
\mu \psi(\x) = \left(- \frac{\hbar^2}{2m} \triangle + V(\x) + g|\psi(\x)|^2 \right)\psi(\x),
\end{equation}
where the trapping potential
\begin{equation}\label{pot}
V(\x) = \frac{1}{2}m\big(\omega_r^2(x^2+y^2) + \omega_z^2 z^2\big),
\end{equation}
and the normalization condition $\int \mbox{d} \x |\psi(\x)|^2 = N$.
Below, we focus on elongated cigar shaped condensate, for which $\omega_z \ll \omega_r$.
The two counterpropagating condensates are created from the stationary one by applying
Bragg pulse and switching off the trapping potential \cite{keterle1}. 
After the pulse, the wavefunction takes the following form
\begin{equation}\label{Cwzor}
\psi(\x,0) = \frac C2\big(\psi(\x)e^{i Qz}\!+\!\psi(\x)e^{-i Qz}\big)\!=\!C \psi(\x) \cos (Qz),
\end{equation} 
where $C$ is the normalization coefficient. This wavefunction represents two wavepackets, $\psi(\x)^{\pm i Qz}$, each propagatin with mean velocity $\pm \hbar Q/m$ alogn $z$-axis. 
The collision takes place along longitudinal axis, the $z$-axis, of the condensate.

The decomposition of the wavefunction into two counterpropagating wavepackes is permissible also for later times.
To this end, we assume that the width of two counterpropagating condensates in the momentum space during the collision are much smaller then their mean momenta equal to $\pm \hbar Q$.
We additionally assume that the mean field potential $gn$ is much smaller than the kinetic energy $\frac{\hbar^2 Q^2}{2m}$.
In Appendix~\ref{timeevolution} we show that these two conditions can be replaced by a single one which reads
\begin{equation}\label{Cond1}
\frac{2 \sigma_r}{Q a_{hor}^2} \ll 1,
\end{equation}
where $\sigma_r$ denotes the radial width of the initial condensate, and $a_{hor} = \sqrt{\frac{\hbar}{m\omega_r}}$ is the radial harmonic oscillator length. 
The above condition combined  with the fact that the system is elongated, implies  $Q \gg \frac{1}{\sigma_z}$,
where $\sigma_z$ denotes the longitudinal size of the initial condensate.
As a consequence, $C \simeq \sqrt{2}$ which makes Eq.~(\ref{Cwzor}) to take the following simple form:
\begin{equation}\label{Cwzor2}
\psi(\x,0) = \sqrt{2} \psi(\x) \cos (Qz). 
\end{equation} 

The assumption the widths of the momentum distribution of both colliding condensates during the whole collision are much smaller than $Q$, leads to two distinct momentum distributions 
centered around $\pm Q {\bf e}_z$.
Therefore, it is natural to decompose the condensate wavefunction $\psi(\x,t)$ into two wavefunctions, denoted by $\psi_{\pm Q}(\x,t)$, in the following way:
\begin{equation} \label{psip}
 \psi(\x,t) = \left(\psi_{+Q}(\x,t) e^{iQz} + \psi_{-Q}(\x,t) e^{-iQz} \right) e^{-i\frac{\hbar Q^2}{2m}t}.
\end{equation}
These functions, $\psi_{\pm Q}$, describe the two counterprogating parts of the condensate. Note that, as implied by Eq.~(\ref{Cwzor2}), 
initial conditions are $\psi_{\pm Q}(\x,0) = \frac{1}{\sqrt{2}} \psi(\x)$ for these components.

The stated assumption leads to great simplification of the solution of the GP equation, Eq.~\eqref{Gp}, on the basis of the slowly varying envelope approximation \cite{slowlyvarying}.
Within this approximation, the GP equation for $\psi(\x,t)$ decouples into set of two equations for $\psi_{\pm Q}(\x,t)$ 
\begin{eqnarray} \nonumber
&& i \hbar \partial_t \psi_{\pm Q}(\x,t) = \left( \mp  i\frac{\hbar^2}{m} Q \partial_z  -   \frac{\hbar^2}{2m}
\triangle \right) \psi_{\pm Q}(\x,t)  
\\ \label{SV}
&&
+  g \left( |\psi_{\pm Q}(\x,t)|^2 + 2|\psi_{\mp Q}(\x,t)|^2  \right)  \psi_{\pm Q}(\x,t). 
\end{eqnarray}
Let us remark, that Eq.~(\ref{SV}) is much better for numerical implementation of our problem than the initial GP equation, Eq.~(\ref{Gp}).
The reason is that in Eq.~(\ref{SV}) the highly oscillatory behavior in position, due to $e^{\pm i Q z}$, as well as in time, due to $e^{-i Q^2 \hbar t/2m}$, is removed. 
Consequently, the window in the momentum representation, required for numerical simulation, needs to take the momentum width of the $\psi_{\pm Q}$ alone, 
and this is much smaller than the momentum window required for the solution of Eq.~(\ref{Gp}), the latter being of the order of $2Q$.

In order to describe the properties of the wavefunctions $\psi_{\pm Q}$, we use variational method, to solve stationary and time dependent GP equations, Eqs.~(\ref{sGP}) and~(\ref{SV}).
The details of the solution are described in Appendix~\ref{ApSol}. Here, we just state the obtained results.

We assume that initially the wavefeunction can be approximated by a gaussian ansatz of the form:
\begin{equation} 
\psi_{\pm Q}(\x,0) =  \sqrt{\frac{N}{2\pi^{3/2} \sigma_z \sigma_r^2 }}
\exp\!\left(\!-  \frac{x^2+y^2}{2\sigma_r^2} - \frac{z^2}{2\sigma_z^2} \!\right).
\end{equation}
After time $t$, the wavefunctions evolve into:
\begin{eqnarray} \nonumber
\psi_{\pm Q}(\x,t) &=&  \sqrt{\frac{N}{2\pi^{3/2} \sigma_z \sigma_r^2(t) }}
\exp\left(-  a_r(t) (x^2+y^2) \right)
\\ \label{vatnowe}
& & \exp \left(- a_z(t) (z \mp v_0t)^2 - i \phi(t) \right),
\end{eqnarray}
where 
\begin{eqnarray} \nonumber
\sigma_r^2(t) &=& \sigma_r^2 (1+\omega_r^2t^2),
\\ \nonumber
a_r(t) &=& \frac{1}{2\sigma_r^2(t)} (1-i\beta \omega_r t),
\\ \nonumber
a_z(t) &=& \frac{1}{2\sigma_z^2}\left( 1 - i \left( \beta -\frac{1}{\beta}  \right) \arctan(\omega_r t) \right),
\\ \label{parwar}
\phi(t) &=& \left( \frac{7\beta}{4} -  \frac{3}{4\beta} \right) \arctan (\omega_r t),
\end{eqnarray}
where $v_0 = \hbar Q / m$, and  $\beta = \frac{\sigma_r^2}{a_{hor}^2} \geqslant 1$. Notice, that the final form is also a gaussian function, but with centres moving in opposite directions
with velocities $\pm v_0 \mathbf{e}_z$, and with time-dependent widths.

The components of the wavefunctions, given by $\psi_{\pm Q}$ in Eq.~(\ref{vatnowe}), can be investigated further in order to determine important timescales in the problem.
First, notice that there is a characteristic time $\tau_c = \frac{\sigma_z}{v_0}$ during which the wavepackets cross each other. 
This can be defined as the time of the collision.
The second characteristic time $\tau_{ex} = \omega_r^{-1}$ is equal to the time needed for $\sigma_r(t)$ to change its width by a factor of $\sqrt{2}$. 
This two times describe the density properties of the system. 
Next, characteristic times correspond to changes in the phase of the wavepacket. 
It is natural to define following three timescales:
$\tau_r =  \frac{1}{\beta \omega_r}$, 
$\tau_z = \frac{1}{\omega_r} \tan \left( \beta -\frac{1}{\beta} \right)^{-1} $,
and 
$\tau_\phi = \frac{1}{\omega_r} \tan \left( \frac{7\beta}{4} -  \frac{3}{4\beta} \right)^{-1}$.
Note, that $\tau_{ex} \geqslant \tau_r$, $\tau_z \geqslant \tau_r $, and $\tau_\phi \geqslant \frac{4}{7} \tau_r$.
Therefore, we see the three times, $\tau_{ex}$, $\tau_z$, and $\tau_r$ are all larger than $\tau_r/2$.

In this section we described the  colliding condensates. In the next one, we characterize the properties of the scattered atoms.

\section{Back to back correlations}\label{Sbb}
 
\subsection{General considerations}

Let us now show that the back to back correlations are  given by the term $\left| M \left( \x_1,\x_2,T \right)  \right|^2$.
To this end, we analyze the structure of the anomalous density $M \left( \x_1,\x_2,T \right)$. 
However, notice first that in the experimental situation the atoms are measured at time $T$ which is usually much larger than the time of the scattering process. 
It is thus permissible to investigate the limit $T \rightarrow \infty$. We introduce new variables: $\x_{1,2} = \frac{\hbar \K_{1,2}}{m} T $, and define
\begin{eqnarray} \nonumber
 M(\K_1,\K_2) &=&  \left( \frac{\hbar T}{m}  \right)^3\lim_{T\rightarrow \infty} \exp \left(- i \frac{\hbar (k_1^2 +k_2^2) }{2m}  T \right)
 \\ \label{Mss}
 & &  M \left( \frac{\hbar \K_1}{m} T,\frac{\hbar \K_2}{m} T;T \right). 
 \end{eqnarray}
In the above we have introduced additional phase 
$\exp[- i \hbar (k_1^2 +k_2^2)T/2m]$ to get finite limit
and additional factor  $(\hbar T/m)^3$ to satisfy normalization condition  $ \int  \mbox{d} \K_1 \mbox{d} \K_2 \, 
|M(\K_1,\K_2)|^2 = \int \mbox{d} \x_1 \mbox{d} \x_2 \, |M(\x_1,\x_2,T)|^2 $.
Let us now continue taking $K$ as free propagator. 
Upon inserting Eq.~(\ref{M}) into Eq.~(\ref{Mss}) and using the explicit form of the free propagator we arrtive at
\begin{eqnarray}\label{Mkw}
M(\K_1,\K_2) 
 &=& \frac{1}{\hbar(2\pi)^3} \int_{0}^\infty \mbox{d} t \int \mbox{d} \x  \, 
\\ \nonumber
&&
 \exp \left( - i (\K_1+\K_2) \x + i \frac{\hbar(k_1^2 + k_2^2)}{2m} t \right)   B (\x ,t).
\end{eqnarray}
Substituting now Eq.~(\ref{psip}) into Eq.~(\ref{B0}) we obtain
\begin{eqnarray} \label{Bdef}
&& B(\x,t) = g \psi^2(\x,t) 
\\ \nonumber
&& = g \left( \psi_{+Q}^2e^{i2Qz} + 2 \psi_{+Q} \psi_{-Q} + \psi_{-Q}^2 e^{-i2Qz} \right)e^{-i\frac{\hbar Q^2}{m}t}.
\end{eqnarray}
We shall now make use of the assumptions stated in Section \ref{mod}.
We assumed there that the widths in the momentum space of each of 
wavepacket $\psi_{\pm Q}$ at any time are much smaller than $Q$ and that the kinetic energy
$\frac{\hbar^2 Q^2}{2m}$ is much larger than than the mean field interaction energy $gn$.
This assumptions have two consequences. First, the dominant temporal phase is given by factor  $\exp( - i \hbar Q^2t/2m)$.
Second, the dominant spatial phase factor in $\psi_{\pm Q}e^{\pm iQz}$ is 
$\exp \left( \pm i Q z  \right)$.
Inserting Eq.~(\ref{Bdef}) into Eq.~(\ref{Mkw}) and using the above stated assumptions we notice that the temporal integral is vanishingly small unless $k_1^2 + k_2^2 \simeq 2 Q^2$.
As the width of $\psi_{\pm Q}$ in momentum space is much smaller than $Q$ the spatial integral gives nonzero values for the term $\psi_{+Q} \psi_{-Q}$ if $|\K_1 + \K_2| \ll  Q $, 
and for the terms $\psi_{\pm Q}^2 $ if  $|\K_1 + \K_2 \mp 2 Q {\bf e}_z | \ll Q $.
Consequently, the term responsible for the scattering of atoms into the observed halo is due to the term $\psi_{+Q} \psi_{-Q}$. 
As we are interested only in the properties of atoms appearing in the halo we neglect the contribution from the other two terms, i.e.,
\begin{eqnarray}\label{Bp}
B(\x,t) = 2 g \psi_{+Q}(\x,t)\psi_{-Q}(\x,t) \exp \left(-i\frac{\hbar Q^2}{m} t \right).
\end{eqnarray}  
The above analysis also shows that the anomalous density with $B$ given by Eq.~(\ref{Bp}) 
leads to $\K_1$ and $\K_2$ being practically antiparallel with length equal approximately to $Q$. 
For other choices of $\K_1$ and $\K_2$ the value of anomalous density is vanishing. 
Thus, as long as it is permissible to exploit free propagator for $K$, we have shown that the back to back correlation is represented by the term 
$G^{(2)}_{bb}(\K_1,\K_2) \equiv \left| M \left( \K_1,\K_2\right)  \right|^2$ in the two particle correlation function in Eq.~\eqref{G2}.

The situation with the true propagator, the one which takes into account the interaction of scattered particles with the atoms from the condensates, 
is much more complicated as there is no analytical formula for $K$.
However, under assumptions stated in Section~\ref{model} together with the additional condition
\begin{equation}\label{condition3n}
 36 \left(\frac{gn}{\frac{\hbar^2 Q^2}{m}} \right)^2  Q\sigma_r  \ll 1,
\end{equation}
a semiclassical approximation can be used, which results in 
\begin{eqnarray} \nonumber
&& M(\K_1, \K_2)  
= \frac{1}{\hbar (2\pi)^3}   \int_0^\infty \mbox{d} t \int \mbox{d} \x \, 
\\ \label{Mtdrugie}
&&
\exp \left(  - i (\K_1+\K_2) \x + i \frac{\hbar(k_1^2+k_2^2)}{2m}  t \right) B(\KK,\x,t),
\end{eqnarray}
where
\begin{eqnarray}\label{Bmf}
B(\KK,\x,t) &=& B(\x,t )\exp \left(- i  \Phi(\x,{\bf e}_{\KK},t) \right),
\\ \label{MFPhi}
\Phi(\x,{\bf e}_{\KK},t) &=&  \frac{m}{\hbar^2 Q} \int^\infty_{-\infty}\!\! \mbox{d} s \, V_{en}(\x  + s  {\bf e}_{\KK},t),
\\ \nonumber
V_{en}(\x,t) &=& 2g \left( |\psi_{+ Q} (\x,t)|^2 + |\psi_{- Q} (\x,t)|^2 \right),
\end{eqnarray}
and $\KK = \frac{\K_1-\K_2}{2}$, ${\bf e}_{\KK} = \frac{\KK}{K}$. 
The details of the derivation are given in Appendix~\ref{appendixAD}.
There we also show that $|M|^2$ represents the back to back correlation, the same result which we obtained for the free propagator case.
The expression in Eq.~(\ref{Mtdrugie}) has exactly the same form as the free propagator formula, Eq.~(\ref{Mkw}), the only difference being the change from $B(\KK,\x,t)$ to  $B(\x,t)$. 

The semiclassical approximation leads to the same results as the free propagator approximation if $|\Phi| \ll 1 $, for which $B(\KK,\x,t) \simeq B(\x,t)$. 
In Appendix~\ref{appendixAD} we show that the contribution from this phase can be neglected if
\begin{equation}\label{conditionNn2}
\frac{16 \sigma_r^3}{Qa_{hor}^4} \ll 1.
\end{equation}
From this condition, it is straightforward to obtain the one given in Eq.~(\ref{Cond1}). 

The set of equations: (\ref{sGP}), (\ref{pot}), (\ref{SV}), (\ref{Bp}), (\ref{Mtdrugie}),  (\ref{Bmf}), and  (\ref{MFPhi}) allow for calculation of the anomalous density $M$. 
This function includes the mean field effects of the interaction between condensates and scattered atoms, if the condition given in Eq.~(\ref{condition3n}) is satisfied. 
For a given system configuration the calculations are to be done numerically. 
However, below we exploit the approximate solution of the colliding condensates wavefunctions, given by Eq.~(\ref{vatnowe}), to calculate and analyze the properties of the
back to back correlation function. 
But before going on, we pause to introduce a semiclassical model of the back to back  correlation function, which will serve as a probe of quantum characteristics of the collision process.

\subsection{Semiclassical model}

From a theoretical point of view, it is interesting to compare the results of a quantum model to a classical one. 
To this end, we consider a semiclassical model of colliding clouds.  We describe the two wavepackets by a single-particle phase space densities, denoted by $W_{\pm Q}(\x,\K,t)$,
and apply all the assumptions stated previously that defined our system, i.e.,
\begin{itemize}
\item dilute gas limit, so that only two body collisions are of importance,
\item neglect secondary collision between the scattered atoms and the atoms in the condensates,
\item neglect depletion of the condensate due to the scattering.
\end{itemize}
As we compare this model with the quantum one we put $\K$ vector instead of velocity in the phase space density definition, these are simply related by $\K = m \mathbf{v}/\hbar$.
For this semiclassical model the formula for the back to back part of the second order correlation function takes the following form:
\begin{eqnarray}\nonumber
&& G^{(2)}_{bb}(\KK,\DK) = 2 \frac{\hbar}{m}\sigma_{tot}\int \mbox{d} \KK' \int_0^\infty \mbox{d} t  \int \mbox{d} \x \, \frac{\delta \left( |K|'-|K| \right)}{4\pi K^2} 
\\ \label{G2classical}
&&
W_{+Q}\left(\x,\KK' \!+\! \frac{\DK}{2},t \right) W_{-Q} \left(\x,-\KK' \!+\! \frac{\DK}{2},t \right) |2K'|,
\end{eqnarray}
where the total cross-section $\sigma_{tot} = 8\pi a^2$, whereas the wavevectors $\KK = \frac{\K_1-\K_2}{2}$ and $\Delta\mathbf{K} = \K_1+\K_2$. 
This formula resembles the production term in the collision integral of the Boltzmann equation, in which apropriate substitutions for the physical quantities are made.
Now, we take $W_{\pm Q}(\x,\K,t)$ to be the Wigner distribution:
\begin{eqnarray} \nonumber
W_{\pm Q}(\x,\K,t) &=& \frac{1}{(2\pi)^3} \int \mbox{d} \Delta \x \, 
\bar \psi_{\pm Q}^*\left( \x + \frac{\Delta \x}{2},t \right) 
\\ \label{Wigner}
& & \exp \left( i \K \Delta \x \right) 
\bar \psi_{\pm Q}\left( \x - \frac{\Delta \x}{2},t \right),
\end{eqnarray}
where we introduced
\begin{equation}\label{barpsi}
\bar \psi_{\pm Q}(\x,t) = \psi_{\pm Q}(\x,t) \exp \left(\pm i Qz  - i \frac{\hbar Q^2}{2m} t\right).
\end{equation}
The back to back correlation function is expressed, by Eq.~\eqref{G2classical} and Eq.~\eqref{Wigner}, in terms of the condensate wavefunctions only.

\subsection{Examples of $G^{(2)}_{bb}$ function}

Let us now calculate the anomalous density for some cases using the approximate analytical form
of $\psi_{\pm Q}$ given by the variational method presented in Appendix \ref{ApSol}. 
However, in the considerations below in this Section we assume that the condition in Eq.~(\ref{conditionNn2}) is satisfied, and so $B(\K,\x,t) \simeq B(\x,t)$. 
Using Eqs.~(\ref{vatnowe}), (\ref{parwar}), (\ref{Bp}), and (\ref{Mtdrugie}),  performing the spatial integral, and introducing the dimensionless time $\tau = \omega_r t$,
we arrive at the following formula
\begin{eqnarray} \nonumber
&& M(\KK, \DK )  
= A  \int_0^\infty \mbox{d} \tau \, 
\frac{\exp \left( -  \alpha^2 \tau^2 c_z(\tau) - i2\phi(\tau) \right)}{  (1-i\beta \tau) \sqrt{ c_z(\tau)}}  
\\ \label{Mgauss2}
&&
\exp \left(  i  \omega \tau - \frac{\Delta K_r^2\sigma_r^2(1+\tau^2)}{4 (1-i\beta \tau)} - 
\frac{\Delta K_z^2\sigma_z^2}{4 c_z(\tau)}\right),
\end{eqnarray}
where $\DK = \K_1+\K_2 $, $A = \frac{gN}{\hbar (2\pi)^3 \omega_r}$, $\alpha = \frac{\tau_{ex}}{\tau_c} = \frac{Qa_{hor}^2}{\sigma_z}$,
$\Delta K_r^2 = \Delta K_x^2 + \Delta K_y^2$; the dimensionless frequency is
\begin{eqnarray*}
\omega = \left( K^2 + \frac{\Delta K^2}{4} - Q^2 \right)a_{hor}^2,
\end{eqnarray*}
and the time dependent functions are
\begin{eqnarray*}
c_z(\tau) &=& 2\sigma_z^2 a_z(\tau) = \left( 1 - i \left( \beta -\frac{1}{\beta}  \right) \arctan \tau \right),
\\
\phi(\tau) &=& \left( \frac{7\beta}{4} -  \frac{3}{4\beta} \right) \arctan \tau.
\end{eqnarray*}
Now, we center $K$ at $Q$, introducing  $\delta K = K - Q$, and rewrite $\omega$ in terms of $\delta K$ as
$\omega = \left( 2Q\delta K \left( 1 + \frac{\delta K}{2Q} \right) +  \frac{\Delta K^2}{4}\right) a_{hor}^2$.
According to general consideration above  $\delta K \ll Q$, which leads to approximate form of 
$\omega $:
\begin{eqnarray*}
 \omega \simeq \left( 2Q \delta K  + \frac{\Delta K_r^2 + \Delta K_z^2}{4}  \right)a_{hor}^2.
\end{eqnarray*}
In Eq. (\ref{Mgauss2}) the term $\exp \left(-\frac{\Delta K_z^2\sigma_z^2}{4 c_z(\tau)} \right)$
implies that $|\Delta K_z \sigma_z|$ is maximally of the order of unity. This gives $ \frac{1}{4} \Delta K_z^2 a_{hor}^2$ 
present in the above formula to be maximally equal to $ \frac{a_{hor}^2}{\sigma_z^2} \ll 1$ and can be neglected.
As a result we end up with 
\begin{equation}\label{omegawzor}
 \omega \simeq \left( 2Q \delta K  + \frac{\Delta K_r^2}{4}  \right)a_{hor}^2.
\end{equation}
Now the anomalous density given by (\ref{Mgauss2}) and (\ref{omegawzor}) is a function of
dimensionless variables $\Delta K_r \sigma_r$, $\Delta K_z \sigma_z$ and $2 Q \delta K a_{hor}^2$ that
depends on two dimensionless parameters $\alpha$ and $\beta$.
In fact a complete analysis should investigate the anomalous density for all values of $\alpha$ and $\beta$.
However, this is in practice impossible. 
We therefore choose few examples for different values of the parameters.

The measurement of correlation function was already performed for metastable helium atoms in the Palaiseau group~\cite{paryz0} and is planned to be performed 
in the Vienna group~\cite{Wieden}.
In Appendix \ref{metahelium} we calculate $\alpha$ and $\beta$ present in the above formulas using the parameters of these experiments. 
We obtain $\alpha \simeq 0.22$, $\beta \simeq 3.3$ for the Palaiseau, and $\alpha \simeq 0.2$, $\beta \simeq 11$ for the Vienna setup.
In the experiment, the number of condensate atoms can be reduced by the use of radio frequency ``knife'' which results in the increase of $\alpha$ and decrease of $\beta$.
Thus, in all the calculations presented below we take $\alpha=0.2$ as the minimal value and $\beta=10$ as the maximal value.

\subsubsection{Fast collision}

\begin{figure}[t!]
  \includegraphics[clip, scale=0.35]{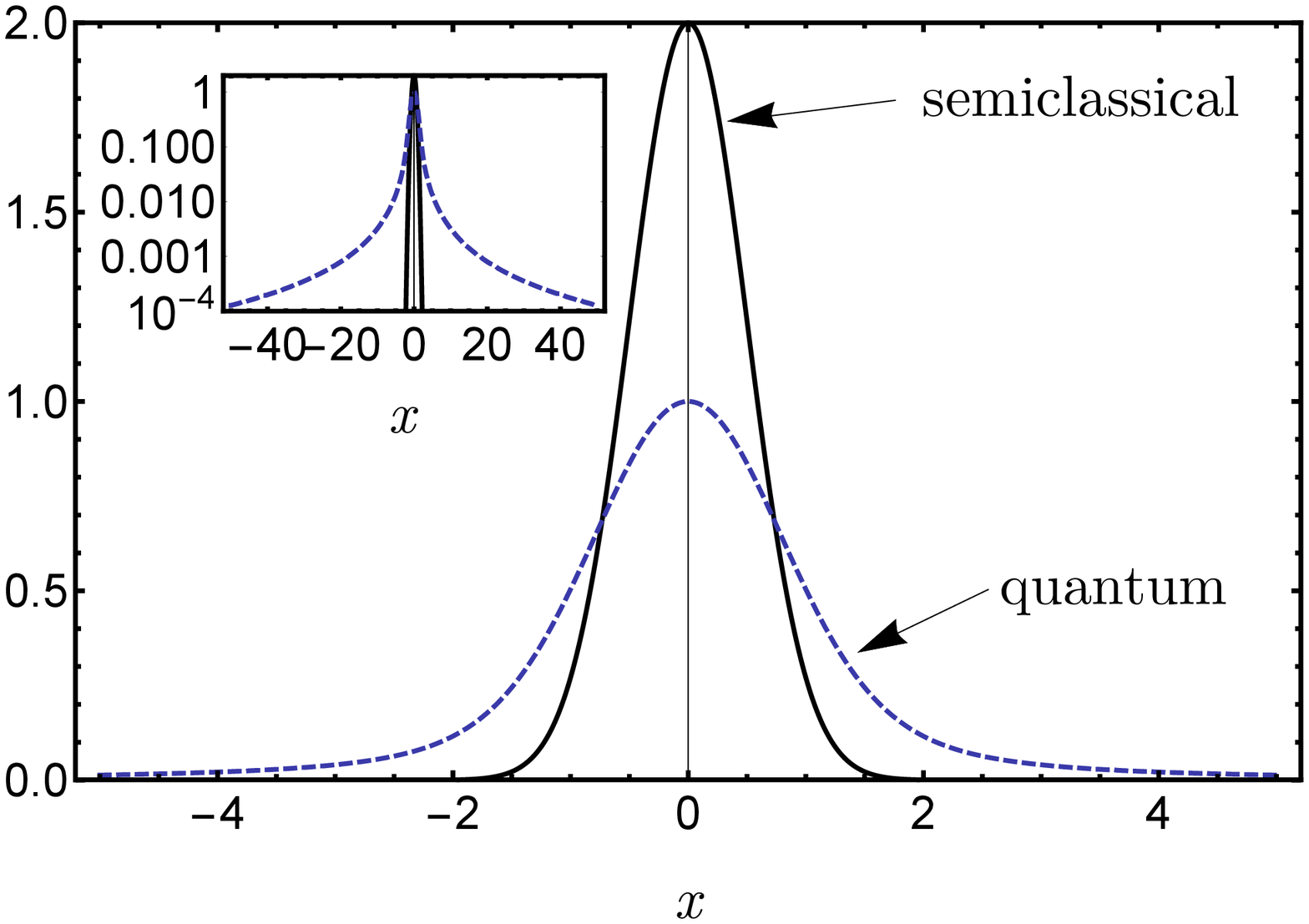}
  \caption{The figure shows $G_{bb}^{(2)}(\KK,\DK)$ (solid black), given by Eq.~\eqref{szybkieklasyczne}, and $|M(\KK,\DK)|^2$ (dashed blue), given by Eq.\eqref{szybkieWzor}, 
    that depends on $\delta K$. 
    The solid black line, ``semiclassical'', is given by $2\exp(-2x^2)$, 
    the dashed blue line, ``quantum'', shows $|\exp(-x^2)(1+\mathrm{erf}(ix))|^2$; the variable $x=\delta K\sigma_z$. 
    The inset presents the same plot but in logarithmic scale to expose the presence of long tails in the quantum case.}
  \label{G2clq}
\end{figure}

The ``fast collision'' case realize when the velocity of the wavepackets is large enough so that the change of the wavefunctions 
$\psi_{\pm Q}$, apart from movement along the collisional axis, is negligible during the whole collision. 
This means that $\tau_c$ is much smaller than all the other characteristic timescales.
As $\tau_r/2$ is the smallest of the characteristic times (apart from $\tau_c$) this condition can be restated as 
\begin{equation}\label{szybkieC}
\tau_r  \gg \tau_c \ \rightarrow \  \alpha \gg \beta   \   \rightarrow \ Q \gg \frac{\sigma_r^2 \sigma_z}{a_{hor}^4}.
\end{equation}
In this case the anomalous density reads
\begin{eqnarray} \nonumber
M(\K_1, \K_2 )  
&=&\frac{A\sqrt{\pi}}{2 \alpha} \exp \left( -\frac{ \Delta K_r^2\sigma_r^2 + \Delta K_z^2 \sigma_z^2}{4}  \right)  
\\ \label{szybkieWzor}
& & \exp \left(   -  \delta K^2 \sigma_z^2   \right) \left( 1+ \mbox{erf}(i \delta K \sigma_z) \right).
\end{eqnarray}
The details of the derivation can be found in Appendix~\ref{bbApp}.
Notice first that the width in $\Delta K_{r,z}$ of the $|M(\KK,\DK)|^2$ function is $\sqrt{2}$ larger than the 
momentum density width of $|\psi_{\pm Q}(\K,t)|^2 \propto \exp \left( - (k_x^2+k_y^2)\sigma_r^2 - k_z^2\sigma_z^2 \right) $.
Also, the width in $K$,  denoted by $\Delta_K$ and equal approxmately to $\sigma_z^{-1}$, is much smaller than the width in $\Delta K_r$, equal approximately to $\sigma_r^{-1}$. 
The analogous semiclassical expression for the $G^{(2)}_{bb}(\KK,\DK)$ function (calculated also in Appendix~\ref{bbApp})
reads:
\begin{eqnarray} \nonumber
G^{(2)}_{bb}(\KK,\DK) 
&=& 2 \left( \frac{A\sqrt{\pi}}{2\alpha} \right)^2  \exp \left( -\frac{\Delta K_r^2 \sigma_r^2+  \Delta K_z^2 \sigma_z^2 }{2}    \right)
\\ \label{szybkieklasyczne}
& & \exp \left( -  2 \delta K^2 \sigma_z^2\right).  
\end{eqnarray}
Comparing the above formula with $|M(\KK,\DK)|^2$ given by Eq.~(\ref{szybkieWzor}), we clearly see that the $\DK$ dependence is the same in both cases.
However, the semiclassical and quantum formulas differ in $K$ dependence. 
In Fig.~\ref{G2clq} we plot both the semiclassical dependence  $2 \exp( -2 x^2 )$, where $x = \delta K \sigma_z$, and the quantum dependence $| \exp(-x^2)(1+\mbox{erf}(ix))|^2$. 
We observe that the quantum dependence is wider with respect to the semiclassical one, and has a long tail which is absent in the semiclassical case. 
What is worth noticing both functions integrated over $\delta K$ give the same result,
i.e., $ \int \mbox{d} x \, 2 \exp( -2 x^2 ) = \int \mbox{d} x \,| \exp(-x^2)(1+\mbox{erf}(ix))|^2  $.

\subsubsection{Strong radial confinement}

Let us now consider another case when the mean-field energy $gn$ is much smaller than the kinetic energy along radial direction,  
$gn \ll \frac{\hbar^2}{2m \sigma_r^2}$.
According to Eq.~(\ref{gn}), in such ``strong radial confinement'' case $\sigma_r \simeq a_{hor}$ or, alternatively, $\beta \simeq 1$.
Then the anomalous density given by (\ref{Mgauss2}) and (\ref{omegawzor}) takes the  form:
\begin{eqnarray*} \nonumber
 M(\KK, \DK )  &=& A \exp\left( - \frac{\Delta K_r^2 a_{hor}^2 + \Delta K_z^2 \sigma_z^2}{4} \right) \times
\\
& & \times f_q(\delta K Q a_{hor}^2),
\end{eqnarray*}
where 
\begin{eqnarray*}
f_q(x,\alpha) =  \int_0^\infty \mbox{d} \tau \, \frac{\exp \left( -  \alpha^2 \tau^2 + i 2 x \tau \right)}{1+
i \tau }. 
\end{eqnarray*}
For this collision configuration, the semiclassical result reads
\begin{eqnarray*}\nonumber
 G^{(2)}_{bb}(\KK,\DK) &=& A^2 \exp\left( - \frac{\Delta K_r^2 a_{hor}^2 + \Delta K_z^2 \sigma_z^2}{2} \right)\times
\\
& & \times f_{cl}(\delta K Q a_{hor}^2),
\end{eqnarray*}
where 
\begin{eqnarray*}
 f_{cl}(x,\alpha) = \frac{\pi}{\alpha} \int_0^\infty \frac{\mbox{d}z }{\sqrt{\alpha^2+z}} \exp \left( -2 z - \frac{1}{2\alpha^2}(2x -z)^2   \right).
\end{eqnarray*}

We observe that the dependence on $\Delta \KK$ and $K$ decouples in the quantum as well as in the semiclassical model.  Additionally, the $\DK$ dependence is the same in
both models. The widths in $\Delta K_r$ and $\Delta K_z$ are equal approximately to $a_{hor}^{-1}$ and $\sigma_z^{-1}$, respectively.  As in the fast collision case, the
widths in $\Delta \KK$ of the $|M(\KK,\DK)|^2$ function are $\sqrt{2}$ larger than the momentum density widths, for which $|\psi_{\pm Q}(\K,t)|^2 \propto \exp \left( -
(k_x^2+k_y^2)a_{hor}^2 - k_z^2\sigma_z^2 \right) $.  The dependence on $\delta K$ of $G^{(2)}_{bb}(\KK,\DK) = | M(\KK, \DK )|^2 $ in both models is determined by
$|f_q(x,\alpha)|^2$ and $f_{cl}(x,\alpha)$ functions (where $x = \delta K Q a_{hor}^2$), respectively. We note here that, as in the fast collision case, $f_q$ and $f_{cl}$ satisfy
normalization condition: $ \int\! \mbox{d} x \, |f_q(x,\alpha)|^2 = \int \mbox{d} x \, f_{cl}(x,\alpha)$.

\begin{figure}[t!]
  \includegraphics[clip, scale=0.35]{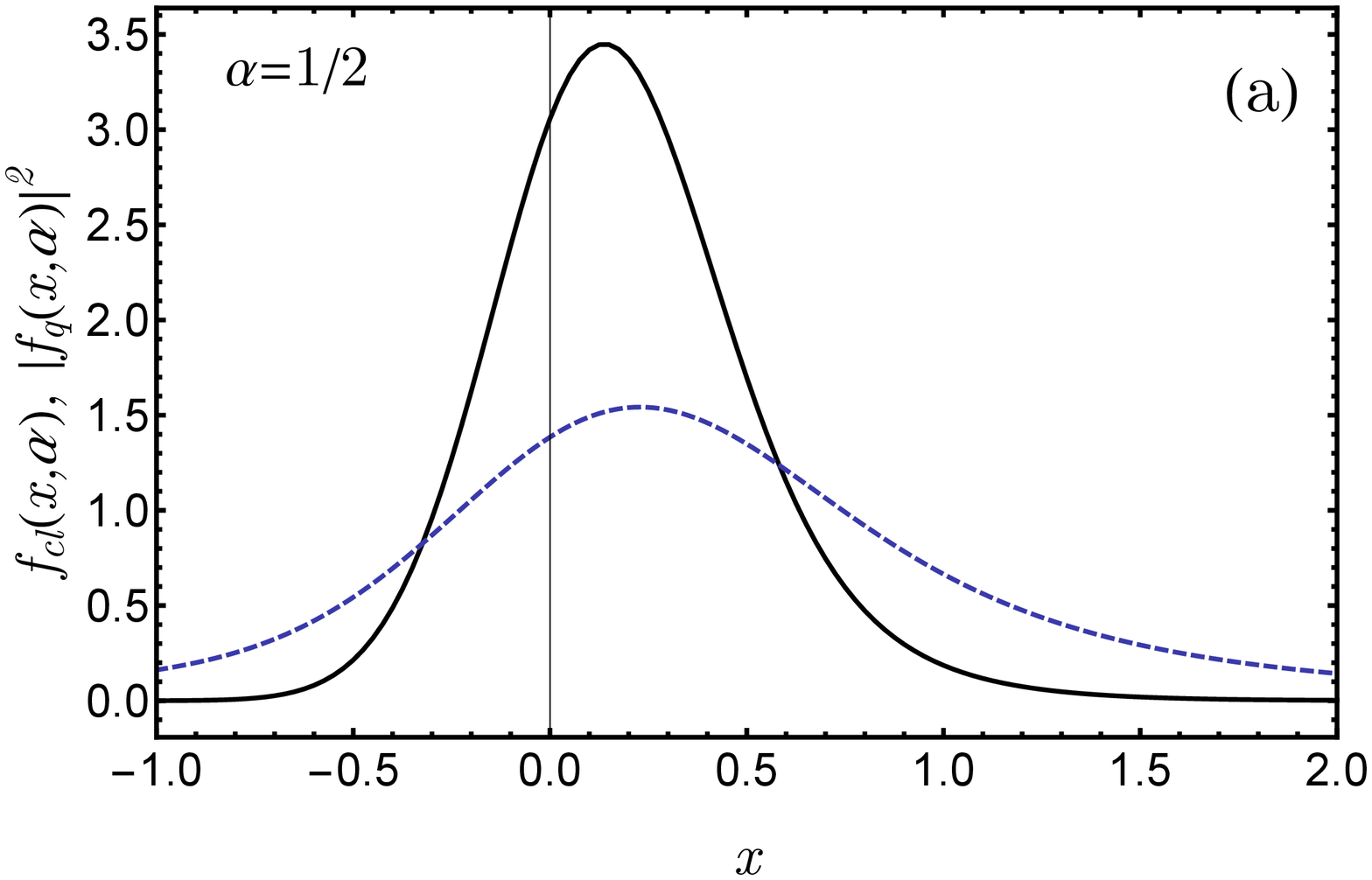}\\
  \includegraphics[clip, scale=0.35]{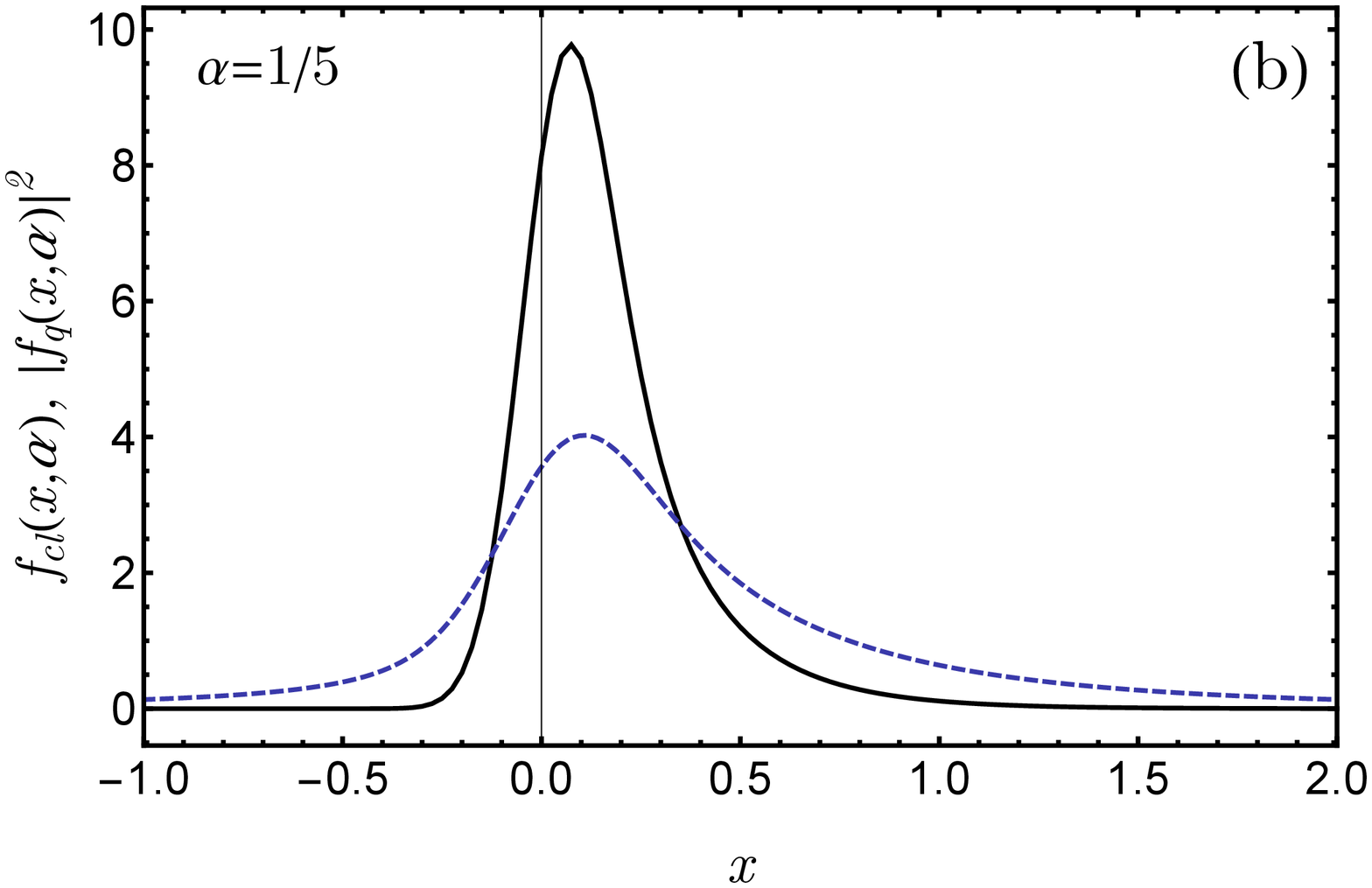}\\
  \includegraphics[clip, scale=0.35]{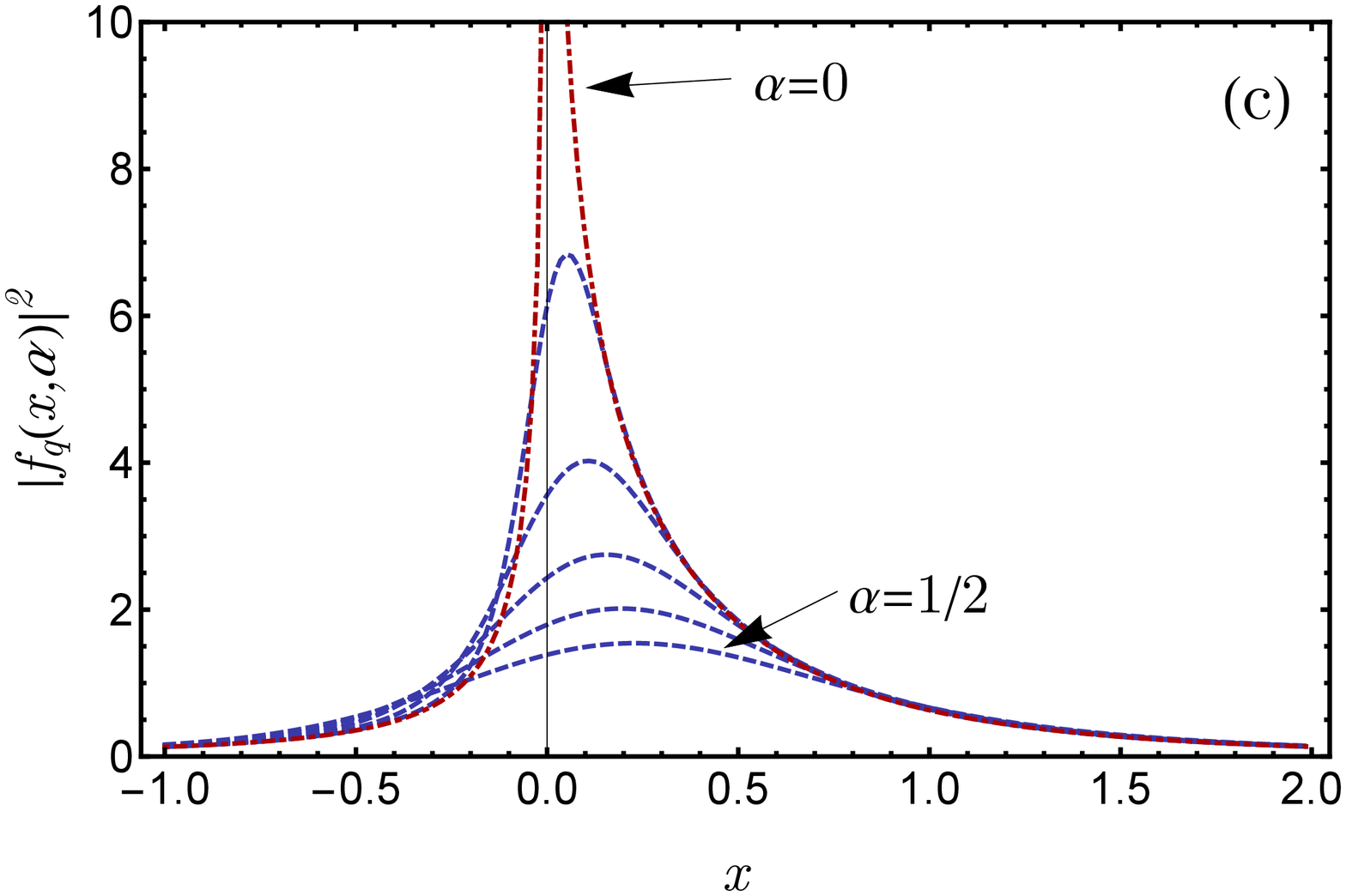}\\
  \caption{The functions $f_{cl}(x,\alpha)$ (solid black) and $|f_q(x,\alpha)|^2$ (dashed blue) as a function of $x$ for two different values of 
    parameter: $\alpha=1/2$ (panel a) and $\alpha=1/5$ (panel b).
    Panel c shows function $|f_q(x,\alpha)|^2$ for $\alpha=0.5,0.4,0.3,0.2,0.1$ (from bottom to top). 
    The limiting function for $\alpha=0$ is shown in dotted-dashed red.
  }
  \label{fclq}
\end{figure}

The two function $f_q$ and $f_{cl}$ depend on parameter $\alpha$, the ratio of the expansion time to the collision time.  The case $\alpha \gg 1$ describes the fast
collision analyzed above, and, therefore, we focus only on $\alpha < 1$.  As mentioned previously, the experimentally important minimal value of $\alpha$ is $0.2$.  In
Fig.~\ref{fclq} (panels a and b), we plot the functions $|f_q(x,\alpha)|^2$ and $f_{cl}(x,\alpha)$ for two values $\alpha =1/2, 1/5$.

As we see from Fig.~\ref{fclq}, in strong radial confinement case the differences between the semiclassical and quantum model are similar to that in the fast collision.
As before the quantum function is wider than the semiclassical one.  However in this case, there is a shift of the position of the maximum to positive values of $\delta K$, which is, however, much smaller than the width in $\delta K$.  
The curves $|f_q(x,0.2)|^2$ and $|f_q(x,0.5)|^2$, as can be seen in Fig.~\ref{fclq} (panel c), approach each other at a certain point.  The universal curve to which all
$|f(q,\alpha)|^2$ converge is $|f_q(x,0)|^2$.  Both curves are almost the same for $x> 1.7 \alpha$, but for smaller $x$ they start to differ. For small value of $x$, 
$|f_q(x,0)|^2 \propto \log^2 (2x)$, and thus tends to infinity as $x \to 0$.  

The maximum of $|f(q,\alpha)|^2$ grows with its position tending to zero as $\alpha$ gets smaller.  Thus, we cannot choose halfwidth as the characteristic width of
$|f(q,\alpha)|^2$.  Instead, we define it as the value of $x_0$ for which the normalized integral under the curve is $1/2$, i.e,
\begin{eqnarray*}
\frac{\int_{-x_0}^{x_0} \mbox{d}x \, |f_q(x,\alpha)|^2 }{ \int_{-\infty}^{\infty} \mbox{d}x \, |f_q(x,\alpha)|^2  } = \frac{1}{2}.
\end{eqnarray*}
This definition is motivated by the fact that the detectors, on which particles fall, have finite sizes. The measurement of two particle correlation function is always
accompanied by integration of that quantity over the size of the detector. As $x \propto \delta K$, which is directly related to the position of the detectors, the
measurement results in integration over $x$.  Using the above formula, we find $x_0=0.26$ for $\alpha = 0$ (universal curve), $x_0=0.32$ for $\alpha=0.2$
(the smallest realistic value considered in the paper), $x_0 = 0.60$ for $\alpha=0.6$, and  $x_0=0.86$ for $\alpha=1$. 
We see that these values are  of the same order. 

As a consequence, the characteristic width $\Delta_K$ for $\alpha < 1$ is approximately equal to $1/2Q a_{hor}^2$ while the position of the maximum, denoted as $\delta
K_{max}$, is always smaller than $1/2 Qa_{hor}^2$.  For $\alpha \rightarrow \infty$, we obtain the fast collision case where the width of $\delta K$ is given by
$1/\sigma_z$, and this turns out to be true for $\alpha > 1$.  According to the condition given by Eq.~(\ref{Cond1}) and the fact that $\sigma_z \gg \sigma_r$ both of
this widths are much smaller than $1/\sigma_r$. Therefore, for all considered values of $\alpha$, the width in $\delta K$ is much smaller than the width in $\Delta K_r$
equals approximately to $1/\sigma_r$.

\subsubsection{Largest mean-field energy impact}

According to Eq. (\ref{gn}) mean field energy divided by characteristic kinetic energy
connected with the radial confinement $\frac{\hbar^2}{m \sigma_r^2}$ is  proportional
to $\beta^2 -1$. Thus the larger the mean field energy (we mean $gn /(\hbar^2/m\sigma_r^2)$) 
the larger is $\beta$. On the other hand the effective time of integration is given by $1/\alpha$.
The mean field energy impact shall be the largest for largest possible value of $\beta$ and
smallest possible value of $\alpha$. Therefore we call such case "largest mean-field energy impact".

\begin{figure}[t!]
  \includegraphics[clip, scale=0.40]{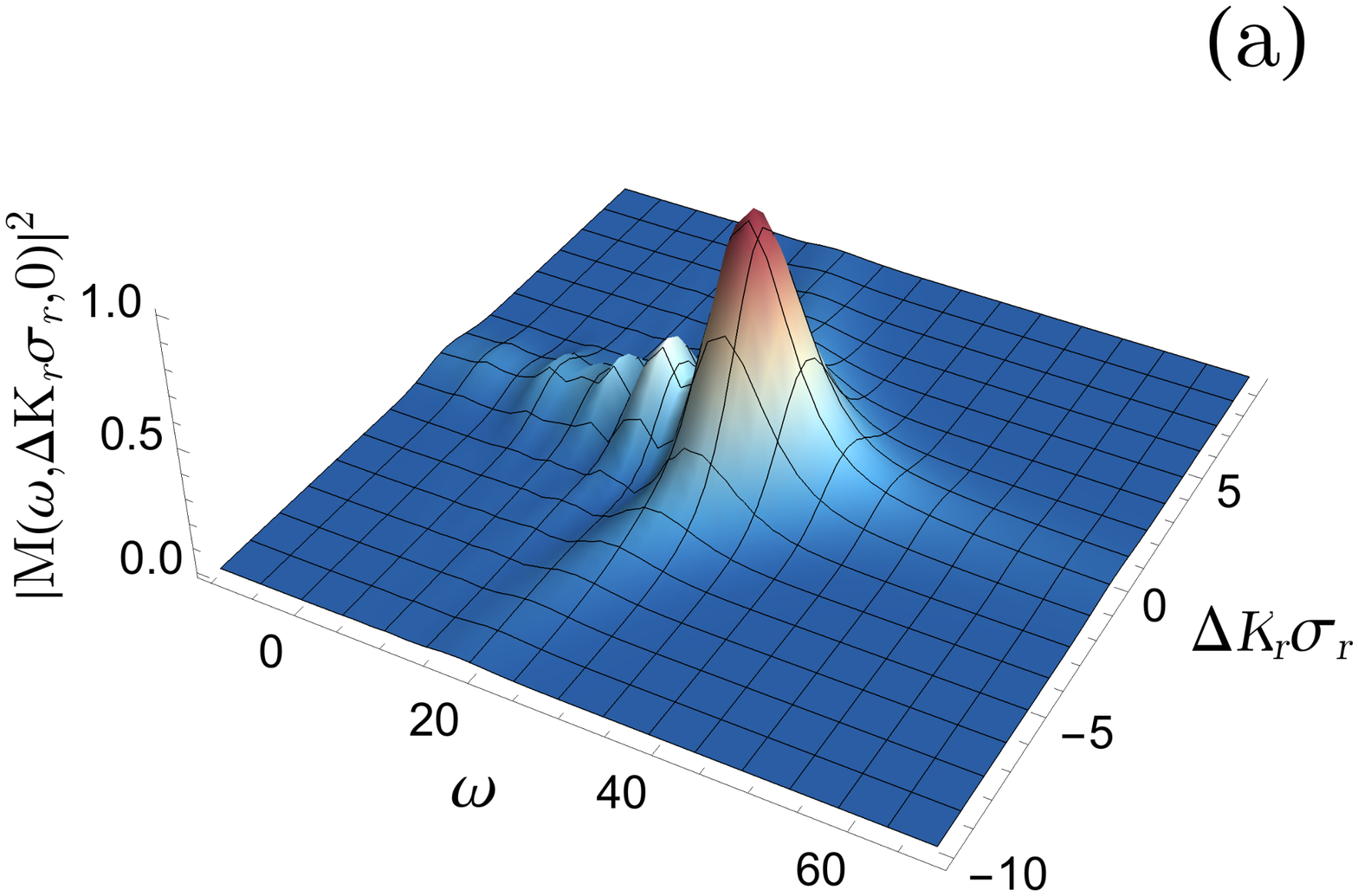}\\
  \includegraphics[clip, scale=0.40]{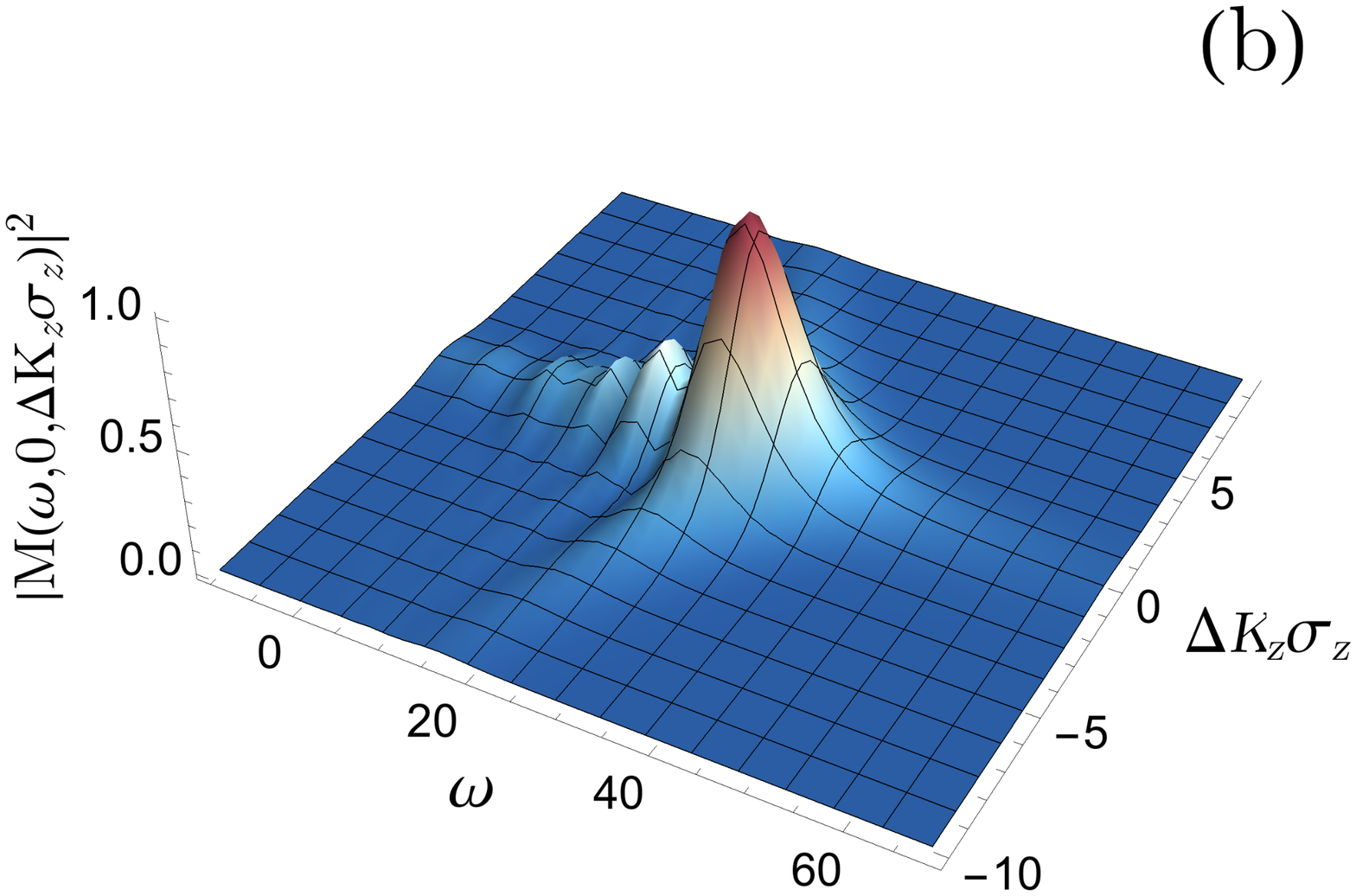}
  \caption{Normalized to maximum cuts of the function $|M(\omega,\Delta K_r \sigma_r,0)|^2$ (panel a) and $|M(\omega,0,\Delta K_z \sigma_z)|^2$ (panel b)
    calculated with formula given by Eq.~\eqref{Mgauss2} for $\alpha=0.2$ and $\beta=10$.
  }
  \label{M2cuts}
\end{figure}

As discussed above, we take specific values of $\alpha=0.2$ and $\beta=10$ as extremal values that are experimentally feasible.  
The anomalous density given by
Eq.~(\ref{Mgauss2}) is a function of $\omega$, $\Delta K_r \sigma_r$ and $\Delta K_z \sigma_z $, and can be written as $|M(\omega,\Delta K_r \sigma_r,\Delta K_z
\sigma_z)|^2$, that explicitly depends on three independent parameters.  In Fig.~\ref{M2cuts} we plot its cuts $|M(\omega,\Delta K_r \sigma_r,0)|^2$ and
$|M(\omega,0,\Delta K_z \sigma_z)|^2$. We observe that the maximum is for $\omega=25$ and $\Delta K_r = \Delta K_z = 0$.  The characteristic width in $\Delta K_z$ and
$\Delta K_r$ is approximately $2\sigma_z^{-1}$ and $2\sigma_r^{-1}$, respectively. The characteristic halfwidth in $\omega$ is roughly equal to $6$.  With this values,
the term $\Delta K_r^2 a_{hor}^2/4$ in Eq.~(\ref{omegawzor}) for $\omega$ can be estimated to be maximally
\begin{eqnarray*}
\frac{\Delta K_r^2 a_{hor}^2}{4} = \frac{a_{hor}^2}{\sigma_r^2} = \frac{1}{\beta} 
\end{eqnarray*} 
As $\beta = 10$  this term is much smaller than unity and, as the width in $\omega$ is $6$, can be neglected. 
This results in $\omega \simeq  2\delta K Q a_{hor}^2$, and thus in the halfwidth in variable $\delta K$ equal to ${3}/{Q a_{hor}^2}$.
According to the condition Eq.~(\ref{Cond1}), it is much smaller than $1/\sigma_r$. Note, that the shift of the maximum in $\omega$ is larger than the width in $\omega$.

\subsubsection{The limiting case of $\alpha \to 0$ and $\beta \to \infty$}

Here, we analyze the limiting case for which $\beta \to \infty$ and $\alpha \to 0$. The most important contribution to the integral in Eq.~(\ref{Mgauss2}) comes from the times
$\tau \ll 1$. The anomalous density, given by Eq.~(\ref{Mgauss2}), takes then the following form
\begin{eqnarray} 
\label{EqM2lim}
 M(\KK, \DK )  
&=& \frac{A}{\beta}  \int_0^\infty \mbox{d} \tau \, 
\frac{\exp \left( i \left( \frac{\omega}{\beta} - \frac{7}{2}  \right) \tau \right)}{  (1-i\tau)^{3/2}}  \times
\\ \nonumber
&&
\times \exp \left( - \frac{\Delta K_r^2\sigma_r^2 + \Delta K_z^2\sigma_z^2}{4 (1-i \tau)}  \right),
\end{eqnarray}
in which we changed the variable from $\beta \tau$  to $\tau$.

The anomalous density $M$ in this situation is a function of $\tilde \omega = \left( \frac{\omega}{\beta} - \frac{7}{2} \right)$ and $\Delta^2 = \Delta K_r^2\sigma_r^2 +
\Delta K_z^2\sigma_z^2$, so the number of parameters were significantly reduced as compared to the general formula given in Eq.~(\ref{Mgauss2}). 

\begin{figure}[t!]
  \includegraphics[clip, scale=0.40]{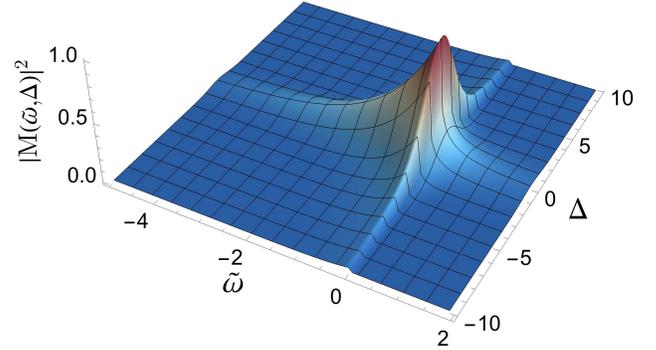}
  \caption{
    The function $|M(\tilde\omega,\Delta)|^2$ calculated with formula given by Eq.~\eqref{EqM2lim}.
  } \label{M2lim}
\end{figure}

In Fig.~\ref{M2lim} we plot $|M(\tilde \omega, \Delta)|^2$, from which we observe that the width in $\Delta$ at the maximum value of $\tilde \omega \simeq 0$ is equal to $2$.
The shape of the function in $\tilde \omega$ is highly asymmetric. For $\tilde \omega > 0$ the function rapidly decays, while for $\tilde \omega < 0$ the decay is much slower, 
with a width of the tail approximately equal to unity.  

Remembering that
\begin{equation}\label{omegawz}
\omega = \beta \tilde \omega + \frac{7}{2}\beta,
\end{equation}
we notice that the maximum of $\omega$ is shifted to $7\beta/2$, and the width approximately equals $\beta$.
In the same way as in largest mean-field energy impact case, we estimate in Eq.~(\ref{omegawzor}) the term ${\Delta K_r^2 a_{hor}^2}/{4} \simeq {1}/{\beta}$.
As we see, it can be neglected yielding
\begin{equation}
\omega \simeq 2\delta K Q a_{hor}^2.
\end{equation}
Substituting these results into Eq.~(\ref{omegawz}), we obtain the position of the maximum, $\delta K_{max}$, and the width $\Delta_K$:
\begin{equation}\label{bmale}
\delta K_{max} \simeq \frac{7}{4} \frac{\sigma_r^2}{Q a_{hor}^4} 
\quad \mathrm{and} \quad
\Delta_K \simeq \frac{\sigma_r^2}{Q a_{hor}^4}.
\end{equation}
Due to condition (\ref{conditionNn2}), both $\delta K_{max}$ and $\Delta_K$ are much smaller than the width in $\Delta K_r \simeq 2\sigma_r^{-1}$.

Finally, now we comment on the results ofthe previous example for which $\beta=10$. The shift of the maximum was equal to $25$, which is smaller than $7\beta/2 = 35$.
The width in $\omega$ was found to be equal to $6$, which is quite close to $10$ as predicted above. The small discrepancies are caused by the fact that $\beta$ is still
not large enough.

\subsection{General correlation properties }

Above we have shown few examples of the anomalous density. 
We saw that in the fast collision and strong radial confinement cases the $K$ and $\DK$ dependence 
decoupled. In the tw other cases as seen in the figures the $K$ and $\DK$ dependence almost decouple.
the general observation (although not strict) is that for constant $\beta$ this dependence 
tends to decouple better while enlarging $\alpha$. It can be see even in the fast collision case
when $\alpha \gg \beta$ and the dependence is fully decoupled.
The additional observation is that the width in $\DK$ gets smaller with enlargement of $\alpha$.
This change is the largest for $\beta \gg 1$ when in the case the case of small $\alpha$
the width in $\Delta K_{r,z}$ is roughly $2/\sigma_{r,z}$ and goes down to $1/\sigma_{r,z}$
for $\alpha \gg \beta$ which is the fast collision case.

\subsubsection{Properties of averaged function $\int\!\! \mbox{d}\KK \ G^{(2)}_{bb}(\KK,\DK) $}

The fact that in the two of the analyzed examples the integrals over $\delta K$ gave the same result for both quantum and semiclassical model suggests that it may be a
general property.  This is indeed the case and in Appendix~\ref{bbApp} we prove that $\int \mbox{d}\KK \ G^{(2)}_{bb}(\KK,\DK) $ is the same for both classical and
quantum models as long as  $B(\K,\x,t) \simeq B(\x,t) $. 

Now, let us investigate the properties of the averaged back to back correlations.  In Appendix~\ref{bbApp} we show that under condition stated in Section~\ref{model} it
takes the following form
\begin{eqnarray} \nonumber
&&\int \mbox{d} \KK \, |M(\KK,\DK)|^2 \simeq  \frac{ \pi m Q }{ (2\pi)^6\hbar^3} 
\\ \nonumber
&&
\int_{0}^\infty \mbox{d} t \int \mbox{d} \Omega_\KK \left| \int \mbox{d} \x \,  
 \exp \left(- i \DK \cdot \x  \right)  B(\KK,\x,t) \right|^2,
\end{eqnarray}
where $\Omega_\KK$ denotes the solid angle coordinates of $\KK$ vector.  Note, that the width in $\Delta \KK$ is given by the width of the $B(\KK,\x,t)$ 
averaged over time and the directions of $\KK$.  
Let us now turn to the regime for which $B(\K,\x,t) \simeq B(\x,t)$. 
Emploing Eq.~(\ref{Bp}) we find that
\begin{eqnarray} \nonumber
&&\int \mbox{d} \KK \, |M(\KK,\DK)|^2 \simeq  \frac{ \pi m Q }{ (2\pi)^6\hbar^3} 4\pi (2g)^2 \times
\\ \nonumber
&& 
\times \int_{0}^\infty \mbox{d} t \left| \int \mbox{d} \x \,  
 \exp \left(- i \DK \cdot \x  \right)  \psi_{+Q}(\x,t)\psi_{-Q}(\x,t) \right|^2.
\end{eqnarray}
From this formula, we see that the width of $\int \mbox{d} \KK \, |M(\KK,\DK)|^2 $ in $\DK$ is directly related to 
the momentum width of $\psi_{+Q}\psi_{- Q}$ averaged over time.
This relation can be evaluated exactly in gaussian ansatz the case, for which we have Eq.~(\ref{vatnowe}).
In Appendix~\ref{bbApp} we prove that
\begin{eqnarray} \nonumber
&& \int\!\! \mbox{d} \KK \, |M(\KK,\DK)|^2  \propto \int_0^\infty \!\!\mbox{d} t \, 
\frac{1}{\sigma_r^2(t)\sigma_z} \exp \left( - 2\frac{v_0^2t^2}{\sigma_z^2}  \right)\times
\\ \label{momden}
&& \quad\quad\quad\quad\quad\quad\quad\times\left|\psi_{+ Q} \left(\frac{\DK}{\sqrt{2}},t \right)\psi_{- Q} \left(\frac{\DK}{\sqrt{2}},t \right)\right|^2,
\end{eqnarray}
where $\psi(\K,t)$ is the Fourier transform of the wavefunction, $\int \mbox{d} \x \, e^{-i\K\x} \psi(\x,t)$.  The width of averaged $|M|^2$ is increased by a factor of $\sqrt{2}$ 
with respect to $\psi_{\pm Q}$ momentum density.  The same effect is visible in the two of the above calculated examples: the fast
collision and the strong axial confinement cases.  

Let us now analyze the shape of the averaged function. After the integration over $\KK$ it depends on $\Delta K_r$ and $\Delta K_z$.  To obtain
dependence on a single parameter only, we perform additional integration over variable $\Delta K_z$.  In Appendix~\ref{bbApp} we show that the final functions takes the following form
\begin{eqnarray}
&& \int \mbox{d} \Delta K_z  \int \mbox{d} \KK \, |M(\KK,\DK)|^2  \simeq C_b \int_{0}^\infty \frac{\mbox{d} \tau}{ 1+ \tau^2  } \times
\nonumber \\ 
&& 
\label{Eq:avrM2}
\quad \times \exp\left( - 2 \frac{\alpha^2}{\beta^2} \tau^2 - \frac{ \Delta K_r^2 \sigma_r^2}{2} 
\frac{1+ \tau^2/\beta^2}{1+ \tau^2 }  \right).
\end{eqnarray}
The evaluation of this integral requires addition (over time domain) of gaussian functions $\exp( - { \Delta K_r^2 \sigma_r^2}/{2 w^2(\tau) })$
with a time depended width $w(\tau) =  \surd[{{(1+\tau^2)}/{(1 + \tau^2/\beta^2)}}]$ 
and weight $\exp( - 2 {\alpha^2 \tau^2}/{\beta^2})/(1+ \tau^2) $. Below, we investigate this function in more details.

Note first, that in the case $\beta \simeq 1$, the width $w(\tau) \simeq 1$ and the averaging of the correlation function yields $ \exp( - { \Delta K_r^2
  \sigma_r^2}/{2})$.  This result can be seen in the calculation of the anomalous density in the fast collision and strong radial confinement cases.  The obtained gaussian
function is the narrowest one of all the possibilities.  In all the other cases $\beta > 1$ and the width $w(t)$ grows in time.  The widest possible functions in
effective variable $\frac{ \Delta K_r^2 \sigma_r^2}{2} $ is attained in the limit $\beta \rightarrow \infty$, and reads
\begin{equation} \label{dawson}
\int_{0}^\infty \!\!\mbox{d} \tau\,
\frac{ e^{  - \frac{ \Delta K_r^2 \sigma_r^2}{2(1+\tau^2)} } }{1+ \tau^2} 
= \frac{\sqrt{2}}{|\Delta K_r|\sigma_r} \mbox{F}\left( \frac{|\Delta K_r|\sigma_r}{\sqrt{2}} \right),
\end{equation}
where $F(x) =  (-i {\sqrt{\pi}}/{2}) \exp (-x^2) \mbox{erf}(ix)$ is the Dawson integral.

\begin{figure}[t!]
  \includegraphics[clip, scale=0.35]{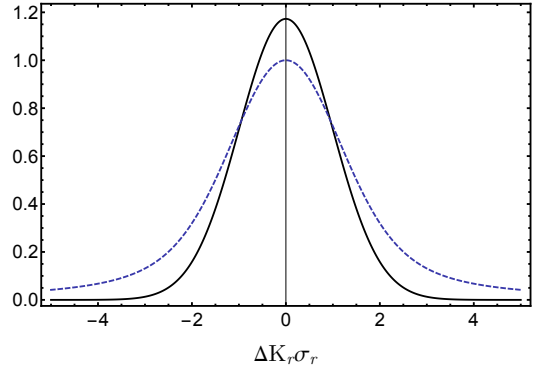}
  \caption{
    The integral given in Eq.~\eqref{Eq:avrM2} for $\alpha=0.2$ and the narrowest case $\beta=1$ as a function of $\Delta K_r \sigma_r$ 
    (solid black line). The widest possible integral given by Eq.~\eqref{dawson} is plotted in dashed blue.
  } \label{fig:avrM2}
\end{figure}

In Fig.~\ref{fig:avrM2} we plot the function given in Eq.~\eqref{dawson}, together with the one in Eq.~\eqref{Eq:avrM2} for the narrowest function case, $\beta = 1$.  We
notice that the width of the function given by (\ref{dawson}) is about $3/2$ larger than that of gaussian function.  The large difference between these functions results
from the tail of the Dawson integral.  In Appendix~\ref{timeevolution} we show that the axial momentum width of $\psi_{\pm Q}$ grows in time from $\frac{1}{\sigma_r} $
to $\frac{1}{\sigma_r} \frac{\sigma_r^2}{a_{hor}^2} = \frac{\beta}{\sigma_r}$. When $\beta \gg 1$, the momentum width grows substantially.  As seen in expression from
Eq.~(\ref{momden}) the shape of $\int\! \mbox{d} \KK \, |M(\KK,\DK)|^2$ is directly related with the momentum density of the wavefunctions $\psi_{\pm Q}$ integrated over
time.  The long tail seen in Fig.~\ref{fig:avrM2} results from the addition of the momentum densities with the width growing in time.

\subsubsection{Properties of $ G^{(2)}_{bb}(K,\DK)$ in variable $K$}

As we have seen above in all of the analyzed examples, the width in $\delta K$ is much smaller than the width in $\Delta K_r \simeq \sigma_r^{-1}$.
Here, we give a simple explanation of this fact by analyzing Eqs.~(\ref{Mkw}) and~(\ref{Bp}).

The  temporal dependence of the integrand in these equations is given by
\begin{eqnarray*}
&& \frac{\hbar( k_1^2 + k_2^2 - 2Q^2)}{2m} t  - 2 \phi(t) =
\\
&& = \frac{\hbar}{m}\left( 2 Q \delta K + \delta K^2 + \frac{\Delta K^2}{4}  \right) t - 2 \frac{\mu}{\hbar} t,
\end{eqnarray*}
where $\phi(t)$ is the spatially independent phase of the $\psi_{\pm Q}$ functions approximately given by $\phi(t) \approx \frac{\mu}{\hbar} t$, where $\mu$ is the
chemical potential.  As analyzed above $|\delta K| \ll Q$ so the term $\delta K^2$ can be neglected.  Additionally, we take $\Delta \KK=0$.  The the shift of the maximum
$\delta K_{max}$ is equal to $\frac{\hbar}{m} 2Q \delta K_{max} - 2 \frac{\mu}{\hbar} =0 $, and the width $\Delta_K$ can be estimated by setting $\frac{\hbar}{m} 2 Q
\Delta_K \tau_d = 1$, where $\tau_d$ is some characteristic time of the process. Consequently, we arrive at
\begin{equation} \label{1234}
\delta K_{max} =  \frac{\mu m}{\hbar^2 Q} 
\quad \mathrm{and} \quad
\Delta_K = \frac{m}{2 \hbar Q \tau_d}.
\end{equation}
The widest possible $\delta K$ is reached for the shortest $\tau_d$.  The characteristic times described in the Section~\ref{model} provide the shortest timescale
represented by 
\begin{equation} \label{taud}
 \tau_d = \mbox{min}\left(\tau_c,\frac{\tau_r}{2} \right).
\end{equation}
Substituting the above in Eq. (\ref{1234}), we arrive at
\begin{eqnarray} \label{shift}
&& \delta K_{max} =  \frac{1}{Q a_{hor}^2} \left( \frac{7}{4} \beta - \frac{3}{4\beta} \right),
\\ \label{widK}
&& \Delta_K =  \mbox{max} \left(  \frac{1}{2\sigma_z}, \frac{\sigma_r^2}{Q a_{hor}^4} \right) 
= \frac{1}{Qa_{hor}^2} \mbox{max} \left( \frac{\alpha}{2},\beta\right),
\end{eqnarray}
where we made use of Eq.~(\ref{muvar}).

Let us see, whether these formulas agree with the examples presented above remembering that they are only rough estimation of true values.
In the fast collision case the approximate formula (\ref{shift}) predicts nonzero value of 
$\delta K_{max}$ while Eq.~(\ref{szybkieWzor}) gives
$\delta K_{max}=0$. However due to condition given by Eq.~(\ref{szybkieC}), $\delta K_{max}
\ll {1}/{\sigma_z}$, which is the $\delta K$ width.
This means that within approximations undertaken in the paper
both formulas give the same value.
The width in $\delta K$ predicted by (\ref{widK})
is $\Delta_K = {1}/{2\sigma_z}$ while the value given by (\ref{szybkieWzor}) is about twice larger.

In the strong radial confinement case the above formulas give $\delta K_{max} =
{1}/{Qa_{hor}^2}$ and $\Delta_K = {1}/{Qa_{hor}^2}$. 
While comparing them with true values we notice that the
scaling  $\frac{1}{Qa_{hor}^2}$ is correct. 
The difference is the prefactors which in fact are smaller than predicted by Eq. (\ref{shift}) and (\ref{widK}).
The same situation happens in the third example where the
width predicted by Eq. (\ref{widK}) is about twice larger then the true value while the true shift is about
30 percent smaller than  given by Eq. (\ref{shift}). 
In the last example, both the shift and width are correctly predicted by expressions in Eq.~(\ref{shift}) and~(\ref{widK}).  
To conclude, the above formulas are in fair agreement with all analyzed examples.

We remark, that the simple derivation of the shift $\delta K_{max}$ can be also interpreted as energy conservation law during the collision of two particles. Specifically, we
have two particles with incoming kinetic energy ${\hbar^2 Q^2}/{2m}$, they collide and leave the condensate. During the process, each particle gain additional energy equal to
the chemical potential. Finally, their kinetic energies (due to the assumption  $\Delta \KK=0$ they are the same for both particles) are ${\hbar^2 (Q + \delta
  K_{max})^2}/{2m}$. The energy conservation law requires
\begin{eqnarray*}
\frac{\hbar^2 (Q + \delta K_{max})^2 }{2m} = \frac{\hbar^2 Q^2}{2m} + \mu,
\end{eqnarray*}
which, after omission of the term $\delta K_{max}^2$, leads to Eq.~(\ref{shift}) as expected.

The width in $\DK$ is given by the spatial Fourier transform, the width in axial direction is $\sigma_r$. Therefore, the minimal width in $\Delta K_r$ is
${1}/{\sigma_r}$.  According to Eq.~(\ref{conditionNn2}), both $\delta K_{max}$ and $\Delta_K$, given by Eqs.~(\ref{shift}) and~(\ref{widK}), are much smaller than the
width in $\Delta K_r$.

The above analysis was based on quantum considerations. It is instructive to present a simple classical reasoning, though. We denote by $\sigma_{kr,kz}$ the widths of
the momentum densities of $\psi_{\pm Q}$.  The momenta of the two atoms, before the collision, may be written as
\begin{eqnarray*}
\K_1' &=& (Q + c_{1,z}\sigma_{kz}) {\bf e}_z +  \sigma_{kr}(c_{1,x} {\bf e}_x + c_{1,y} {\bf e}_y),
\\
\K_2' &=& - (Q + c_{2,z}\sigma_{kz}) {\bf e}_z +  \sigma_{kr}(c_{2,x} {\bf e}_x + c_{2,y} {\bf e}_y),
\end{eqnarray*}
where the coefficients $|c|\leqslant 1$. We define $\KK' = \frac{\K_1'-\K_2'}{2}$ and $\DK' = \K_1'+\K_2'$.
The energy and momentum conservation laws require
\begin{eqnarray*}
\DK' = \DK \ \ \ \ \ \ |\KK| = |\KK'|.
\end{eqnarray*}
Substituting $\K_1'$ and $\K_2'$ into the above equations, we obtain
\begin{eqnarray*}
\DK &=& C_x\sigma_{kr}{\bf e}_x + C_y \sigma_{kr} {\bf e}_y +C_z \sigma_{kz} {\bf e}_z    
\\
4K^2 &=&  \left(2Q + \Delta c_z \sigma_{kz}  \right)^2 + \left( \Delta c_x^2  + \Delta c_y^2 \right) \sigma_{kr}^2
\end{eqnarray*}
where the new coefficients are $C_j = c_{1,j} + c_{2,j}$ and $\Delta c_j = c_{1,j} - c_{2,j}$, with $j=x,y,z$.
The second of the equations can be rewritten as
\begin{eqnarray*}
8 Q \delta K + 4 \delta K^2 =  4 Q \Delta c_z \sigma_{kz} + \Delta c_z^2\sigma_{kz}^2 + \left( \Delta c_x^2  + \Delta c_y^2 \right) \sigma_{kr}^2,
\end{eqnarray*}
where $\delta K = K-Q$. Neglecting $\delta K^2$ term we obtain 
\begin{eqnarray*}
\delta K  \simeq  \frac{\Delta c_z}{2} \sigma_{kz} +  \frac{1}{8Q} \left(\Delta c_z^2\sigma_{kz}^2 + \left( \Delta c_x^2  + \Delta c_y^2 \right) \sigma_{kr}^2 \right).
\end{eqnarray*}
The constraints $|\Delta c_j|\leqslant 2$ restrict the values of $\delta K$, yielding the width in $\delta K$ to be approximately equal to
\begin{eqnarray*}
\Delta_K  \simeq  \sigma_{kz} +  \frac{2\sigma_{kr}^2 + \sigma_{kz}^2 }{2Q}. 
\end{eqnarray*}
As $Q \gg \sigma_{kr} $ and $\sigma_{kr} \gg \sigma_{kz}$, the with is much smaller than the width in $\Delta K_r \simeq 2 \sigma_{kr}$. Thus, as in the quantum model the
width in $\delta K$ is much smaller than the width in $\Delta K_r$.

\subsubsection{Correlation volume}\label{velw}

Here we show that the fact $\Delta_K$ is much smaller than the width in $\Delta K_r$ has crucial consequences for the particle correlation properties, i.e., for the correlation volume.
To this end, we analyze measurement of two atoms, one at $\K$ and another at $\K'$. We define the correlation volume, as the volume for which $\K'$, with $\K$ being constant, 
the particles are still significantly correlated. 
In order to calculate this quantity, we first find $\K_0'$ for which $G^{(2)}_{bb}(\KK_0,\DK_0) $,
where $\KK_0 = \frac{\K-\K_0'}{2} $, $\DK = \K + \K_0'$,  takes maximum value.
By changing the variables to $ \delta \K' = \K' -\K_0'$, from which we have
\begin{eqnarray} \nonumber
 \KK &=& \KK_0 - \frac{\delta \K'}{2},
\\ \label{dkwid}
\DK &=& \DK_0 + \delta \K',
\end{eqnarray}
we arrive at
\begin{equation} \label{Kwid}
 K \simeq K_0 - \frac12 {\bf e}_{\KK_0}\cdot \delta \K'
\end{equation}
 where ${\bf e}_{\KK_0} = \frac{ \KK_0}{K_0} $.  As found above, the widths in $\DK$ are approximately equal to ${1}/{\sigma_r}$ and ${1}/{\sigma_z}$ in the axial and
 longitudinal direction, respectively. Thus, from Eq.~(\ref{dkwid}) we find the same for $\delta \K'$.  On the other hand, from Eq.~(\ref{Kwid}) we find that the width
 in ${\bf e}_{\KK_0} \delta \K' $ is approximately equal to $2 \Delta_K$.  As ${\bf e}_{\KK_0} \simeq {\bf e}_\K$, we find that the width in $\delta k'$ in
 a given direction ${\bf e}_{ \delta \K'} = (\cos \tilde \phi \sin \tilde \theta,\sin \tilde \phi \sin \tilde \theta,\cos \tilde \theta)$ can be estimated as
\begin{equation}
\label{widthsdelta}
 \delta_{\delta k'} = \mbox{min} \left( \frac{1}{\sigma_r \sin \tilde \theta} , \frac{1}{\sigma_z |\cos \tilde \theta|}, \frac{2 \Delta_K}{{\bf e }_\K {\bf e}_{\delta \K'} }    \right),
\end{equation}
where ${\bf e}_\K = (\sin \theta,0,\cos \theta)$.

Now, let us briefly analyze Eq.~\eqref{widthsdelta}. As ${1}/{\sigma_r} \gg \Delta_K$ the largest width $\frac{1}{\sigma_r}$ is possible only for a small region around
$\cos \tilde \theta =\cos \tilde \phi = 0$.  In the remaining area, we have a competition between second and third term of the above formula.  The third term is smaller
than the second as long as $2\Delta_K \sigma_z \frac{|\cos \tilde \theta|}{ {\bf e }_\K {\bf e}_{\delta \K'} } < 1 $.  Thus, the value $2 \Delta_K \sigma_z$, which is
not smaller than unity, cf. Eq.~(\ref{widK}), defines the range of angles where the above inequality holds.  But, independently of this value, for $\cos \tilde \theta
=0$ and $ {\bf e }_\K {\bf e}_{\delta \K'} \neq 0$ the inequality is satisfied.  For example, for ${\bf e }_\K = {\bf e}_{\delta \K'} = {\bf e}_x$ ) the width in $\delta
k' = 2 \Delta_K$.  In Section~\ref{DENSITY} we show that in the radial direction the minimal width of the single particle density equals approximately $1/\sigma_r$.
Therefore, the correlation width along $x$-axis is much smaller than the density width.

\section{Local correlations and single particle density} \label{lc}

\subsection{General considerations}\label{genloc}

In this section we show that the term $ \left|G^{(1)} \left( \x_1,\x_2,T \right) \right|^2 $ in the two particle correlation function, Eq.~\eqref{G2}, represents collinear correlations
of the particles with aligned velocities.

To this end, we investigate the limit $T \rightarrow \infty$  and, along the lines of the study of the back to back correlations, we define:
\begin{eqnarray} \nonumber
&& G^{(1)}(\K_1,\K_2) = \lim_{T \rightarrow \infty }
\exp \left( i \frac{\hbar(k_1^2 - k_2^2)}{2m}T \right)
\left( \frac{\hbar T}{m}  \right)^3 \times
\\ \label{G1def}
&& \quad \times G^{(1)} \left(\frac{\hbar \K_1}{m}T, \frac{\hbar \K_2}{m} T, T \right). 
\end{eqnarray}
Now, we insert Eq.~(\ref{pertG1}) and~(\ref{Mss}) in Eq.~(\ref{G1def}) arriving at
\begin{eqnarray}\label{Gkn}
G^{(1)}(\K_1,\K_2) = \int \mbox{d} \K \, M^*(\K_1,\K) M(\K,\K_2).
\end{eqnarray}
This formula is the central subject of the analysis in this section.

According to previous considerations, the anomalous density $M(\K_1,\K_2)$ is nonzero only in the region where $\K_1$ and $\K_2$ are practically antiparralel with length
approximately equal to $Q$.  The Eq.~(\ref{Gkn}) implies that $\K_1$ and $\K_2$ are practically parallel, with the length approximately equal to $Q$.  This shows that
the correlations represented by $G^{(2)}_{loc}(\K_1,\K_2) = \left|G^{(1)}(\K_1,\K_2) \right|^2 $ are the ``local'' ones, i.e., nonzero only if $\K_1 \approx \K_2$.  As a
consequence, the width of the scattered atoms halo, described by the single particle density $G^{(1)}(\K_1,\K_1)$, is much smaller than its radius being close to $Q$.

Below we derive an approximate formula for the first order correlation function $G^{(1)}$ that is well suited for numerical treatment.  First, however, it is convenient to introduce
into the quantum problem the quantity that in the semiclassical limit describes the source of atoms. This object, denoted by $ f\left(\x,\K,t\right)$, characterizes the
distribution of atomic momenta $\hbar \K$ at every point in space $\x$ and time $t$ emitted by the source.  The function $f$ attains its semiclassical meaning through
the relation with the single particle Wigner function $W(\x,\K,T)$ of atoms emitted by the source:
\begin{eqnarray}\label{Wc}
W(\x,\K,T) = \int_0^T \mbox{d} t \, f\left(\x - \frac{\hbar \K}{m}(T-t),\K, t  \right).
\end{eqnarray}
This formula is classical in a sense that it  assumes the particles to travel with a velocity $\hbar \K/m$.
In Appendix~\ref{derf} we show that $G^{(1)}$ can be expressed in terms of the source $f$ function, by the following formula:
\begin{eqnarray} \nonumber
&& G^{(1)}\left(\K +\frac{\Delta \K}{2},\K -\frac{\Delta \K}{2} \right) = 
\\ \label{G1formula0}
&&  \int \mbox{d} \x \int_0^\infty \mbox{d} t \,  \exp \left(i \Delta \K \left( \x -\frac{\hbar}{m} \K t  \right) \right)  f\left(\x,\K, t  \right).
\end{eqnarray}
This equation is useful, because it is the source $f$ that is well suited for various approximations. Below, taking this equation as a starting point, we exploit specific properties of the
system to simplify the formula for $G^{(1)}$.

First, we shall approximate $k \simeq Q$ in Eq.~\eqref{G1formula0}, and find
\begin{equation}\label{G1formula}
 G^{(1)}\left(\K,\Delta \K\right) \!\!=\!\!   \int\!\! \mbox{d} \x \int_0^\infty\!\! \mbox{d} t \,  
\exp \left(i \Delta \K \left( \x   \!-\!  v_0 t {\bf e}_\K \right)  \right)
f\left(\x,\K, t  \right),
\end{equation}
where we renamed the variables $G^{(1)}\left(\K, \Delta \K \right) = G^{(1)}\left(\K +\frac{\Delta \K}{2},\K -\frac{\Delta \K}{2} \right)$, which should not lead to
confusion.  In Appendix~\ref{wypf} we show that, as long as Eq.~(\ref{conditionNn2}) is satisfied, the source function $f$ takes the approximate form
\begin{eqnarray} \nonumber
&& f(\x,\K, t ) \simeq   \frac{2 \sigma_{tot} \hbar^2}{ \pi^2 m^2 v_0}  \int  \mbox{d} \delta r    \, \exp \left( - i4 \delta r \delta k  \right) \times
\\ \label{sourcef7n}
&&
\quad \times \psi_Q^*\psi_{-Q}^*\left( \x - \delta r {\bf e}_\K ,t\right)  \psi_Q\psi_{-Q}\left( \x + \delta r {\bf e}_\K,t\right),
\end{eqnarray}
where $\delta k = k-Q$, and $\sigma_{tot} = 8 \pi a^2$ is the total cross section for collisions of two atoms.

If the source function is a semiclassical quantity, one would expect to find it from classical considerations.
In Appendix~\ref{wypfclas} we show that under the condition
\begin{equation} \label{condclas}
8 \frac{\sigma_r^3}{a_{hor}^2 \sigma_z} \ll 1
\end{equation}
this turns out to be true, i.e., the expression in Eq.~(\ref{sourcef7n}) can be derived from semiclassical model presented in Section~\ref{Sbb}.  Furthermore, in Appendix~\ref{wypf} 
we show that in a such case the formula for the source function can be simplified to
\begin{eqnarray} \nonumber
&& f(\x , \K, t ) =   \frac{2 \hbar \sigma_{tot}}{\pi^2 m Q \sin \theta} \int \mbox{d} \delta x  \, 
\exp \left( - i  \frac{4 \delta k \delta x}{\sin \theta} \right) \times
\\ \label{sourcef5n}
&& \quad \times \psi_{+Q}^*\psi_{-Q}^*\left(\x - \delta x {\bf e}_x, t \right)  
\psi_{+Q}\psi_{-Q}
\left(\x + \delta x {\bf e}_x, t \right),
\end{eqnarray}
where we took $\K = k (\sin \theta,0,\cos \theta)$ without the lost of generality.
Now, we insert Eq.~(\ref{sourcef5n}) into Eq.~(\ref{G1formula}), use the identity
\begin{eqnarray*}
&&\int \mbox{d} \delta x  \, 
\exp \left( - i  \frac{4 \delta k \delta x}{\sin \theta} \right) = \left( \frac{2}{\pi \sin \theta} \right)^2
\\
&&
  \int    \mbox{d} \delta k_y  \mbox{d}\delta k_z   \int \delta \x  \,
\exp \left( - i  \frac{4 \delta \K \delta \x}{\sin \theta} \right),
\end{eqnarray*}
where $\delta \K = (\delta k, \delta k_y,\delta k_z)$, introduce
$\x_1 = \x + \delta \x $, $\x_2 = \x - \delta \x$, and, finally, we obtain
\begin{eqnarray*}
&& G^{(1)}\left(\K,\Delta \K \right) =   \frac{\hbar \sigma_{tot}}{\pi^4 m Q\sin^3 \theta}  \int_0^\infty \mbox{d} t \, 
\exp \left( - i \Delta \K {\bf e}_\K v_0 t \right)
\\
&&
  \int    \mbox{d} \delta k_y  \mbox{d} \delta k_z 
\int \mbox{d} \x_1 \mbox{d}\x_2 \, \exp \left( - i \frac{2}{\sin \theta} \delta \K (\x_1 -\x_2)  \right)
\\
&& 
 \exp \left(  i \Delta \K \frac{\x_1 + \x_2}{2} \right) 
\psi_{+Q}^*\psi_{-Q}^*(\x_2, t )  \psi_{+Q}\psi_{-Q} (\x_1, t ).
\end{eqnarray*}
This expression be be further simplified, when we substitute $\tilde \Psi(\K,t) = \int \mbox{d} \x \, \exp \left( - i \K \x \right) \psi_{+Q}\psi_{-Q} (\x, t )  $. As a result, we obtain
\begin{eqnarray*}
&& G^{(1)}\left(\K, \Delta \K \right) =   \frac{\hbar \sigma_{tot}}{\pi^4 m Q\sin^3 \theta}  \int_0^\infty \mbox{d} t \, 
\exp \left( - i \Delta \K {\bf e}_\K v_0 t \right)
\\
&&
\int    \mbox{d} \delta k_y \mbox{d} \delta k_z \,
\tilde \Psi^* \left( \frac{2\delta \K}{\sin \theta}  +\frac{\Delta \K}{2},t \right)
\tilde \Psi \left( \frac{2\delta \K}{\sin \theta}  - \frac{\Delta \K}{2},t \right).
\end{eqnarray*}

From this equation, we obtain particularly simple form of the single particle density, $\rho(\K) = G^{(1)}\left(\K,0 \right) $, which reads
\begin{equation} \label{rhowzor10}
 \rho(\K) \!=\!  \frac{\hbar \sigma_{tot}}{\pi^4 m Q\sin^3 \theta}  \int_0^\infty \!\!\mbox{d} t 
\int  \!\!  \mbox{d} \delta k_y  \mbox{d} \delta k_z \,
\left| \tilde \Psi \left( \frac{2\delta \K}{\sin \theta},t \right) \right|^2\!.
\end{equation}
The above formulae are well suited for numerical computation using FFT routines.

Finally, let us notice an important property of single particle density given by (\ref{G1formula}) and 
(\ref{sourcef5n}). It has the functional form 
$\rho(\delta k) = \frac{1}{\sin\theta} g\left( \frac{\delta k}{\sin \theta}  \right) $, where $g$ is a function.
It follows that upon integration over radial variable $k$ we obtain the spherical angle density,
\begin{eqnarray*}
&& \int_0^\infty k^2 \mbox{d} k \, \rho(\K)  \simeq \int_{-\infty}^\infty Q^2 \mbox{d} \delta k \, \rho(\K) 
\\ 
&& = Q^2 \int_{-\infty}^\infty \mbox{d}\left(\frac{\delta k}{\sin \theta}\right)  \, g\left( \frac{\delta k}{\sin \theta} \right),
\end{eqnarray*}
that is an $\theta$ angle independent value. 
This is caused by the fact that atoms scatter only in the s-wave, which has an angle independent differential cross section.

\subsection{Single particle density and number of scattered atoms}\label{DENSITY}

The formulae derived above can be used to calculate the total number of scattered atoms, and to investigate the properties of the density of the scattered atoms.

First, let us concentrate on total number of scattered atoms, $N_{sc} = \int \mbox{d}\K \, \rho(\K)$. The source $f$ function depends only on $\delta k$, and is
nonvanishing only if $|\delta k| \ll Q$, so we approximate $\int\! \mbox{d} \K \simeq 4\pi Q^2 \int_{-\infty}^\infty \mbox{d} \delta k$.  From Eqs.~(\ref{G1formula})
and~(\ref{sourcef7n}) we obtain that
\begin{eqnarray} \nonumber
N_{sc} \simeq 4 \sigma_{tot} v_0 \pi   \int_0^\infty \mbox{d}t \int \mbox{d} \x \,  |\psi_{+Q}(\x,t)|^2|\psi_{-Q}(\x,t)|^2.
\end{eqnarray}
This result shows the production rate of scattered atoms is directly proportional to the product of the densities of the counterpropagating clouds.  In the case of
a gaussian ansatz, see Eq.~(\ref{vatnowe}), the number of scattered atoms is
\begin{equation}
N_{sc} = \frac{2\sqrt{2} (Naa_{hor})^2 v_0}{\sqrt{\pi}\sigma_z \sigma_r^2 } \int_0^\infty \mbox{d}t \, \frac{\exp \left( - 2 \frac{v_0^2}{\sigma_z^2} t^2 \right) }{1+\omega_r^2t^2}.
\end{equation}
Therefore, the number of atoms scattered is an integral of a time dependent rate of production of atoms, which vanishes for times larger than  min$(2\tau_{ex},\tau_c)$. 
This timescale can thus be interpreted as an effective production time, during which particles are scattered from the colliding clouds. 

Let us now analyze the density $\rho(\K)$ of scattered atoms in the case of a gaussian ansatz.
Inserting formula from Eq.~(\ref{vatnowe}) into Eq.~(\ref{rhowzor10}), and performing the gaussian integral we obtain 
\begin{eqnarray} \label{rhowzor000} 
  \rho(\K) &=& 
  \frac{D_f}{\sin\theta }  \int_0^\infty   \frac{ \mbox{d} \tau}{\sqrt{(1+\tau^2)(1+\beta^2 \tau^2)}} \times
  \\ \nonumber
& & 
  \times \exp \left(   - 2 \alpha^2 \tau^2 - 2\frac{ \delta k^2\sigma_r^2}{ \sin^2 \theta}  \frac{1+\tau^2}{1+\beta^2 \tau^2} \right),
\end{eqnarray}
where $\delta k = k-Q$, $\tau = \omega_r t$ and, due to cylindrical symmetry, we took specific $\K = k(\sin \theta,0,\cos\theta)$.  As seen from the formula, the integral
above is a function of $\frac{\delta k \sigma_r}{|\sin \theta|}$ and parameters $\alpha$ and $\beta$.  In the fast collision $\alpha \rightarrow \infty$ 
or strong axial confinement $\beta=1$ cases, the density takes the form $\rho(\K) \propto \frac{1}{\sin \theta} \exp
\left( - 2\frac{ \delta k^2\sigma_r^2}{ \sin^2 \theta} \right) $.  In all the other cases the above formula shows that we sum gaussian functions $ d(\tau) \exp \left( -
2\frac{x^2}{w^2(\tau)} \right)$ with the widths, $ w^2(\tau) = {(1+\beta^2 \tau^2)}/{(1+\tau^2)}$, growing in time, and weights, $d(\tau)
=[(1+\tau^2)(1+\beta^2 \tau^2)]^{1/2} \exp \left( - 2 \alpha^2 \tau^2 \right)$, decreasing in time.  

\begin{figure}[t!]
  \includegraphics[clip, scale=0.35]{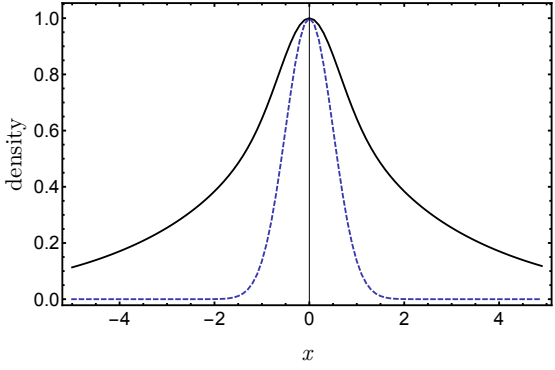}
  \caption{The density $\rho(\mathbf{k})$ of scattered atoms, normalized to maximum value, as a function of $x=\delta k \sigma_r/\sin\theta$ 
    calculated with Eq.~\eqref{rhowzor000}. 
    The solid black line is for $\alpha = 0.2$ and $\beta=10$. The dashed blue line is the narrowest case for $\alpha=0.2$ and $\beta=1$.}
  \label{fig:dens}
\end{figure}

Therefore, the density in the fast collision and strong radial confinement cases takes the possible narrowest shape and the the widest density is for largest possible value of
$\beta$ and smallest possible value of $\alpha$. As discussed in Section~\ref{Sbb} this is the case for which $\alpha=0.2$ and $\beta=10$.  The density $\rho(\K)$
normalized to its maximal value is a function of $x =\frac{\delta k \sigma_r}{\sin \theta}$, and is plotted in Fig.~\ref{fig:dens} for these values of $\alpha$ and
$\beta$.  In the plot, we show additionally the narrowest possible case.  As we see for $\beta =10$ the density distribution has a long tail coming from the latest
$\tau$, where $w(\tau)$ is the largest.  However, the halfwidth changed only roughly twice with respect to the $\beta=1$ case.

\subsection{Examples of $G^{(2)}_{loc}$ function }

Let us now analyze the properties of the source function $f$ and $ G^{(1)}$ in the case of gaussian ansatz given by Eq.~(\ref{vatnowe}).  Inserting the gaussian ansatz
into Eq.~(\ref{sourcef5n}) and performing the integral one arrives at
\begin{eqnarray} \nonumber 
&& f(\x,\K, t ) \simeq   \frac{C_f }{ (1+\omega_r^2t^2)^{3/2} \sin \theta}
\exp\! \left(-  \frac{2(x^2+y^2)}{\sigma_r^2(t)} \right)\times
\\ \label{fvat}
&&
 \times \exp\! \left[ - 2\!\left(\! \beta \omega_r t  \frac{x}{\sigma_r(t)}   \!-\! \frac{\delta k \sigma_r(t)}{\sin \theta}\!\right)^2 
  \!-\! \frac{2(z^2+v_0^2t^2)}{\sigma_z^2} \right]
\end{eqnarray}
where $C_f = \frac{\sigma_{tot} \hbar^2 N^2 }{ 2^{3/2}\pi^{9/2} m^2 v_0 \sigma_z^2 \sigma_r^3}$ and, without lost of generality, we took $\K = k(\sin
\theta,0,\cos\theta)$.  Note that $\frac{\delta k \sigma_r(t)}{\sin\theta}$ dependence is given by the gaussian with a shift of the maximum equal to $\beta \omega_r
t {x}/{\sigma_r(t)} $.

With the source function $f$ at hand, we can now calculate single particle correlation function $G^{(1)}$. To this end, we insert $f$ from Eq.~(\ref{fvat}) into
Eq.~(\ref{G1formula}), and, after performing the spatial integral, we finally obtain
\begin{eqnarray} \nonumber 
&& G^{(1)}(\K,\Delta \K) \simeq \frac{D_f}{  \sin \theta} \int_0^\infty \mbox{d} \tau\frac{1}{\sqrt{(1+\tau^2)(1+\beta^2 \tau^2)} } \times
\\ \nonumber
&&
\times \exp \left(  - i \omega \tau  - 2 \alpha^2 \tau^2 - \frac{  \Delta k_y^2 \sigma_r^2(1+\tau^2) + \Delta k_z^2 \sigma_z^2 }{8}  \right) \times
\\
&& \label{Gogolne0}
\times \exp \left( - \left( 2\frac{ \delta k^2\sigma_r^2}{ \sin^2 \theta}  + \frac{\Delta k_x^2 \sigma_r^2}{8} \right) 
\frac{1 + \tau^2}{1 + \beta^2 \tau^2 } 
 \right),
\end{eqnarray}
where $D_f = \frac{(N a  a_{hor})^2}{\pi^2 Q \sigma_z \sigma_r}$, and
\begin{equation}\label{omegaloc0}
\omega = Qa_{hor}^2 \Delta \K {\bf e}_\K  - \beta \frac{\Delta k_x \delta k \sigma_r^2}{\sin \theta } \frac{1 + \tau^2}{1 + \beta^2 \tau^2 },
\end{equation}
with $\Delta \K {\bf e}_\K = \Delta k_x \sin \theta + \Delta k_z \cos \theta$.

From the plot of the density shown in Fig.~\ref{fig:dens} in the case of $\beta=10$ and $\alpha=0.2$,
we see its halwidth is reached for $\left|\frac{\delta k \sigma_r}{\sin \theta} \right| \simeq 2 $.
As we are not interested in the structure of the tails of the  $G^{(1)}$, and, therefore,  we shall
restrict  our considerations to the region where $\left|\frac{\delta k \sigma_r}{\sin \theta} \right| \leqslant 1$,
which is a bulk region of high density for all values of $\beta$ and $\alpha$.

Let us now analyze Eq.~(\ref{Gogolne0}) together with Eq.~(\ref{omegaloc0}) in the bulk density region.  Note first that the condition in Eq.~(\ref{conditionNn2}) together with
$\left|\frac{\delta k \sigma_r}{\sin \theta} \right| \leqslant 1 $ and $\sin \theta > \frac{1}{2}$ implies that $|Q a_{hor}^2 \sin \theta| \gg \left|\beta \frac{\delta k
  \sigma_r^2}{\sin \theta } \right| $.  Consequently, as long as  $|{\bf e}_{\Delta \K} {\bf e}_\K| \gg \frac{\sigma_r^3}{Q a_{hor}^4}$ 
(which according to Eq.~(\ref{conditionNn2}) is much smaller than unity) we can neglect the term $\beta \frac{\Delta k_x \delta k \sigma_r^2}{|\sin \theta| } \frac{1 + \tau^2}{1 + \beta^2 \tau^2 } $  in $\omega$.
Other values of $|{\bf e}_{\Delta \K} {\bf e}_\K|$ are much smaller than unity.  Therefore, if we set ${\bf e}_\K = (\sin \theta,0,\cos \theta)$, we can approximate
${\bf e}_{\Delta \K} \simeq (\cos \theta \cos \tilde \varphi, \sin \tilde \varphi ,-\sin \theta \cos \tilde \varphi) $.  
We note that in such a case the width in $\Delta k$, given by the exponent $ \exp
\left( - {\Delta k_z^2 \sigma_z^2 }/{8} \right)$ term, is equal to ${1}/{(\sigma_z\sin \theta \cos \tilde \varphi)}$. 
Therefore, we have the following equality
\begin{eqnarray*}
 \beta \frac{\Delta k_x \delta k \sigma_r^2}{\sin \theta } \frac{1 + \tau^2}{1 + \beta^2 \tau^2 } =
\beta \frac{ \cot \theta  \delta k \sigma_r^2}{\sin \theta \sigma_z } \frac{1 + \tau^2}{1 + \beta^2 \tau^2 }.
\end{eqnarray*}
The condition in Eq.~(\ref{condclas}) together with $\left|\frac{\delta k \sigma_r}{\sin \theta} \right| \leqslant 1 $ results in the following inequality: $ \left| \beta
\frac{\Delta k_x \delta k \sigma_r^2}{\sin \theta } \frac{1 + \tau^2}{1 + \beta^2 \tau^2 } \right| \ll 1$.  The largest characteristic time $\tau$ is equal to
unity. Thus, the term $\beta \frac{\Delta k_x \delta k \sigma_r^2}{\sin \theta} \frac{1+ \tau^2}{1+ \beta^2 \tau^2} \tau $ can be neglected. We have thus shown that for any
direction of the vectors ${\bf e}_{\Delta \K}$ and ${\bf e}_{\K}$ we can approximate
\begin{equation}\label{omegaloc2}
\omega \simeq Q a_{hor}^2 \Delta \K\, {\bf e}_\K.
\end{equation}

In what follows, we show that $ {\Delta k_x^2 \sigma_r^2}/{8} \ll 1$, which further simplifies the single particle correlation function. 
We take ${\bf e}_{\Delta \K} = (\cos \tilde \theta \cos \tilde \varphi, \sin \tilde \varphi ,-\sin \tilde \theta \cos \tilde \varphi)$. 
The width in $\Delta k$ given by the the exponent in $ \exp( -{\Delta k_z^2 \sigma_z^2 }/{8} )$ term is equal to 
${1}/{(\sigma_z\sin \tilde \theta \cos \tilde \varphi )}$, which yields
\begin{eqnarray*}
\Delta k_x^2 \sigma_r^2 \approx  \frac{\sigma_r^2}{\sigma_z^2} \cot^2 \tilde \theta. 
\end{eqnarray*}
This is much smaller than unity for $\tilde \theta \gg \frac{\sigma_r}{\sigma_z}$, which, for $\sigma_r \ll \sigma_z$, is satisfied for almost all $\tilde \theta$ except
an excluded region around $\tilde \theta = 0$.  There we note that $|\Delta k_z| = |\sin \tilde \theta \cos \tilde \varphi | \ll |\cos \tilde \theta \cos \tilde \varphi |
=|\Delta k_x| $, and so we approximate $\omega \simeq Qa_{hor}^2 \Delta k_x \sin \theta $.  By inspecting expression in Eq.~(\ref{Gogolne0}), we notice that the minimal
characteristic time $\tau$ is approximately equal to $\mbox{min}\!\left( {1}/{2\alpha}, {1}/{\beta} \right)$.  Consequently, the width in $\omega$ is equal to
$\mbox{max}(2\alpha,\beta)$. As a result, the maximal width in $\Delta k$ can be estimated as $\frac{1}{\sin \theta} \left( \frac{\sigma_r^2}{Q a_{hor}^4},
\frac{2}{\sigma_z} \right) $. According to the condition in Eq.~(\ref{conditionNn2}), this is much smaller than ${1}/{\sigma_r}$, which results in $ \Delta k_x^2
\sigma_r^2 \ll 1 $.  Finally, the single particle correlation function takes the following form
\begin{eqnarray} \nonumber 
&& G^{(1)}(\K,\Delta \K) \simeq \frac{D_f}{  \sin \theta} \int_0^\infty \mbox{d} \tau\frac{1}{\sqrt{(1+\tau^2)(1+\beta^2 \tau^2)} } \times
\\ \nonumber
&&
\times\exp \left(  - i \omega \tau  - 2 \alpha^2 \tau^2 - \frac{  \Delta k_y^2 \sigma_r^2(1+\tau^2) + \Delta k_z^2 \sigma_z^2 }{8}  \right)\times
\\ \label{Gogolne02}
&& 
\times\exp \left( - 2\frac{ \delta k^2\sigma_r^2}{ \sin^2 \theta}  
\frac{1 + \tau^2}{1 + \beta^2 \tau^2 } 
 \right),
\\ \label{omegaloc02}
&& \mathrm{with}\ \ \omega \simeq Qa_{hor}^2 \Delta \K \, {\bf e}_\K .
\end{eqnarray}

The formula for the single particle correlation function in Eq.~\eqref{Gogolne02} can be further simplified. To this end, we note that from Eq.~(\ref{Gogolne0}) the term
$(1+\tau^2)$ in $ \exp \left( -{\Delta k_y^2 \sigma_r^2(1+\tau^2)}/{8} \right)$ may effectively vary between $1$ and $5$, always giving the width in $\Delta k$ being of
the order of ${1}/{\sigma_r}$. Then, we may approximate $\Delta k_y^2 \sigma_r^2(1+\tau^2) \simeq \Delta k_y^2 \sigma_r^2 $.  Furthermore, to restore the cylindrical
symmetry of the correlation function, we multiply the right hand side in Eq.~\eqref{Gogolne02} by the factor $\exp \left( -{\Delta k_x^2 \sigma_r^2}/{8} \right)
\simeq 1$,   Therefore, the correlation function can be written as a product of two factors:
\begin{equation} \label{G1sim0} 
 G^{(1)}(\K,\Delta \K) \simeq G_1\! \left( \frac{\delta k}{\sin \theta},\Delta \K\, {\bf e}_\K \right) G_2(\Delta \K),
\end{equation}
where the two functions can be conveniently defined according to:
\begin{eqnarray}\nonumber
&&  G_1\! \left(\frac{\delta k}{\sin \theta},\Delta \K\, {\bf e}_\K \right) =
\frac{D_f}{  \sin \theta} \int_0^\infty\!\! \mbox{d} \tau\frac{1}{\sqrt{(1+\tau^2)(1+\beta^2 \tau^2)} }\times
\\ \label{G1sim1}
&&
\ \ \times \exp\!\left[  - i  Qa_{hor}^2 \Delta \K {\bf e}_\K \tau  \!-\! 2 \alpha^2 \tau^2 \!-\! 2\frac{ \delta k^2\sigma_r^2}{ \sin^2 \theta}  
\frac{1 + \tau^2}{1 + \beta^2 \tau^2 } 
 \right]\!,
\\ \label{G2wzor}
&& G_2(\Delta \K) = \exp \left(- \frac{ (\Delta k_x^2 + \Delta k_y^2) \sigma_r^2 + \Delta k_z^2 \sigma_z^2  }{8} \right).
\end{eqnarray}
In the paragraph preceding Eq.~\eqref{Gogolne02}, we estimated the width in $\Delta \K {\bf e}_\K$ to be maximally given by
\begin{equation}\label{wid0}
\Delta_k = \max \left( \frac{2}{\sigma_z}, \frac{\sigma_r^2}{Q a_{hor}^4} \right).
\end{equation}
Below, we analyze few examples to see how to apply the obtained formulas, and investigate the width of the correlation function.

\subsubsection{Fast collision and strong radial confinement}

\begin{figure}[t!]
  \includegraphics[clip, scale=0.35]{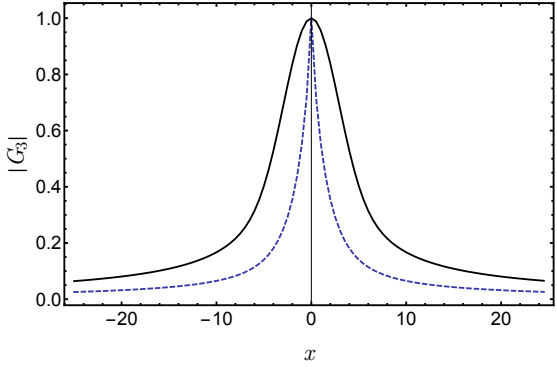}
  \caption{The plot of $|G_3|$ in two scenarios.
    The solid black line -- the fast collision case; here  $x = \Delta \K \cdot {\bf e}_\K \sigma_z$, and $G_3$ is given by Eq.~\eqref{Eq:G3}.
    The dashed blue line -- the strong confinement configuration; here  $x = \Delta \K\, {\bf e}_\K Qa_{hor}^2 $, and $G_3$ is given by Eq.~\eqref{Eq:G3strong} with $\alpha=0$.
  }
  \label{fig:G3}
\end{figure}

In the case of fast collision and strong axial confinement case, the
function $G_1$  decouples into a product of two factors:
\begin{eqnarray*}
G_1\! \left( \frac{\delta k}{\sin \theta},\Delta \K\, {\bf e}_\K \right) = \rho\! \left( \frac{\delta k}{\sin \theta} \right)
G_3(\Delta \K {\bf e}_\K),
\end{eqnarray*}
where the density is given by the expression in Eq.~(\ref{rhowzor000}), which in this case is:
\begin{eqnarray*}
\rho \left( \frac{\delta k}{\sin \theta} \right)  \propto \frac{1}{\sin \theta} \exp \left(-2\frac{ \delta k^2\sigma_r^2}{ \sin^2 \theta}   \right).
\end{eqnarray*}
In the fast collision case, we have
\begin{eqnarray}
\label{Eq:G3}
G_3(\Delta\, \K {\bf e}_\K) &=& \exp \left( - \frac{ (\Delta \K \cdot {\bf e}_\K)^2 \sigma_z^2  }{8}  
\right) \times
\\
& & \times \left( 1 - \mbox{erf}\left( \frac{i\Delta \K \cdot {\bf e}_\K \sigma_z }{2 \sqrt{2}} \right)  \right). \nonumber
\end{eqnarray}
In Fig.~\ref{fig:G3} we plot $|G_3|$ as an universal function of variable $\Delta \K \cdot {\bf e}_\K \sigma_z$ . From the figure, we observe that the halfwidth in
 $ \Delta \K \cdot {\bf e}_\K$ is equal to ${4.2}/{\sigma_z}$. 

In the strong radial confinement configuration, the function $G_3$ takes the following form:
\begin{eqnarray}
\label{Eq:G3strong}
G_3(\Delta \K\, {\bf e}_\K) \propto  \int_0^\infty \frac{ \mbox{d} \tau }{1+ \tau^2 } 
 e^{  - i  Qa_{hor}^2 \Delta \K {\bf e}_\K \tau  - 2 \alpha^2 \tau^2 }.
\end{eqnarray}
If  $\alpha \gg 1$, the function reduces to the one in the fast collision case.  In the other limiting case, if  $\alpha \ll 1$
the term  $ 2 \alpha^2 \tau^2$ can be neglected resulting in a simple integral.
In Fig.~\ref{fig:G3} we plot that function, i.e. $|G_3|$, normalized to unity at the maximum given by Eq.~\eqref{Eq:G3strong} with 
the term $2\alpha^2 \tau^2$ dropped.
From the figure, we observe that the halfwidth in $\Delta \K\cdot {\bf e}_\K $ is equal to ${1.2}/{Qa_{hor}^2}$. 

\subsubsection{Largest mean-field energy impact considered} 

\begin{figure}[t!]
  \includegraphics[clip, scale=0.35]{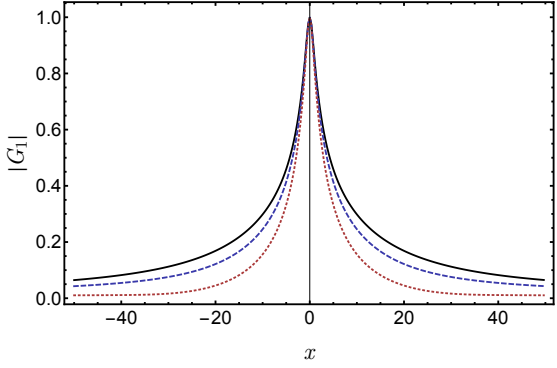}
  \caption{The plot of the $|G_1|$, normalized to unity in maximum, as a function of $x=Qa_{hor}^2 \Delta \K {\bf e}_\K$ for different values of 
    $\delta k\sigma_r/ \sin \theta = 0$ (solid black), $0.5$ (dashed blue), $1$ (dotted red). The function is calculated with Eq.~(\ref{G1sim1}).
  }\label{fig:G1}
\end{figure}

As in the back to back correlation we consider the largest mean-field energy impact case $\alpha=0.2$, $\beta = 10$.
It is additionally motived by the fact that in such case the density takes the widest possible form
as found in Subsection \ref{DENSITY}. 
In Fig.~\ref{fig:G1} we plot $ \left|G_1 \left( \frac{\delta k}{\sin \theta},\Delta \K\, {\bf e}_\K \right) \right|$ 
for $\left| \frac{\delta k \sigma_r}{\sin \theta} \right| < 1$.
From the figure we obseve that the halfwidth in $ Q a_{hor}^2 \Delta \K\, {\bf e}_\K $ depends monotonically on the value of $ \frac{\delta k \sigma_r}{\sin \theta} $ 
taking values equal to $4.25$ and 2.91 
for zero and unity, respectively. 
Note, that it is about $3$ times larger than for $\beta = 1$.
The rough estimation gives the width to be $\beta = 10$ times larger. Thus, we see that, although overestimated, it is still a correct upper estimate.

\subsection{General correlation properties}

\subsubsection{The $\Delta \K$ and $\Delta \K {\bf e}_\K$ widths }

The gaussian ansatz provided us with Eq.~(\ref{G1sim0}), which expressed the single particle correlation function as a product of two factors, $G_1$ and $G_2$.
The function $G_2$ is the Fourier transform of the initial condensate density squared.
Basing on the analysis of the presented examples, we can improve the formula in Eq.~(\ref{wid0}) for the halwidth in  $\Delta \K {\bf e}_\K$ to the following form:
\begin{equation} \label{wid1}
\Delta_k = \max \left( \frac{4}{\sigma_z}, \frac{\sigma_r^2}{Q a_{hor}^4} \right)
\end{equation}
where the term $4/\sigma_z$ is an approximation of $4.2/\sigma_z$ obtained in the fast collision case.

It is interesting to note that the same result can be obtained by the inspection of Eq.~(\ref{G1formula}),
which we rewrite below for convenience,
\begin{equation}\label{G1formula2}
 G^{(1)}\left(\K,\Delta \K\right)\! =\!\!   \int\!\! \mbox{d} \x \int_0^\infty\!\! \mbox{d} t \,  
e^{i \Delta \K \cdot ( \x   -  v_0 t {\bf e}_\K )  } f(\x,\K, t).
\end{equation}
From this expression, we see the $\Delta \K$ dependence enters through $\Delta \K \cdot {\bf e}_\K v_0 t$ and $\Delta \K \cdot \x$.  Intuitively, the source size is of
the order of the condensate size.  We can therefore estimate the width in $\Delta \K$ given by the spatial integral as equal to the inverse size of the condensate width
in respective direction.  On the other hand, the width in $\Delta \K\, {\bf e}_\K$ can be estimated from expression in Eq.~(\ref{G1formula2}); it is equal to
$\frac{1}{v_0 \tau_d}$, where $\tau_d$ is the characteristic time given by (\ref{taud}) which results in   
 the width of $\Delta \K\, {\bf e}_\K$ given by:
\begin{eqnarray*}
 \Delta_k = \max\! \left( \frac{1}{\sigma_z}, \frac{2\sigma_r^2}{Q a_{hor}^4} \right).
\end{eqnarray*}
The appearance of $\tau_d$ in this derivation is not surprising. As discussed in Section~\ref{mod}, the time $\tau_d$ is the minimal characteristic time present in the
evolution of functions $\psi_{\pm Q}$, which, through Eq.~(\ref{sourcef5n}), are directly related to the source function $f$.
Note that the above formula is the same estimate as given by Eq. (\ref{wid0}) up to factor $2$ (which can appear in such rough estimates).

It is interesting to notice, that the width in $\Delta \K {\bf e}_\K$ given by Eq.~(\ref{wid1}) is close to the width in $K$ given by Eq.~(\ref{widK}).
In fact, this can be understood from the analysis of Eq.~(\ref{Gkn}), which can be rewritten as
\begin{eqnarray*}
G^{(1)}(\K_1,\K_2) = \int \mbox{d} \K' \, M^*(\K_1,\K') M(\K_2,\K'),
\end{eqnarray*}
where we used the symmetry $M(\K',\K_2) = M(\K_2,\K')$. We introduce now a convenient representation of variables, i.e., we rewrite the above in 
($\K$, $\Delta \K$, $\KK$, $\DK$): $ M(\KK,\DK) = M(\K_1,\K_2) $, $ G^{(1)}\left( \K,\Delta \K \right) = G^{(1)}\left( \K_1,\K_2 \right)$.
In addition to this, we change variable $\K'$ into $\delta \K$ according to: $\K' = -\K - \delta \K$. We then obtain the following expression:
\begin{eqnarray*}
G^{(1)}(\K,\Delta \K) &=& \int \mbox{d} \delta \K \, 
M^* \left( \K + \frac{\Delta \K}{4} + \frac{\delta \K}{2} , \frac{\Delta \K}{2} - \delta \K \right) 
\\
& &
M \left( \K - \frac{\Delta \K}{4} + \frac{\delta \K}{2} , -\frac{\Delta \K}{2} - \delta \K \right).
\end{eqnarray*}
The two anomalous densities are taken for the vectors which magnitudes are given by:
\begin{eqnarray*}
K_{1,2} = \left| \K \pm \frac{\Delta \K}{4} + \frac{\delta \K}{2}  \right| \simeq k + \left( \frac{\delta \K}{2} \pm \frac{\Delta \K}{4}  \right) {\bf e}_\K.
\end{eqnarray*}
The difference of the two vectors is then  $K_1-K_2 \simeq  \Delta \K\, {\bf e}_\K/2 $.
If $|K_1-K_2|$ is larger than $2\Delta_K$,  the above integral is vanishingly small.
Therefore, the width in $ \Delta \K\, {\bf e}_\K$, denoted as $\Delta_k$, 
is approximately $4 \Delta_K$.
This relation is confirmed by equations 
(\ref{widK}) and (\ref{wid1}) when we take into account that $\Delta_K$ and $\Delta_k$
given by these equations are rough estimates.

\subsubsection{Correlation volume}

In analogy to the definition from the previous Section, we define here the correlation volume as the volume in $\Delta \K$ for which two particles are still significantly
correlated.

The correlation volume is in fact related to the width in $\Delta k$ of $|G^{(1)}|$ function, calculated for arbitrary direction $(\sin \tilde \theta \cos \tilde \phi,
\sin \tilde \theta \sin \tilde \phi, \cos \tilde \theta)$.  Combining all the results at hand regarding the local correlations, we can estimate this width as
\begin{equation}\label{deltakeq}
\delta_{\Delta k}  = \mbox{min} \left(  \frac{2.5}{\sigma_r \sin \tilde \theta}, \frac{2.5}{\sigma_z |\cos \tilde \theta |}, 
\frac{\Delta_k}{ {\bf e}_{\Delta \K} \cdot  {\bf e}_\K }  \right),
\end{equation}
where ${\bf e}_{\Delta \K} \cdot {\bf e}_\K = \sin \tilde \theta \cos \tilde \phi \sin \theta + \cos \tilde \theta \cos \theta $ and $\Delta_k$ is given by
Eq.~(\ref{wid1}). The factor $2.5$ comes directly from the halfwidth of the $G_2$ function given by (\ref{G2wzor}).

Let us briefly analyze this formula. Due to the fact that ${1}/{\sigma_r} \gg \Delta_k$, the largest possible width, known to be ${2.5}/{\sigma_r}$, is possible only for
small region around $\cos \tilde \theta = 0$ and $\cos \tilde \phi = 0 $.  In the remaining area we have a competition between second and third term of the above formula.
However, as $\Delta_k \sigma_z \geqslant 4$, which is implied by Eq.~(\ref{wid1}), the second term of the above formula is less than the third one if ${|\cos \tilde
  \theta |}/{ {\bf e}_{\Delta \K} \cdot {\bf e}_\K } \geqslant {2.5}/{4} $.  For $|\cos \tilde \theta | \geqslant {2.5}/{4}$, the condition is satisfied, which gives us
the region where $\delta_{\Delta k} ={2.5}/{\sigma_z|\cos \tilde \theta |} $.  On the other hand, for $\cos \tilde \theta \simeq 0$, excluding the region for which ${\bf
  e}_{\Delta \K} \cdot {\bf e}_\K \simeq 0 $, we obtain $\delta_{\Delta k} = {\Delta_k}/{ {\bf e}_{\Delta \K} \cdot {\bf e}_\K } $.

Note that  $\Delta_k$ is similar to $\Delta_K$ and furthermore the formulas for 
$\delta_{\delta k'}$ and $\delta_{\Delta k} $  given by Eqs.~(\ref{widthsdelta}) and (\ref{deltakeq}) are similar as well.
Thus the back to back and local correlation volumes take similar values.  

Let us finally remark, that in the case ${\bf e}_{\Delta \K} \cdot {\bf e}_\K =0 $ the local correlation width is given by the spatial Fourier transform of the source,
which is given by the inverse of the condensate widths in respective directions.  This is in fact the famous Handbury Brown and Twiss effect \cite{HBT0}. 
In the original work this effect was associated with the measurement of the star diameter \cite{HBT}. Here, the star is replaced by a condensate -- a source of particles.

\section{Summary}

In this paper we analyzed the elastic scattering of atoms from elongated Bose-Einstein condensates colliding in the direction of the long axis which we choose to be $z$ axis.  Our theory is valid in
the collisionless regime, where multiple scattering processes are negligible.  Additionally we focused our considerations on the spontaneous regime, in which the bosonic enhancement
effect is negligible and the use of perturbation theory is justified. We showed that the two particle correlation functions decomposes into ``back to back'' and ``local''
parts which describe different aspects of the system.  The ``back to back'' part $G^{(2)}_{bb}$ describes the correlation of two particles with almost opposite velocities.  It reflects
the fact that particles, due to binary collisions from counter propagating condensates, are scattered in pairs with almost opposite velocities.  The ``local'' part $G^{(2)}_{loc}$
characterizes correlation of particles with velocities being almost the same, and is related to the bosonic bunching effect.  Within perturbation theory, we derived
approximate expressions for the one and two particle correlations functions of scattered atoms connecting these functions with the wavefunctions of the colliding
condensates.

The formulas that we obtained are convenient for numerical computation when time dependent Gross-Pitaevskii equation is solved numerically.  Furthermore, we introduced
time dependent variational approach and obtained approximate form of the condensate wavefunctions.  Having those, we calculated and analyzed back to back and local parts
of the two particle correlation function in different regimes of parameters of the system.
We found that the correlation function depends on two dimensionless parameters of the system:
$\alpha =\frac{Qa_{hor}^2}{\sigma_z}$ and $\beta = \frac{\sigma_r^2}{a_{hor}^2}$ where $a_{hor} = \sqrt{\frac{\hbar}{m \omega_r^2}}$ 
is the radial harmonic oscillator length, Q is the mean wavevector of the colliding clouds and $\sigma_{r,z}$
denotes the radial and longitudinal sizes of the initial condensate.

In the chosen  regime of parameters, the back to back part, $G^{(2)}_{bb}$, is a function of $\DK = {(\K_1 + \K_2)}/{2}$ and $K = |\KK| \equiv {|\K_1 - \K_2|}/{2}$,
where $\K_1$ and $\K_2$ are the wavevectors of scattered atoms. 
Analyzing few examples we have found that the $\DK$ and $K$ dependence of the $G^{(2)}_{bb}(K,\DK)$ 
almost decouples. 
For $\beta =1$ and $\alpha \gg \beta$ cases the width in 
$\Delta K_{x,y}$ and $\Delta K_z$ are approximately equal to ${1}/{\sigma_r}$ and
${1}/{\sigma_z}$. This widths broaden maximally by factor of two 
with growing value of $\beta$ and decreasing value of $\alpha$.
We found that $G^{(2)}_{bb}$ has the maximum at $K=Q+ \delta K_{max} $ where 
$\delta K_{max} \approx \frac{m \mu}{\hbar^2 Q}$ where $\mu$ is the chemical potential of the initial
condensate. The width in $K$ of $G^{(2)}_{bb}$ can be estimated as $\frac{m}{2\hbar Q \tau_d}$
where $\tau_d$ is the characteristic time on which the wavefunctions of the colliding clouds changes substantially.

The local correlation, $G^{(2)}_{loc}(\K_1,\K_2)$, of the two particle correlation function is directly related to a single particle correlation function, $G^{(2)}_{loc} =
|G^{(1)}|^2$.  In this study the convenient variables are $\K = {(\K_1+\K_2)}/{2} $ and $\Delta \K = \K_1-\K_2$.  
 Exploiting the condensate wavefunctions given
by the variational ansatz we have calculated $G^{(1)}(\K,\Delta \K)$. 
Analyzing the problem on the single particle level, we found that the 
density is the narrowest in the $\beta=1$ and $\alpha \gg \beta$ cases.
In other cases, the bulk region of the density stays practically the same while the tails of the density distribution start to
grow together with the increase of the $\beta$ and decrease of $\alpha$. 
We have analyzed the dependence of $G^{(1)}(\K,\Delta \K)$ in variable $\Delta \K$ in the bulk region of the density. 
We found that  the $\K$ and $\Delta \K$ dependence decouples.  We found that the dependence in $\Delta \K$ is given by two contributions.  First comes from spatial Fourier
transform of the source of particles, and is in fact the famous Hanbury Brown and Twiss effect.  As a result the width in $\Delta \K$ is given by the inverse size of the
initial condensate density in the respective direction. 
The second arises as a dependence in $\Delta \K \cdot {\bf e}_\K$ variable. 
We estimated the width in  
$\Delta \K \cdot {\bf e}_\K$ of $|G^{(1)}|$ to be roughly equal to $\frac{m}{\hbar Q \tau_d}$ and showed its direct relation to the width in $K$
of the $G^{(2)}_{bb}$.

Having all these results we found the back to back and local correlation volume properties and showed that they are similar in both cases.

Finally, we presented semiclassical models of both types of correlations.  In the case of the back to back correlations we employed Wigner functions as a mean to describe
the phase space densities of the colliding clouds. We showed that $G^{(2)}_{bb}$, taken from the semiclassical model, averaged over $\KK$ yields the same results as the
quantum model. Furthermore, we showed that large differences between quantum and semiclassical model appear in $\KK$ dependence.  In the case of the local correlations,
we showed that, under specific requirements, the semicalssical formula for the source function is the same as the quantum one.

\begin{acknowledgments}
We acknowledge the discussions with Denis Boiron, Jan Chwede\'nczuk, Piotr Deuar, Karen Kheruntsyan Marek Trippenbach and Chris Westbrook.
Pawe\l{} Zi\'n acknowledges the support of the Mobility Plus programme. 
Tomasz Wasak acknowledges the support of the Polish Ministry of Science and Higher Education programme ``Iuventus Plus'' (2015-2017) Project No. IP2014 050073, 
and the National Science Center Grant No. 2014/14/M/ST2/00015.
\end{acknowledgments}

\appendix



\section{Bogoliubov method: perturbative approach}\label{AppBog}

Let us introduce the propagator defined by the equation:
\begin{equation}\label{BBBGreen}
\left( i \hbar \partial_t - H_0(\x ,t) \right) K(\x,t;\x',t') = 0,
\end{equation}
with the boundary condition:
\begin{equation}\label{BBBboundary}
K(\x,t;\x',t) = \delta(\x'-\x).
\end{equation}
We further introduce the operator $\hat \delta'(\x,t)$ by:
\begin{equation}\label{BBBdeltaprim}
\hat \delta(\x,t) =  \int \mbox{d} \x' \, K(\x ,t; \x',0)  \hat \delta'(\x',t).
\end{equation}
The time $t=0$ is chosen in the propagator since at that time the evolution of the system starts.
Substituting Eq.~(\ref{BBBdeltaprim}) into Eq.~(\ref{glowne}) we obtain:
\begin{eqnarray*}
&& \int \mbox{d} \x' \,  K(\x ,t; \x',0) i \hbar \partial_t    \hat \delta'(\x',t) 
\\
&&= B (\x ,t)  \int \mbox{d} \x' \, K^{*}(\x ,t; \x',T_0)  \hat \delta'^\dagger(\x',t).
\end{eqnarray*}
We multiply both sides of the above equation by $K(\x'',0;\x,t)$ and integrate over $\x$ obtaining
\begin{eqnarray} \label{BBBr1}
&&\int\mbox{d} \x \, K(\x'',0;\x,t)  \int \mbox{d} \x' \, K(\x ,t; \x',0) i \hbar \partial_t   \hat \delta'(\x',t) 
\\ \nonumber
&& = \int\mbox{d} \x \, K(\x'',0;\x,t)  B (\x ,t)  \int \mbox{d} \x' \, K^{*}(\x ,t; \x',0)  \hat \delta'^\dagger(\x',t).
\end{eqnarray}
With help of the property
\begin{eqnarray*}
\int\mbox{d} \x \, K(\x'',0;\x,t)  K(\x ,t; \x',0) = \delta (\x''-\x'),
\end{eqnarray*}
Eq.~(\ref{BBBr1}) simplifies to
\begin{eqnarray}\label{BBBglowne2}
&& i \hbar \partial_t   \hat \delta'(\x'',t) 
\\ \nonumber
&&
= \int\mbox{d} \x' \int \mbox{d} \x \, K(\x'',0;\x,t) B (\x ,t)  K^{*}(\x ,t; \x,0)  \hat \delta'^\dagger(\x',t).
\end{eqnarray}
We expand the field operator $\hat\delta'(\x,t)$  in the following series:
\begin{equation}\label{BBBexpansion}
\hat\delta'(\x,t) = \hat \delta'^{(0)}(\x,t)  + \hat \delta'^{(1)}(\x,t) + ...
\end{equation}
By substituting the above expansion into Eq.~(\ref{BBBglowne2}), we obtain an infinite hierarchy of equations. The first two of them reads:
\begin{eqnarray}\label{BBBdelta0}
i \hbar \partial_t   \hat \delta'^{(0)}(\x'',t) &=& 0,
\\ \nonumber
i \hbar \partial_t   \hat \delta'^{(1)}(\x'',t) &=& \int\mbox{d} \x' \int \mbox{d} \x \, K(\x'',0;\x,t) B (\x ,t) \times
\\ \label{BBBdelta1}
& & \times K^{*}(\x ,t; \x',0)  \hat \delta'^{(0)\dagger}(\x',t).
\end{eqnarray}
Eq.~(\ref{BBBdelta0}) can be solved straightforwardly:  $\hat \delta'^{(0)}(\x,t) =  \hat \delta'^{(0)}(\x,0)$.
Substituting this into Eq.~(\ref{BBBdelta1}) and integrating over $t$ we obtain
\begin{eqnarray} \nonumber
\hat \delta'^{(1)}(\x'',T) &=&  \frac{1}{i \hbar} \int_{0}^T \mbox{d} t \int\mbox{d}\x' \int \mbox{d} \x \, K(\x'',0;\x,t) B (\x ,t)
\\ \label{BBBdelta1sol}
& & \times K^{*}(\x ,t; \x',0)  \hat \delta'^{(0)\dagger}(\x',0).
\end{eqnarray}
Subsituting from Eq.~(\ref{BBBdelta1sol}) into Eq.~(\ref{BBBdeltaprim}) we obtain the formal expression for the field operator $\hat \delta $, valid to the first order of
expansion:
\begin{equation}\label{BBBdeltasol}
\hat \delta(\x''',T) \simeq \hat \delta^{(0)}(\x''',T) + \hat \delta^{(1)}(\x''',T),
\end{equation}
where
\begin{eqnarray} \nonumber
\hat \delta^{(0)}(\x''',T) &=& \int \mbox{d} \x'' \, K(\x''' ,T; \x'',0)   \hat \delta'^{(0)}(\x'',0),
\\ \nonumber
\hat \delta^{(1)}(\x''',T) &=&  \frac{1}{i \hbar} \int \mbox{d} \x'' \int_{0}^T \mbox{d} t \int\mbox{d}\x' \int \mbox{d}\x \, K(\x''' ,T; \x'',0)
\\ \nonumber
&\times&  
 K(\x'',0;\x,t) B (\x ,t)  K^{*}(\x ,t; \x',0)  \hat \delta'^{(0)\dagger}(\x',0).
\end{eqnarray}
Now, let us find quantum averages of the field operator on the vaccum state from Eq.~(\ref{stan}).
From Eqs.~(\ref{BBBdeltaprim}), (\ref{BBBboundary}), (\ref{BBBexpansion}) and (\ref{BBBdelta1}) we obtain that
\begin{eqnarray*}
\hat \delta(\x,0) = \hat \delta'(\x,0) = \hat \delta'^{(0)}(\x,0). 
\end{eqnarray*}
Using the above, together with the definition of the vacuum state, see Eq.~(\ref{stan}), we obtain
\begin{equation}\label{BBBr2}
\hat \delta'(\x,0)|0\rangle = \hat \delta'^{(0)}(\x,0)|0 \rangle = 0,
\end{equation}
which implies
\begin{equation}\label{BBBr4}
\langle 0| \hat \delta'^\dagger(\x,0) = \langle 0 |\hat \delta'^{\dagger(0)}(\x,0) = 0.  
\end{equation}
Combining these formulas, with the bosonic commutation relation, 
$[\hat \delta(\x,0),\hat \delta^\dagger(\x',0)] = \delta(\x-\x')$,
we arrive at
\begin{equation}\label{BBBr3}
\langle 0| \hat \delta'^{(0)}(\x,0) \hat \delta'^{(0)\dagger}(\x',0)|0 \rangle = \delta(\x-\x').
\end{equation}
Now, we substitute expression from Eq.~(\ref{BBBdeltasol}) into  Eq.~(\ref{Mdef}), which define anomalous density, and obtain
\begin{eqnarray*}
\langle 0 | \hat \delta(\x_1,T) \hat \delta(\x_2,T) |0 \rangle  &=& \langle 0 | (\hat \delta^{(0)}(\x_1,T) +\hat \delta^{(1)}(\x_1,T)  ) \times \\
&& \times (\hat \delta^{(0)}(\x_2,T) +\hat \delta^{(1)}(\x_2,T)) |0 \rangle .   
\end{eqnarray*}
The right hand side of this equation consists of four terms.  As we see from Eq.~(\ref{BBBdeltasol}) $\hat \delta^{(0)}(\x,T) $ is a function of $\hat
\delta'^{(0)}(\x',0) $ while $\hat \delta^{(0)}(\x,T) $ a function of $\hat \delta'^{(0)\dagger}(\x',0) $. This fact, together with Eqs.~(\ref{BBBr2}) and
(\ref{BBBr4}), implies that
\begin{eqnarray*}
&& \langle 0 | \hat \delta^{(0)} (\x_1,T)  \hat \delta^{(0)} (\x_2,T) |0 \rangle  
= \langle 0 | \hat \delta^{(1)} (\x_1,T)  \hat \delta^{(0)} (\x_2,T) |0 \rangle 
\\
&&
=  \langle 0 | \hat \delta^{(1)} (\x_1,T)  \hat \delta^{(1)} (\x_2,T) |0 \rangle  =0.
\end{eqnarray*}
As a consequence, we obtain the following form for the anomalous dnsity: 
\begin{eqnarray*}
\langle 0 | \hat \delta(\x_1,T) \hat \delta(\x_2,T)|0 \rangle  =\langle 0 | \hat \delta^{(0)}(\x_1,T) \hat \delta^{(1)}(\x_2,T) |0\rangle .
\end{eqnarray*}
Substituting now the forms of $\hat \delta^{(0)}$ and $\hat \delta^{(1)}$, given by Eq.~(\ref{BBBdeltasol}), we obtain
\begin{eqnarray*}
&&  \langle 0 | \hat \delta(\x_1,T) \hat \delta(\x_2,T)|0 \rangle  = \frac{1}{i \hbar} \int\mbox{d}\x''' \mbox{d}\x'' \mbox{d}\x' \mbox{d}\x \int_{0}^T \mbox{d} t \, B (\x ,t) \\
&&  K(\x_1 ,T; \x''',0)  K(\x_2 ,T; \x'',0)K(\x'',0;\x,t)   K^{*}(\x ,t; \x',0) \\
&&  \langle 0| \hat \delta'^{(0)}(\x''',0) \hat \delta'^{(0)\dagger}(\x',0) |0 \rangle.
\end{eqnarray*}
Using now the property of the operator $\delta'^{(0)}$ at the initial time, see Eq.~(\ref{BBBr3}), we can perform one integral over $\x'''$, which yields
\begin{eqnarray*}
&& \langle \hat \delta(\x_1,T) \hat \delta(\x_2,T) \rangle  =  \frac{1}{i \hbar} \int  \mbox{d}\x''\mbox{d}\x' \mbox{d}\x \int_{0}^T \mbox{d} t \, B (\x ,t)
\\
&&  K(\x_1 ,T; \x',0) K(\x_2 ,T; \x'',0)    K(\x'',0;\x,t)   K^{*}(\x ,t; \x',0).
\end{eqnarray*}
In the final step we invoke the properties of the propagator:
\begin{eqnarray} \nonumber
&& K^*(\x',t';\x,t) = K(\x,t;\x',t')
\\ \nonumber
&& \int \mbox{d}\x' \, K(\x,t;\x',t')  K(\x',t';\x'',t'') = 
K(\x,t;\x'',t''),
\end{eqnarray}
which results in the following for the anomalous density:
\begin{eqnarray} \nonumber
&& M(\x_1,\x_2,T) = \langle \hat \delta(\x_1,T) \hat \delta(\x_2,T) \rangle  =
\frac{1}{i \hbar} \int_{0}^T \mbox{d} t 
\\ \label{BBBmain}
&&  \int \mbox{d}\x \, K(\x_1 ,T; \x,t) K(\x_2 ,T; \x,t)     B (\x ,t).
\end{eqnarray}
It can be interpreted in the following way. The probability amplitude of observing a pair of particles at $\x_1$ and $\x_2$ is the superposition of amplitudes of emitting a pair
of particles, $B(\x,t)$, at position $\x$ and time $t$, which further propagate to final positions $\x_1$ and $\x_2$ and time $T$.

Let us calculate the one body correlation function given by Eq~.(\ref{G1}). Substituting the form of $\hat \delta $ given by Eq.~(\ref{BBBdeltasol}) into
Eq.~(\ref{G1}) we obtain:
\begin{eqnarray*}
\langle 0| \hat \delta^\dagger(\x_1,T) \hat \delta(\x_2,T) |0\rangle &=& \langle 0| (\hat \delta^{\dagger(0)}(\x_1,T)+\hat \delta^{\dagger(1)}(\x_1,T)  ) \times \\
&\times&  (\hat \delta^{(0)}(\x_2,T)+\hat \delta^{(1)}(\x_2,T)  ) |0\rangle .
\end{eqnarray*}
Now, using the same reasoning which led us to the anomalous density, we obtain that the only nonzero term in the above equation is 
$\langle 0| \hat \delta^{\dagger(1)} (\x_1,T) \hat \delta^{(1)}  (\x_2,T) |0\rangle $. We substitute the form of
$\hat \delta^{(0)} $ and  $\hat \delta^{(1)} $ given by Eq.~(\ref{BBBdeltasol}) to obtain:
\begin{widetext}
\begin{eqnarray*}
&&\langle 0| \hat \delta^\dagger(\x_1,T) \hat \delta(\x_2,T) |0\rangle 
= \langle 0| \hat \delta^{\dagger(1)} (\x_1,T) \hat \delta^{(1)}  (\x_2,T) |0\rangle  =
\\
&& = \langle 0 | \int \mbox{d} \x_1' \, K^*(\x_1 ,T; \x_1',0)  \frac{1}{(-i) \hbar} \int_{0}^T \mbox{d} t_1 \int\mbox{d}\x_1'' \int \mbox{d}\x_1''' \, 
K^*(\x_1',0;\x_1'',t_1) B^* (\x_1'' ,t_1)  K(\x_1'' ,t_1; \x_1''',0)  \hat \delta'^{(0)}(\x_1''',0) \times
\\
&&  \times \int \mbox{d} \x_2' \, K(\x_2 ,T; \x_2',0)  \frac{1}{i \hbar} \int_{0}^T \mbox{d} t_2 \int\mbox{d}\x_2'' \int \mbox{d}\x_2''' \, 
K(\x_2',0;\x_2'',t_2) B (\x_2'' ,t_2)  K^{*}(\x_2'' ,t_2; \x_2''',0)  \hat \delta'^{(0)\dagger}(\x_2''',0)|0 \rangle . 
\end{eqnarray*}
\end{widetext}
With the help of the Eqs.~(\ref{BBBr3}) and (\ref{BBBmain}) we obtain the final formula for the single particle correlation function:
\begin{equation} \label{G1pertA}
G^{(1)}(\x_1,\x_2,T) = \int \mbox{d} \x \ M^*(\x_1,\x,T) M(\x,\x_2,T).
\end{equation}

\section{Approximate solution of the GP equation}\label{ApSol}
In this appendix we find approximate initial state of the Bose-Einstein condensate  and its subsequent time evolution.

\subsection{Initial state}\label{Av}

To find the initial state of cloud we rely on the variational method. To this end, we assume a gaussian profile:
\begin{equation} \label{an1}
 \psi(\x) = \sqrt{\frac{N}{\pi^{3/2} \sigma_z \sigma_r^2 }}\exp \left( - \frac{x^2+y^2}{2\sigma_r^2} -\frac{z^2}{2\sigma_z^2} \right),
\end{equation}
with the norm equal to $N$, and search for parameters $\sigma_r$ and $\sigma_z$ that minimize the Hamiltonian of the Gross-Pitaevskii Eq.~(\ref{sGP}):
\begin{equation}\label{Ham}
H = \int \mbox{d} \x \left( \frac{\hbar^2}{2m} |\nabla \psi(\x)|^2 +V(\x)|\psi(\x)|^2 + \frac{g}{2}|\psi(\x)|^4 \right),
\end{equation}
where $V(\x)$ is given by Eq.~(\ref{pot}). Since the trapping frequency in radial direction is much larger than in the axial,$\omega_r \gg \omega_z$, the ground state
wavefunction is very elongated.  This allows us to approximate the Hamiltonian by neglecting the kinetic energy along $z$-direction:
\begin{eqnarray}\nonumber
H &=& \int \mbox{d} \x \, \frac{\hbar^2}{2m} \left(| \partial_x \psi(\x)|^2 + | \partial_y \psi(\x)|^2 \right) 
\\ \label{Ham2}
&+& \int \mbox{d} \x \,\left( V(\x) |\psi(\x)|^2  + \frac{g}{2}|\psi(\x)|^4 \right)
\end{eqnarray}
Now we insert the ansatz from Eq.~(\ref{an1}) into the Hamiltonian in Eq.~(\ref{Ham2}, which now becomes a function of two parameters $\sigma_r$ and $\sigma_z$. 
Taking the first derivatives, equating to zero, and solving, yields
\begin{eqnarray} \label{sr0}
 \sigma_r &=& a_{hor}\left(1 + \sqrt{\frac{2}{\pi}}\frac{Na}{\sigma_z} \right)^{1/4},
\\ \nonumber
\\ \label{sz0}
 \left( \frac{\sigma_z \sigma_r}{a_{hoz}^2} \right)^2  &=& \sqrt{\frac{2}{\pi}} \frac{ Na}{\sigma_z},
\end{eqnarray}
where $a$ is the scattering length, $a_{hor} = \sqrt{\frac{\hbar}{m \omega_r}}$, and $a_{hoz} = \sqrt{\frac{\hbar}{m \omega_z}}$.
Note, that according to Eq.~(\ref{sr0}) 
\begin{equation} \label{srahor2}
\sigma_r \geqslant a_{hor}.
\end{equation}

Finally, Let us calculate a quantity that often appears in this paper, namely the meanfield energy $gn$, $n$ being the maximal density of the cloud.
According to Eq.~(\ref{an1}) and Eq.~(\ref{sr0}) this quantity is
\begin{equation} \label{gn}
 gn = \frac{gN}{\pi^{3/2}\sigma_z \sigma_r^2 } = \frac{2\hbar^2}{m\sigma_r^2} \left( \frac{\sigma_r^4}{a_{hor}^4} -1 \right).
\end{equation}
We also need chemical potential $\mu$, which by definition is the derivative of the ground state energy $E_0$ with respect to the number of atoms, $\mu = \frac{\partial
  E_0}{\partial N}$.  Inserting the variational ansatz from Eq.~(\ref{an1}) into Eq.~(\ref{Ham2}), and using the formulas
obtained above, we arrive at:
\begin{equation} \label{muvar}
\mu = \frac{\hbar^2}{m a_{hor}^2} \left( \frac{7 \sigma_r^2}{4 a_{hor}^2}  - \frac{3a_{hor}^2}{4 \sigma_r^2} \right).
\end{equation}

\subsection{Time evolution}\label{timeevolution}

Due to the symmetry of the trapping potential $V(\x)$, see Eq.~(\ref{pot}), we have $\psi(x,y,z) = \psi(x,y,-z)$.  This fact supplemented by Eq.~(\ref{psip}) implies that
$\psi_{+Q}(\x,0) = \psi_{-Q}(\x,0)$. Using now Eq.~(\ref{SV}), we find the following property of the initial state:
\begin{equation}\label{symetry}
\psi_{+Q}(x,y,z,t) = \psi_{-Q}(x,y,-z,t).
\end{equation}
The term $\mp  i\frac{\hbar^2}{m} Q \partial_z $ in Eq.~(\ref{SV}) is responsible for the movement of the wave-packet with a constant velocity $\pm v_0 = \pm  \hbar Q/m$.
It is thus convenient to define a new function by
\begin{equation}\label{tildepsi1}
\tilde \psi(x,y,z,t) = \psi_{+Q}(x,y,z+v_0t,t). 
\end{equation}
Exploiting  Eqs.~(\ref{symetry}) and (\ref{tildepsi1}), we find that Eq.~(\ref{SV}) can be rewritten as
\begin{eqnarray} \nonumber
&& i \hbar \partial_t \tilde \psi(\x,t) = \left( -   \frac{\hbar^2}{2m} \triangle + g |\tilde \psi(\x,t)|^2  \right) 
\tilde \psi(\x,t) +
\\ \label{svea}
&&   + 2g|\tilde \psi(x,y,-z-2v_0t,t)|^2  \tilde \psi(\x,t).
\end{eqnarray}

This equation describes the expansion of the wave-packet $\tilde \psi$ in the presence of the mean-field potential $g\left(|\tilde \psi(\x,t)|^2+ 2|\tilde
\psi(x,y,-z-2v_0t,t)|^2 \right)$. The term $g|\tilde \psi(\x,t)|^2$ has the same symmetry as $\tilde \psi$ and, thus, it increases the rate of the expansion.  However,
the second term in this potential is caused by the cloud $-Q$ which is moving with velocity $-2v_0$ with respect to the $+Q$ wavepacket, which results in the asymmetry in
the expansion of the $\tilde \psi$ function.  The goal of the present analysis is to obtain an approximate form of $\tilde \psi$.  We therefore approximate the term
$|\tilde \psi(x,y,-z-2v_0t,t)|^2$ by $|\tilde \psi(x,y,z,t)|^2$ restoring the symmetry. 
This approach leads to an ``upper'' limit on the expansion rate. 
Conversely, a ``lower'' limit can be obtained by completely neglecting the term $|\tilde \psi(x,y,-z-2v_0t,t)|^2$. 
In both cases, Eq.~(\ref{svea}) takes the simple form
\begin{eqnarray} \label{sveaN}
i \hbar \partial_t \tilde \psi(\x,t) = \left( -   \frac{\hbar^2}{2m} \triangle + \tilde g |\tilde \psi(\x,t)|^2  \right)
\tilde \psi(\x,t),
\end{eqnarray}
where $\tilde g = 3g$ or $\tilde g = g$ in the ``upper'' or ``lower''  limiting case, respectivelly.

We now use the variational method to solve the above equation, assuming that the norm $\tilde N = \int\!\! \mbox{d} \x |\psi(\x,t)|^2$; we dropped tilde from the wavefunction
for the sake of notational clarity.
The Eq.~\eqref{svea} can be formally derived from minimizing the action
\begin{eqnarray*}
S[\psi] = \int \mbox{d} t \,L(\psi,t),
\end{eqnarray*}
with the Lagrangian
\begin{equation}\label{lan}
L = \int\!\! \mbox{d} \x \!  \left[ i \frac{1}{2}\hbar 
\left( \psi^* \partial_t \psi - \psi \partial_t \psi^* \right)
- \left( \frac{\hbar^2}{2m} |\nabla \psi|^2 
+ \frac{g}{2} |\psi|^4 \right)\right].
\end{equation}
To obtain approximate evolution we assume a time-dependent variational ansatz:
\begin{eqnarray} \nonumber
 \psi(\x,t) &=&  \sqrt{\frac{\tilde N}{\pi^{3/2} \sigma_z(t) \sigma_r^2(t) }}
\exp\left(-  a_r(t) (x^2+y^2) \right)\times
\\ \label{vat}
& & \times\exp \left(- a_z(t) z^2 - i \phi(t) \right),
\end{eqnarray}
where $ a_{r,z}(t) = {1}/{2\sigma_{r,z}^2(t)} - i b_{r,z}(t)$, and $\sigma_{r,z}(t)$, $b_{r,z}(t)$ and $\phi(t)$ are real time dependent variational parameters.  
The norm of the profile is equal to $\tilde N$.  We insert the above profile into the Lagrangian given by Eq.~(\ref{lan}), and integrate over space variables to obtain
\begin{eqnarray*}
&& L =  i \frac{\hbar}{2} ((\dot a_r - \dot a_r^*)  \sigma_r^2 + (\dot a_z - \dot a_z^*)  \frac{1}{2}\sigma_z^2 ) 
\\ 
&&
 + 
\frac{\hbar^2}{2m} ( 2 |a_z|^2 \sigma_z^2 + 4 |a_r|^2 \sigma_r^2) + \frac{\tilde g \tilde N}{2} \frac{1}{(2\pi)^{3/2} \sigma_z \sigma_r^2}.  
\end{eqnarray*}
The Euler-Lagrange equations of motion,
\begin{eqnarray*}
\frac{\mbox{d}}{\mbox{d} t} \frac{\partial L}{\partial \dot b_{r,z}} - \frac{\partial L}{\partial b_{r,z} } = 0
\ \ \mathrm{and} \ \
\frac{\mbox{d}}{\mbox{d} t} \frac{\partial L}{\partial \dot \sigma_{r,z}} - \frac{\partial L}{\partial \sigma_{r,z} } = 0,
\end{eqnarray*}
lead to
\begin{eqnarray*}
&& \hbar 2 \sigma_r \dot \sigma_r  = 4 \frac{\hbar^2}{m} b_r \sigma_r^2, \ \ \ \ \hbar \sigma_z \dot \sigma_z  = 2 \frac{\hbar^2}{m} b_z \sigma_z^2,
\\
\\
&& \hbar \dot b_r 2 \sigma_r - \frac{\hbar^2}{m \sigma_r^3}  + \frac{\hbar^2}{m} 4 b_r^2 \sigma_r
- 2\frac{\tilde g \tilde N}{2} \frac{1}{(2\pi)^{3/2} \sigma_z \sigma_r^3} = 0,
\\
\\
&&   \hbar \dot b_z \sigma_z - \frac{\hbar^2}{2 m \sigma_z^3}   + \frac{\hbar^2}{m} 2 b_z^2 \sigma_z
- \frac{\tilde g \tilde N}{2} \frac{1}{(2\pi)^{3/2} \sigma_z^2 \sigma_r^2} = 0. 
\end{eqnarray*}
These can be transformed into more useful form:
\begin{eqnarray}\label{err}
&& b_r = \frac{m \dot \sigma_r}{2\hbar \sigma_r}, \ \ \ \
\ddot \sigma_r = \frac{\hbar^2}{m^2 \sigma_r^3} \left( 1 + \sqrt{\frac{2}{\pi}}\tilde N \tilde a  \frac{1}{\sigma_z}  \right)  
\\ \nonumber
\\ \label{ezr}
&& b_z = \frac{m \dot \sigma_z}{2\hbar \sigma_z}, \ \ \ \ \ 
 \ddot \sigma_z = \frac{\hbar^2}{m^2}  \left( \frac{1}{\sigma_z^3}  + \sqrt{\frac{2}{\pi}}  \tilde N \tilde a \frac{1}{\sigma_z^2 \sigma_r^2} \right),\quad\quad
\end{eqnarray}
where we substituted $\tilde g = \frac{4\pi \hbar^2 \tilde a}{m}$. The phase $\phi(t)$ is obtained from integrating GP equation:
\begin{eqnarray}\label{eqphase}
 i \hbar \int \psi^* \partial_t \psi = \int \left( \frac{\hbar^2}{2m}  |\nabla \psi|^2 + \tilde g |\psi|^4 \right).
\end{eqnarray}
The primary object of our study is a strongly elongated system for which initially $\sigma_z(0) \gg \sigma_r(0)$.  We can expect that in the course of time $\sigma_r(t)$
changes substantially, whereas the change of $\sigma_z(t)/\sigma_z(0)$ is small. We therefore approximate $\sigma_z(t)$ by $\sigma_z(0)$ in Eqs.~(\ref{err}) and
(\ref{ezr}), which supports analytical solutions of the following form:
\begin{eqnarray} \nonumber
\sigma_r^2(t) &=&   \frac{\hbar^2}{m^2} \left( 1+ \sqrt{\frac{2}{\pi}}\frac{\tilde N \tilde a}{\sigma_z(0)} \right) \frac{t^2}{\sigma_r^2(0)} 
+ \sigma_r^2(0) 
\\ \label{sr}
&=& \sigma_r^2(0) (1+ \tilde \omega^2 t^2),
\\ \nonumber
\\ \nonumber
 b_r(t) &=& \frac{\hbar}{2m} \left( 1+ \sqrt{\frac{2}{\pi}}\frac{\tilde N \tilde a}{\sigma_z(0)} \right) 
\frac{t}{\sigma_r^2(t) \sigma_r^2(0)} 
\\ \label{br}
&=&  \frac{m}{2\hbar} \tilde \omega^2 t \frac{\sigma_r^2(0)}{\sigma_r^2(t)}
= \frac{1}{2 \tilde a_{hor}^2} \tilde \omega t \frac{\sigma_r^2(0)}{\sigma_r^2(t)},
\\ \nonumber
\sigma_z(t) &=&  \sigma_z(0) + \frac{\tilde a_{hor}^4}{2\sigma_z^3(0)} \tilde \omega^2 t^2 +   \sqrt{\frac{2}{\pi}} \tilde N \tilde a  
\frac{\tilde a_{hor}^4}{\sigma_z^2(0) \sigma_r^2(0)}\times
\\ \nonumber
\\ \label{sz}
& & \times \left( \tilde \omega t \arctan(\tilde \omega t) - \log \sqrt{1+\tilde \omega^2t^2}  \right),
\\ \nonumber
\\ \label{bz}
b_z(t) &=& \frac{\tilde a_{hor}^2}{4 \sigma_z^4(0)} \!\left[  \tilde \omega t
+ \sqrt{\frac{8}{\pi}}  
\frac{\tilde N \tilde a \sigma_z(0) }{\sigma_r^2(0)}  \arctan (\tilde \omega t) \right],
\end{eqnarray}
where
\begin{equation} \label{omega2}
\tilde \omega^2 =\frac{\hbar^2}{m^2 \sigma_r^4(0)} \left( 1+ \sqrt{\frac{2}{\pi}}\frac{\tilde N \tilde a}{\sigma_z(0)} \right),
\end{equation}
and
\begin{equation} \label{arrow}
 \tilde a_{hor} = \sqrt{\frac{\hbar}{m \tilde \omega}}.
\end{equation}

The phase is derived from Eq.~(\ref{eqphase}) using Eqs.~(\ref{err}) and (\ref{ezr}), and takes the form:
\begin{eqnarray*}
 \hbar \dot \phi(t) = \frac{\hbar^2}{2m \sigma_z^2(t)} +\frac{\hbar^2}{m \sigma_r^2(t)}
 + \frac{7 \tilde g \tilde N}{8 \sqrt{2} \pi^{3/2} \sigma_z(t)\sigma_r^2(t)}.
\end{eqnarray*}
After neglecting the kinetic energy along $z$-direction in the above equation, and taking $\sigma_z(t)\simeq \sigma_z(0)$, 
we obtain
\begin{eqnarray*}
  \phi(t) = \frac{1+\frac{7}{4}\sqrt{\frac{2}{\pi}} \frac{\tilde N \tilde a}{\sigma_z(0)}}{ \sqrt{1+\sqrt{\frac{2}{\pi}} \frac{\tilde N \tilde a}{\sigma_z(0)}  } } 
\arctan \left( \tilde \omega t \right),
\end{eqnarray*}
or its alternative form 
\begin{equation}\label{phiwynik}
  \phi(t) = \left( \frac{7}{4} \frac{\sigma_r^2}{\tilde a_{hor}^2} -\frac{3}{4} \frac{\tilde a_{hor}^2}{\sigma_r^2}  \right)
\arctan \left( \tilde \omega t \right).
\end{equation}
To derive the latter, we used Eqs.~(\ref{omega2}) and (\ref{arrow}).

As can be seen from Eq.~(\ref{sr}), the characteristic time on which $\sigma_r(t)$ changes its width is equal to $1 /\tilde \omega $. To be consistent with the assumption
stated in the derivation of the above formulas, the change of $\sigma_z(t)$ during that time has to be much smaller than $\sigma_z(0)$. Invoking Eq.~(\ref{sz}), this
condition takes the form:
\begin{equation}\label{condition22}
\frac{\tilde a_{hor}^4}{2\sigma_z^4(0)} +   \sqrt{\frac{2}{\pi}} \tilde N \tilde a  
\frac{\tilde a_{hor}^4}{\sigma_z^3(0) \sigma_r^2(0)} \ll 1. 
\end{equation}

Let us now concentrate on the upper limit and take $\tilde N = \frac{N}{2}$, $\tilde a = 3 a$, with the initial condition $\sigma_{r,z}(0) = \sigma_{r,z}$ given by
Eqs.~(\ref{sr0}) and (\ref{sz0}).  Then, the condition from Eq.~(\ref{condition22}) reads
\begin{equation}\label{condition22n}
\frac{\tilde a_{hor}^4}{2\sigma_z^4} +   \sqrt{\frac{2}{\pi}} \frac{3}{2} Na  
\frac{\tilde a_{hor}^4}{\sigma_z^3 \sigma_r^2} \ll 1. 
\end{equation} 
Now, from Eqs.~(\ref{sr0}), (\ref{omega2}) and (\ref{arrow}) we obtain
\begin{eqnarray}\label{tildea1}
\frac{a_{hor}^4}{\tilde a_{hor}^4} 
= \frac{ 1+ \sqrt{\frac{2}{\pi}} \frac{3}{2} \frac{Na}{\sigma_z} }{1+ \sqrt{\frac{2}{\pi}}\frac{Na}{\sigma_z}}.
\end{eqnarray}
Due to the following inequality,  
\begin{eqnarray*}
\frac{3}{2} > \frac{ 1+ \sqrt{\frac{2}{\pi}} \frac{3}{2} \frac{Na}{\sigma_z} }{1+ \sqrt{\frac{2}{\pi}}\frac{Na}{\sigma_z}} > 1,
\end{eqnarray*}
we obtain that 
\begin{equation}\label{similar}
 a_{hor} \geqslant \tilde a_{hor} \geqslant \left( \frac{2}{3}\right)^{\frac14}\! a_{hor} 
\ \ \mathrm{and} \ \
\sqrt{\frac{3}{2}}\omega \geqslant \tilde \omega \geqslant \omega,
\end{equation}
where we have used Eq.~(\ref{arrow}).  
The high anisotropy, for which $\sigma_r \ll \sigma_z$, combined with Eq.~(\ref{srahor2}) results in $a_{hor} \ll \sigma_z$.
As a consequence, the first factor in condition from Eq.~(\ref{condition22n}) is small,  $\frac{\tilde a_{hor}^4}{2\sigma_z^4} \ll 1$.  
From Eqs.~(\ref{sr0}) and (\ref{sz0}), we obtain
\begin{eqnarray*}
\sqrt{\frac{2}{\pi}} \frac{3}{2} Na  
\frac{a_{hor}^4}{\sigma_z^3 \sigma_r^2} = \frac{3}{2} \frac{a_{hor}^4}{a_{hor}^4} = \frac{3}{2} \frac{\omega_z^2}{\omega_r^2} \ll 1.
\end{eqnarray*}
Thus, we have shown that the condition in Eq.~(\ref{condition22}) is satisfied, making the above derivation self-consistent.

Now, we simplify the expression for $b_z(t)$ given in Eq.~(\ref{bz}).
To this end, notice that the maximal phase that appears in the variational ansatz is roughly equal to $b_z(t)\sigma_z^2$.
Then, the first term in expression (\ref{bz}) takes the form $\frac{\tilde a_{hor}^2}{4 \sigma_z^2} \tilde \omega t$. 
As $\omega t$ is maximally of the order of few, and due to elongation of the system, $\tilde a_{hor} \ll \sigma_z$, this term can be neglected.
As a result, we obtain
\begin{eqnarray*}
b_z(t) = \frac{\tilde a_{hor}^2}{2 \sigma_r^2 \sigma_z^2}  \sqrt{\frac{2}{\pi}}  
\frac{\tilde N \tilde a}{\sigma_z}  \arctan (\tilde \omega t),
\end{eqnarray*}
with the alternative form 
\begin{equation}\label{bzwynik}
b_z(t) = \frac{1}{2\sigma_z^2} \left( \frac{\sigma_r^2}{\tilde a_{hor}^2}  -\frac{\tilde a_{hor}^2}{\sigma_r^2}  \right) \arctan (\tilde \omega t),
\end{equation}
where Eqs.~(\ref{omega2}) and (\ref{arrow}) were used.

Let us notice that in the case of ``lower'' limit, for which $\tilde g = g$, we obtain different values of $\tilde a_{hor}$
and $\tilde \omega$, that satisfy following chain of inequalities:
\begin{equation}\label{similar2}
 2^{1/4} a_{hor} \geqslant \tilde a_{hor} \geqslant  a_{hor} 
\ \ \mathrm{and} \ \
\omega \geqslant \tilde \omega \geqslant \frac{1}{\sqrt{2}} \omega.
\end{equation}
We may conclude that there is no substantial qualitative difference between these two cases. Therefore, in the rest of the paper we take $\tilde g = 2g$ as a
interpolation between the lower and upper limits.  In such a case, we have $\tilde a_{hor} = a_{hor}$ and $\tilde \omega = \omega_r$.  Finally, using
Eqs.~(\ref{symetry}), (\ref{tildepsi1}), (\ref{vat}) together with Eqs.~(\ref{sr}), (\ref{br}), (\ref{phiwynik}) and (\ref{bzwynik}) we obtain approximate forms for
$\psi_{\pm Q}$, which are given by:
\begin{eqnarray} \nonumber
&&  \psi_{\pm Q}(\x,t) =  \sqrt{\frac{N}{2\pi^{3/2} \sigma_z \sigma_r^2(1+\omega_r^2 t^2) }}\times
\\ \nonumber
&& \times\exp \left(- \frac{(z \mp v_0t)^2}{2\sigma_z^2}\left( 1 - i \left(  \beta - \frac{1}{\beta}  \right) \arctan(\omega_r t) \right)
\right) \times
\\ \nonumber
&& \times \exp\left(-   \frac{x^2+y^2}{2\sigma_r^2 } \frac{ 1 - i\beta \omega_r t}{1+\omega_r^2 t^2 }  
- i  \left( \frac{7\beta}{4} -  \frac{3}{4\beta} \right) \arctan (\omega_r t)\right),
\end{eqnarray}
where $\beta = \frac{\sigma_r^2}{a_{hor}^2}$, which, according to Eq.~(\ref{srahor2}), cannot be less than unity, $\beta \geqslant 1$.

The momentum density corresponding to the wavefunction is given by
\begin{eqnarray*}
|\psi_{\pm Q}(\K,t) |^2 \propto \exp \left( - \frac{k_r^2}{\sigma_{kr}^2(t)} - \frac{k_z^2}{\sigma_{kz}^2(t)}  \right),
\end{eqnarray*}
where $k_r^2 = k_x^2 + k_y^2$, and
\begin{eqnarray}\label{sigmakr}
\sigma_{kr}^2(t) &=&  \frac{1+\beta^2 \omega_r^2 t^2}{\sigma_r^2(1+ \omega_r^2 t^2)},
\\ \label{sigmakz}
\sigma_{kz}^2(t) &=& \frac{1}{\sigma_z^2} \left(1 +\left(  \beta - \frac{1}{\beta}  \right)^2 \arctan^2(\omega_r t)  \right).
\end{eqnarray}
We note following useful relations: initially $\sigma_{kr,kz}(0) = {1}/{\sigma_{r,z}}$, the final value of the width in radial is
$\sigma_{kr}(\infty) = {\beta}/{\sigma_r} = {\sigma_r}/{a_{hor}^2}$, and the axial width is
$ \sigma_{kz}(\infty) = ({1}/{\sigma_z}) \sqrt{ 1 +\left(  \beta - {1}/{\beta}  \right)^2 {\pi^2}/{4} } $.

In Section~\ref{model} we stated two assumptions. Now, with help of the derived formulas, they can be written as
\begin{equation} \label{dwacond}
 \frac{\hbar^2 Q^2}{2m} \gg gn   \ \ \mathrm{and} \ \ Q \gg \sigma_{kr}(\infty).
\end{equation}
Using Eq.~(\ref{gn}), the first of the above conditions can be rewritten as
\begin{eqnarray*}
\frac{4}{Q^2 \sigma_r^2} \left( \frac{\sigma_r^4}{a_{hor}^4} -1 \right) \ll 1.
\end{eqnarray*}
Thus, if 
\begin{eqnarray*}
\frac{2\sigma_r}{Qa_{hor}^2} \ll 1
\end{eqnarray*}
both of the conditions present in Eq.~(\ref{dwacond}) are satisfied.

\section{Inclusion of the mean field propagator}\label{APMF}

\subsection{Construction of the mean-field propagator} \label{MeanApp}
Here we describe how to construct approximate formula for the propagator $K$ of the Hamiltonian given in Eq.~(\ref{H0}).
The mean field potential present $\hat H_0$ equals $2 g|\psi(\x,t)|^2$.
Using the decomposition from Eq.~(\ref{psip}) we arrive at
\begin{eqnarray*}
2g|\psi(\x,t)|^2 &=& 2g (|\psi_Q (\x,t)|^2 + |\psi_{-Q} (\x,t)|^2) \
\\
& & +2g (\psi_{+Q}^*\psi_{-Q}e^{-2iQz}+c.c. ).
\end{eqnarray*}
The mean field potential decomposes into two parts: a slowly varying envelope part $ V_{en}(\x,t) = 2g (|\psi_Q (\x,t)|^2 + |\psi_{-Q} (\x,t)|^2) $, and an oscillating
part $V_{osc}(\x,t)$ with the fringes oscillating as  $\cos (2Qz)$.  

We begin with neglecting the oscillating part of the
potential $V_{osc}$ in the mean field potential, the scattered particles are then influenced only by $V_{en}$. We comment on this approximation below.
To remind, we restricted our analysis to the collision of highly elongated cigar shaped condensates along the longitudinal $z$-direction.
Also, we are interested only in the atoms that scattered away from the condensates with velocities distant from the $z$-axis.
The time needed for these atoms to leave the cloud is approximately equal to ${\sigma_r}/{v_0}$. During this interval each of the condensates moves by the
distance equal to $v_0({\sigma_r}/{v_0}) = \sigma_r$, which is much smaller than the longitudinal size $\sigma_z$ of the cloud.  
On the other hand, during this time the radial width of the condensates, given by Eq.~(\ref{parwar}), increases by a factor
\begin{eqnarray*}
\frac{\sqrt{1+ \omega_r^2 \left( t + \frac{\sigma_r}{v_0} \right)^2}}{\sqrt{1+ \omega_r^2 t^2} } \simeq 
1 + \frac{ \omega_r t  }{1+\omega_r^2t^2} \frac{\sigma_r}{Q a_{hor}^2} \simeq 1,
\end{eqnarray*}
which follows directly from the condition in Eq.~(\ref{Cond1}). Thus, the condensate densities, and similarly $V_{en}$, do not change appreciably during the time that
takes the scattered atoms to escape the clouds. In such a case, we are legitmate to approximate the time dependent scattering one-body problem
\begin{eqnarray*}
i \hbar \partial_t \varphi(\x,t) = \left( -\frac{\hbar^2}{2m}\triangle +V_{en}(\x,t)  \right) \varphi(\x,t)  
\end{eqnarray*}
by 
\begin{eqnarray}\label{sch}
\left( -\frac{\hbar^2}{2m} \triangle + V_{en}(\x,t) \right) \varphi_\K(\x,t) = \hbar \omega_\K \varphi_\K(\x,t),
\end{eqnarray}
where $\hbar \omega_\K = \frac{\hbar^2 k^2}{2m}$ and $\varphi(\x,t)$ is a plane wave far away for the potential
$\lim_{r \rightarrow \infty } \varphi(\x,t)  = \frac{1}{(2\pi)^{3/2}} e^{i\K\x} $. 
The solution of the above equation is not unique.  However, in scattering theory we identify two solutions denoted as $\varphi_\K^{(+)}$ and $\varphi_\K^{(-)}$ 
with the boundary condition
\begin{equation} \label{boundary}
\lim_{r \rightarrow \infty} \varphi_\K^{(\pm)}(\x) = 
\frac{1}{(2\pi)^{3/2}} e^{i\K\x}.
\end{equation}
The function $\varphi_\K^{(+)}$ describes a physical situation where an incident particle comes from $r=-\infty$ and scatters on the potential, resulting in the scattered
waves that are directed outside the potential.  The $\varphi_\K^{(-)}$ describes the time-reversed situation where the scattered waves are directed toward the potential.
For a given $t$ both of these sets of functions form a complete and orthogonal set \cite{zderzenia}. Thus, from both of these sets we can construct a propagator
\begin{eqnarray} \nonumber
 K(\x_1,t_1;\x_2,t_2)  &=& \int\!\! \mbox{d} \K \,e^{-i\omega_\K(t_1-t_2)} \times
\\ \label{prop0}
& & \times
 \varphi_\K^{(\pm)}(\x_1,t) (\varphi_\K^{(\pm)}(\x_2,t))^*. 
\end{eqnarray}
This is our approximation to the true propagator of the Hamiltonian $\hat H_0$ with neglected $V_{osc}$. 
As we will see below, $\x_2$ and $t_2$ is the position and time the scattered particle is produced in the condensate
whereas $\x_1$ and $t_1$ is the time and position of the measurement.
As the detection is far away from the condensate then $\varphi(\x_1,t) \simeq \frac{1}{(2\pi)^{3/2}} e^{i\K\x} $,
and it does not depend on $t$. The time $t$ has to be taken as some mean time between the time of the birth of the scattered particle
and the time the particle leaves the cloud.
We have shown above that on the time the particle leaves the cloud $V_{en}$ practically does not change.
Thus, we can take $t=t_2$, which results in
\begin{eqnarray} \nonumber
 K(\x_1,t_1;\x_2,t_2)  &=& \int \mbox{d} \K \,e^{-i\omega_\K(t_1-t_2)} \times
\\ \label{prop}
& & \times
 \varphi_\K^{(\pm)}(\x_1,t_2) (\varphi_\K^{(\pm)}(\x_2,t_2))^*. 
\end{eqnarray}

We are now ready to derive an approximate analytical formula for $\varphi_\K^{(+)}$.  In Section~\ref{lc} we showed that under presented approximations the width of the
halo of scattered atoms is much smaller than its radius being close to $Q$.  Thus, we search for $\varphi_\K^{(+)}$ only for $k$ close to $Q$.  The characteristic length
on which the potential $V_{en}$ changes, which is of the order of $\sigma_r$, is much larger than the mean wavelength of the scattered atom $\frac{2\pi}{Q}$ (this fact
follows from Eq.~(\ref{Cond1}) and the fact that $\sigma_r > a_{hor}$ as stated in Eq.~(\ref{srahor2})). In such a situation the use of semiclassical approximation is
justified.  In subsection~\ref{sss} of this Appendix we show that under condition given in Eq.~(\ref{condition3}) the wavefunction $\varphi_\K^{(+)}$ takes the approximate form
\begin{eqnarray} 
 \varphi^{(+)}_{\K}(\x,t) &=& \frac{1}{(2\pi)^{3/2}}  \exp \left( i  \K \x - i\Phi(\x,{\bf e}_\K,t) \right),
\\ \label{phip}
 \Phi(\x,{\bf e}_\K,t)  &=&    \frac{m}{\hbar^2 Q}\int^0_{-\infty} \mbox{d} s \,  V_{en}(\x  + s  {\bf e}_\K,t),
\end{eqnarray}
where ${\bf e}_\K =\frac{\K}{k} $.  The above formula is derived in \cite{zderzenia} and is known as the ``eikonal approximation''.  The expression for $
\varphi^{(+)}_{\K}(\x) $ is a correct approximation in the part of space before, understood as $\x$ that satisfy $\K\cdot \x < 0$, as well as on the potential.  After the
potential the scattering part appears which is clearly not present in the above formula. This means that the form of $ \varphi^{(+)}_{\K}(\x) $ is no longer given by
Eq.~(\ref{phip}). The problem is that in our calculations we need to know the form of $ \varphi^{(+)}_{\K}(\x)$ on and also after the potential.  The way to overcome this
issue is to realize that in the construction of the propagator $K$ we can use the states $\varphi^{(-)}$ instead of $\varphi^{(+)}$. These states satisfy
\begin{equation} \label{pm} 
\varphi_\K^{(+)*} = \varphi_{-\K}^{(-)},
\end{equation}
so $\varphi_\K^{(+)}$ {\it before the potential} is $\varphi_\K^{(-)}$ {\it after the potential}. 
As a result we take
\begin{eqnarray} \nonumber
 K(\x_1,t_1;\x_2,t_2)  &=& \int \mbox{d} \K \,e^{-i\omega_\K(t_1-t_2)} \times
\\ \label{prop2}
& & \times
 \varphi_\K^{(-)}(\x_1,t_2) (\varphi_\K^{(-)}(\x_2,t_2))^*,
\end{eqnarray}
together with 
\begin{eqnarray} \nonumber 
 \varphi^{(-)}_{\K}(\x,t) &=& \frac{1}{(2\pi)^{3/2}}  \exp \left( i  \K \x \right) \times
\\ \label{phim}
&\times&   \exp \!\left[ i\int^0_{-\infty} \mbox{d} s \,  \frac{m}{\hbar^2 Q} V_{en}(\x  - s  {\bf e}_\K,t)  \right],
\end{eqnarray}
which is defined on and after the potential $V_{en}$. In deriving the above, we used the fact that ${\bf e}_{-\K} = - {\bf e}_\K$.

Finally, let us comment on the omission of $V_{osc}$ part of the total mean-field potential.
This potential has fringes represented by the term  $\cos 2Qz$. Thus, in terms of the time dependent perturbation theory
the potential couples incoming plane wave $e^{i\K \x}$ to the plane wave $\exp \left( i \K \x \pm i2Qz \right)$.
The matrix coupling elements of $V_{osc} \propto gn$, and so the probability amplitude
of such coupled wavefunction should be proportional to 
\begin{eqnarray*}
 \frac{gn}{\hbar(\omega_{\K + 2Q {\bf e}_z} - \omega_\K) } \approx \frac{gn}{\frac{\hbar^2Q^2}{2m}},
\end{eqnarray*}
which is much less that unity. Therefore, we neglect $V_{osc}$ in our considerations.

\subsection{Anomalous density}\label{appendixAD}

Here we analyze the formula for the anomalous density. 
Inserting Eq.~(\ref{prop2}) into Eq.~(\ref{M}), omitting the superscript $(-)$, we obtain
\begin{widetext}
\begin{eqnarray} \nonumber
&& M(\x_1,\x_2;T)   =   \frac{1}{i \hbar} \int_{0}^T \mbox{d} t \int \mbox{d} \x \,  
 \int \mbox{d}\K_1' \mbox{d} \K_2' \, \varphi_{\K_1'}(\x_1,t) \varphi^*_{\K_1'}(\x,t)  \varphi_{\K_2'}(\x_2,t) \varphi^*_{\K_2'}(\x,t) 
 \exp \left( - i (\omega_{\K_1'}+\omega_{\K_2'})(T-t)  \right) B(\x,t)
\\ \nonumber
\\ \label{Mps}
&& =  \frac{1}{i \hbar} \int \mbox{d}\K_1' \mbox{d} \K_2' \,  \varphi_{\K_1'}(\x_1,t)  \varphi_{\K_2'}(\x_2,t) \int_0^T \mbox{d} t \int \mbox{d} \x \,
\varphi^*_{\K_1'}(\x,t) \varphi^*_{\K_2'}(\x,t)\exp \left( - i (\omega_{\K_1'}+\omega_{\K_2'}) (T-t)  \right) B(\x,t).
\end{eqnarray}
\end{widetext}
Taking this as a starting point, we derive a formula for $M(\K_1,\K_2)$ defined in Eq.~(\ref{Mss}). 
For large $\x_1$ and $\x_2$ the wavefunctions $\varphi_{\K_1'}(\x_1,t)$ and $\varphi_{\K_2'}(\x_2,t) $ are plane waves 
(close to the potential the scattered part of the wave-functions may still be present):
\begin{eqnarray*}
&& \lim_{r_1,r_2 \rightarrow \infty}  \varphi_{\K_1'}(\x_1,t)  \varphi_{\K_2'}(\x_2,t)  \exp \left( - i  (\omega_{\K_1'}+\omega_{\K_2'}) T  \right)
\\
&& = \frac{1}{(2\pi)^3} \exp \left(i\K_1'\frac{\hbar \K_1}{m}T  + i \K_2' \frac{\hbar \K_2}{m}T  - i  \frac{\hbar}{2m} ({k_1'}^2+{k_2'}^2) T  \right),
\end{eqnarray*}
where we have used $\x_{1,2} = \frac{\hbar \K_{1,2}}{m}T$.
Introducing $\K_1' = \K_1 + \delta \K_1$ and $\K_2' = \K_2 + \delta \K_2$ we obtain
\begin{eqnarray*}
&& \frac{1}{(2\pi)^3} \exp \left(i\K_1' \x_1 + i \K_2' \x_2 \right) \exp \left( - i  \frac{\hbar}{2m} ({k_1'}^2+{k_2'}^2) T  \right)
\\
&&
=  \frac{1}{(2\pi)^3}\exp \left( i  \frac{\hbar}{2m} (k_1^2+ k_2^2 - \delta k_1^2 - \delta k_2^2) T   \right). 
\end{eqnarray*}
Using the above together with Eqs.~(\ref{Mps}) and (\ref{Mss}) we obtain
\begin{eqnarray} \nonumber
&& M(\K_1,\K_2)   =  \left( \frac{\hbar T}{m}  \right)^3\lim_{T\rightarrow \infty}   \frac{1}{i \hbar} \int \mbox{d} \delta \K_1 \mbox{d} \delta \K_2 
\\ \nonumber
&& \times \frac{1}{(2\pi)^3} \exp \left( - i  \frac{\hbar}{2m} ( \delta k_1^2 + \delta k_2^2) T   \right)
\\ \nonumber
&&
\int_0^T \mbox{d} t \int \mbox{d} \x \,
\varphi^*_{\K_1'}(\x,t) \varphi^*_{\K_2'}(\x,t)\exp \left( i (\omega_{\K_1'}+\omega_{\K_2'}) t  \right) B(\x,t).
\end{eqnarray}
Here the term $\exp \left( - i  \frac{\hbar}{2m} (\delta k_1^2 +\delta k_2^2) T   \right) $  
for $T \rightarrow \infty$ serves as an effective Dirac delta function:
\begin{eqnarray*}
 \exp \left( - i  \frac{\hbar}{2m} \delta k^2 T   \right)  = \left(\frac{2\pi m}{i\hbar T} \right)^{3/2} \delta (\delta \K).
 \end{eqnarray*}
As a result, we obtain
\begin{eqnarray} \nonumber
M(\K_1,\K_2)  &=&    \frac{1}{\hbar} 
\int_0^\infty \mbox{d} t \int \mbox{d} \x \,
\varphi^*_{\K_1}(\x,t) \varphi^*_{\K_2}(\x,t) \times
\\ \label{Mmm}
& & \times \exp \left( i \frac{\hbar(k_1^2+k_2^2)}{2m}t  \right) B(\x,t).
\end{eqnarray}
Now, from Eq.~(\ref{phim}) we obtain
\begin{eqnarray}\label{wyr}
&& \varphi^*_{\K_1}(\x,t) \varphi^*_{\K_2}(\x,t) = \frac{1}{(2\pi)^3} \exp \left( - i (\K_1 + \K_2) \x  \right)
\\ \nonumber
&& 
\exp \!\left[ - i \frac{m}{\hbar^2 Q}\int^0_{-\infty}\!\!\! \mbox{d} s \,  
\left(V_{en}(\x  - s  {\bf e}_{\K_1},t)  + V_{en}(\x  - s  {\bf e}_{\K_2},t)\right) \right].
\end{eqnarray}
The phase in square brackets  is an integral over two straight lines meeting at point $\x$.
In Section~\ref{Sbb} we showed that when the free propagator is used the wavevctors $\K_1$ and $\K_2$
for which $M(\K_1,\K_2)$ has non-vanishing value are practically anti-parallel, with the length approximately equal to 
$Q$. Let us show that the same applies here as well. To this end, let us analyze the temporal and spatial dependence 
of the integrand in Eq.~(\ref{Mmm}). We have two terms with such a dependence: $B(\x,t) $ and
the phase $\frac{m}{\hbar^2 Q}\int^0_{-\infty}\!\! \mbox{d} s \,  
\left(V_{en}(\x  - s  {\bf e}_{\K_1},t)  + V_{en}(\x  - s  {\bf e}_{\K_2},t)\right)  $.
As found in the Section~\ref{Sbb}, the dominant temporal phase present in $B(\x,t)$ is given by $\exp \left( -{\hbar^2Q^2t}/{m} \right)$. 
On the other hand, the integral $\int_{-\infty}^0 \mbox{d} s \, \left( V_{en}(\x  - s  {\bf e}_{\K_1},t) + V_{en}(\x  - s  {\bf e}_{\K_2},t) \right) $ can be estimated as $gn \sigma_r$. 
The temporal change of this integral can be estimated as $gn \sigma_r \frac{t}{t_c}$, where
$t_c$ is the characteristic time of the change of the potential $V_{en}$.
As we have shown in subsection~\ref{MeanApp} of this Appendix, the change of the potential $V_{en}$ takes place during
time which is much larger than $\frac{\sigma_r}{v_0}$. Thus, we have  $ t_c = \frac{1}{\epsilon} \frac{\sigma_r}{v_0}$
with $\epsilon \ll 1$. As a result, the temporal dependence of 
$\frac{m}{\hbar^2 Q}\int^0_{-\infty} \mbox{d} s \,  \left(V_{en}(\x  - s  {\bf e}_{\K_1},t)  + V_{en}(\x  - s  {\bf e}_{\K_2},t)\right)  $  can be approximated by 
\begin{eqnarray*}
\frac{m gn}{\hbar^2 Q} \sigma_r \frac{t}{t_c}  =  \epsilon \frac{gn}{\hbar} t.
\end{eqnarray*}
As  $\frac{\hbar^2 Q^2}{2m} \gg gn$, which is the assumption stated in Section \ref{model}, the above is much smaller than
the temporal phase of $B$ equal to $- i{\hbar^2Q^2t}/{m} $. 
When we integrate this two terms with $\exp \left( i {\hbar(k_1^2+k_2^2)t}/{2m}  \right)$,
present in Eq.~(\ref{Mmm}), we obtain $k_1^2 + k_2^2 \simeq 2 Q^2$. 

Let us now analyze the spatial dependence of the integral 
$\int_{-\infty}^0 \mbox{d} s \, \left( V_{en}(\x  - s  {\bf e}_{\K_1},t) + V_{en}(\x  - s  {\bf e}_{\K_2},t) \right) $.
We approximate it as a drop from the maximal value $gn$ to zero on a distance equal to $\sigma_r$. For simplicity, we take $x$ in the direction of the drop 
which results in $\int_{-\infty}^0 \mbox{d} s \, \left( V_{en}(\x  - s  {\bf e}_{\K_1},t) + V_{en}(\x  - s  {\bf e}_{\K_2},t) \right) 
\approx gn \sigma_r \frac{x}{\sigma_r}$. Thus, the phase factor of the analyzed term can be approximated by
$\exp \left(- i {m gn x}/{\hbar^2 Q}  \right) $.
As $gn \ll \frac{\hbar^2Q^2}{2m}$, the above term results in a much smaller shift of the wavevector than $Q$.
On the other hand, according to the assumption stated in Section~\ref{model} the momentum width
of the function $\psi_{\pm Q}$ is much smaller than $Q$. This two facts together with the definition
of $B(\x,t)$, given by Eq.~(\ref{Bp}), imply that $|\K_1 +\K_2| \ll Q$. 
Thus, we have shown that indeed the mean-field term coming from the propagator does not change 
the fact that $\K_1$ and $\K_2$ have lengths close to $Q$ and are approximatelly anti-parallel.

We can now simplify the expression for $M(\K_1,\K_2)$, given by Eqs.~(\ref{Mmm}) and~(\ref{wyr}). 
As $\K_1$ and $\K_2$ are almost anti-parallel, we approximate
\begin{eqnarray*}
&& \int_{-\infty}^0 \mbox{d} s \, \left( V_{en}(\x  - s  {\bf e}_{\K_1},t) + V_{en}(\x  - s  {\bf e}_{\K_2},t) \right)
\\
&& \simeq \int^0_{-\infty} \mbox{d} s \, V_{en}(\x  + s  {\bf e}_{\KK},t),
\end{eqnarray*} 
where $\KK  = \frac{\K_1-\K_2}{2}$. Consequently, from Eqs.~(\ref{Mmm}) and (\ref{wyr})
we finally obtain
\begin{eqnarray} \label{Mtdrugie00}
&& M(\K_1, \K_2)  
= \frac{1}{\hbar (2\pi)^3}   \int_0^\infty \mbox{d} t \int \mbox{d} \x \, 
\\ \nonumber
&&
\exp \left(  - i (\K_1+\K_2) \x + i \frac{\hbar(k_1^2+k_2^2)}{2m}  t \right) B(\KK,\x,t),
\end{eqnarray}
where
\begin{eqnarray}\label{Bmf00}
B(\KK,\x,t) &=& B(\x,t )\exp \left(- i \Phi (\x,{\bf e}_\KK,t)   \right),
\\ \nonumber
\Phi (\x,{\bf e}_\KK,t) &=& \frac{m}{\hbar^2 Q} \int^\infty_{-\infty} \mbox{d} s \,  V_{en}(\x  + s  {\bf e}_{\KK},t).
\end{eqnarray}

From the results of subsection~\ref{sss} of this Appendix, the maximal value of $|\Phi|$ can be estimated as 
$ \frac{m 2gn}{\hbar^2 Q} 4 \sigma_r$. Therefore, if 
\begin{equation}\label{con1111}
4 \frac{gn}{\frac{\hbar^2 Q^2}{2m}} Q \sigma_r \ll 1,
\end{equation}
the phase can be neglected and then we have $ B(\KK,\x,t) \simeq B(\x,t )$.
With help of Eq.~(\ref{gn}) this condition can be rewritten in the following form
\begin{eqnarray*}
 \frac{16}{Q\sigma_r} \left( \frac{\sigma_r^4}{a_{hor}^4} -1 \right) \ll 1.
\end{eqnarray*}
If the condition
\begin{eqnarray*}
 \frac{16\sigma_r^3}{Qa_{hor}^4}  \ll 1
\end{eqnarray*}
is true, the one in Eq.~(\ref{con1111}) is satisfied.

\subsection{Approximate solution of the scattering problem} \label{sss}

Let us now analyze the scattering problem given by Eq.~(\ref{sch}). It can be rewritten as
\begin{eqnarray*}
\left( -\frac{\hbar^2}{2m} \triangle + V_{en}(\x,t) - \frac{\hbar^2 k^2}{2m} \right) \varphi_\K^{(+)}(\x,t) =0.
\end{eqnarray*}
To simplify the notation, in this subsection omit superscript $(+)$, subscript $\K$ and time $t$ in $\varphi_\K^{(+)}(\x,t)$ and $V_{en}(\x,t)$.  To find the approximate
solution of this Schr\"odinger equation we substitute $\varphi(\x) =  e^{i\K \x + i\phi(\x)}/{(2\pi)^3} $. 
Then, the Schr\"odinger equation takes the following form
\begin{eqnarray*}
- i \triangle \phi + 2 \K \nabla \phi + (\nabla \phi)^2 + \frac{2m}{\hbar^2}V_{en} = 0.
\end{eqnarray*}
We solve this equation by a perturbation series $\phi = \phi^{(0)}+ \phi^{(1)} + ...$. We have the set of equations
\begin{eqnarray*}
2 \K \nabla \phi^{(0)}  &=&  -\frac{2m}{\hbar^2}V_{en}
\\ 
 2 \K \nabla \phi^{(1)} &=& - (\nabla \phi^{(0)})^2 + i \triangle \phi^{(0)}.
\end{eqnarray*}
The solution of the above is
\begin{equation}\label{solphikoniec} 
\phi^{(j)}(\x) = - \frac{1}{2 k} \int_{-\infty}^0 \mbox{d} s \, W_j(\x + s {\bf e}_\K),
\end{equation}
where $W_0 = \frac{2m}{\hbar^2}V_{en}$ and $W_1 =(\nabla \phi^{(0)})^2 - i \triangle \phi^{(0)} $.
The maximal value of $| \phi^{(j)}|$ can be estimated as
\begin{equation}\label{estp}
| \phi^{(j)}| \leq \frac{1}{2Q} 4 \sigma_r |W_j|_{max},
\end{equation}
where $|W_j|_{max}$ is the maximal value of the function $|W_j|$.
In the above, we have taken the path through the center of the condensate
with the  smallest possible value of $\sin \theta = \frac{1}{2}$ and the effective width
equal to $\frac{2\sigma_r}{|\sin \theta|} = 4 \sigma_r$.
In this way we arrive at
\begin{equation}\label{estphi0}
| \phi^{(0)}| \leqslant 4 \frac{gn}{\frac{\hbar^2Q^2}{2m}}  Q\sigma_r,
\end{equation}
where we took $|V_{en}|_{max} = 2 gn$. 
As the width of $ \phi^{(0)}_1$  is of the order of  $\sigma_r$ in the axial direction, we estimate 
$\nabla $ and $\triangle$ operators acting on $ \phi^{(0)}$ as $\frac{1}{\sigma_r}$ and $\frac{1}{\sigma_r^2}$.
Then, using Eqs.~(\ref{estp}) and (\ref{estphi0}), we obtain
\begin{equation}\label{estphi1}
| \phi^{(1)}| \leq 32 \left(\frac{gn}{\frac{\hbar^2Q^2}{2m}} \right)^2  Q\sigma_r + 8 \frac{gn}{\frac{\hbar^2Q^2}{2m}}.
\end{equation}
According to the assumption stated in Section~\ref{model}, $ \frac{\hbar^2 Q^2}{2m} \gg gn$,  so the second term
on the righthand side of the above equation can be neglected.
Thus, from Eq.~(\ref{solphikoniec}), approximating  $\frac{1}{2k} \simeq \frac{1}{2Q} $, we obtain
\begin{equation} \nonumber 
 \varphi(\x) = \frac{1}{(2\pi)^{3/2}}  e^{ i  \K \x  - i \frac{m}{\hbar^2 Q}\int^0_{-\infty} \mathrm{d} s \,  V_{en}(\x  + s  {\bf e}_\K,t) },
\end{equation}
as long as
\begin{equation}\label{condition3}
32 \left(\frac{gn}{\frac{\hbar^2 Q^2}{2m}} \right)^2  Q\sigma_r  \ll 1
\end{equation}
is satisfied.

\section{Derivation of expressions used in Section~\ref{Sbb}}\label{bbApp}

In this Appendix we calculate the expressions presented in Section~\ref{Sbb}, in the order they appear in the text.

First, we investigate the semiclassical model.
The Wigner function in the case of gaussian ansatz, see Eq.~(\ref{vatnowe}),
takes the form
\begin{widetext}
\begin{eqnarray} \nonumber 
&& W_{\pm Q}(\x,\K) = \frac{4 N}{(2\pi)^3}   \exp \left( - \frac{x^2+y^2}{\sigma_r^2(t)}  - \frac{(z \mp v_0t)^2}{\sigma_z^2}
-  \left( k_x\sigma_r(t) - \beta \frac{\omega_rt x}{\sigma_r(t)} \right)^2 +
\left( k_y\sigma_r(t) - \beta \frac{\omega_rt y}{\sigma_r(t)} \right)^2   \right) \times
\\ \label{WignerGauss}
&&  \times\exp \left( -  \left( (k_z \mp Q )\sigma_z - \left( \beta - \frac{1}{\beta} \right) \arctan(\omega_r t) \frac{(z \mp v_0 t)}{\sigma_z} \right)^2 \right).
\end{eqnarray}
With this Wigner function
 we calculate the function $G^{(2)}_{bb}$, given by  Eq.~(\ref{G2classical}),
\begin{eqnarray} \nonumber
&& G^{(2)}(\KK,\DK) = (Naa_{hor} \sigma_r)^2 \sigma_z\frac{2^{3/2}}{\pi^{7/2}}
 \int_0^\infty \mbox{d} \tau
 \int \mbox{d} \KK'    \, \frac{\delta \left( |K|'-|K| \right)}{4\pi K^2} 
\exp \left(  - 2\alpha^2 \tau^2   - 2 {K_r'}^2\sigma_r^2(1+\tau^2)   \right)
\\ \label{G2cl1}
&& \frac{1+\tau^2}{(1+\beta^2 \tau^2)|c_z(\tau)|}
 \exp \left( -\frac{\Delta K_r^2 \sigma_r^2(1+\tau^2)}{2(1+\beta^2\tau^2)}   -\frac{\Delta K_z^2 \sigma_z^2}{2 |c_z(\tau)|^2 }
- 2\left((K_z'-Q)\sigma_z - \alpha \tau \left( \beta - \frac{1}{\beta} \right) \arctan \tau \right)^2
\right) K',
\end{eqnarray}
\end{widetext}
where we introduced dimensionless time $\tau = \omega_r t$.

Now, we derive the anomalous density, given by Eq.~(\ref{Mgauss2}), in the fast collision case.
As $\tau_c$ is much smaller than all the other characteristic times 
we can approximate $c_z(\tau) \simeq 1 $, and $1+\tau^2 \simeq 1 $, and $1-i\beta \tau \simeq 1$.
Additionally, $\phi(\tau) \ll 1$ and we can neglect it.
Thus,  performing the temporal integral we arrive at
\begin{eqnarray} \nonumber
&& M(\KK, \DK )  
= \frac{A \sqrt{\pi}}{2\alpha}  \exp \left( -\frac{\omega^2}{4\alpha^2}\right)
\left( 1 + \mbox{erf}\left( \frac{i\omega}{2\alpha}\right) \right)\times
\\ \label{szy1}
&& 
\times\exp \left( - \frac{\Delta K_r^2\sigma_r^2 + \Delta K_z^2\sigma_z^2 }{4}  \right),
\end{eqnarray}
where
\begin{eqnarray*}
\frac{\omega}{\alpha} = 2 \delta K \sigma_z +   \frac{\Delta K_r^2 \sigma_z}{4Q}.
\end{eqnarray*}
The term $\exp \left( - \frac{\Delta K_r^2\sigma_r^2}{4} \right)$
gives the width $\Delta K_r$ to be approximately equal to $1/\sigma_r$.
Then, we have $\frac{\Delta K_r^2 \sigma_z}{4Q} \approx \frac{ \sigma_z}{4 Q \sigma_r^2}  $.
According to condition in Eq.~(\ref{szybkieC}), and the facts that $\sigma_r \geqslant a_{hor}$, see Eq.~\ref{srahor2},
the above term is much smaller than unity and can be neglected.
As a result we obtain 
\begin{eqnarray} \nonumber
M(\K_1, \K_2 )  
&=&\frac{A\sqrt{\pi}}{2 \alpha} \exp \left( -\frac{ \Delta K_r^2\sigma_r^2 + \Delta K_z^2 \sigma_z^2}{4}  \right)  \times
\\ \nonumber
& & \times \exp \left(   -  \delta K^2 \sigma_z^2   \right) \left( 1+ \mbox{erf}(i \delta K \sigma_z) \right).
\end{eqnarray}

Now, we derive the semiclassical expression in the fast collision case.
In the formula from Eq.~(\ref{G2cl1}) all the temporal dependence, apart from $- 2\alpha^2 \tau^2$, can be neglected.
We therefore obtain
\begin{eqnarray*}
&& G^{(2)}(\KK,\DK) \simeq  \frac{ (N a  \sigma_z\sigma_r)^2}{\pi^3} 
\exp \left( -\frac{\Delta K_r^2 \sigma_r^2 + \Delta K_z^2 \sigma_z^2}{2}  \right)
 \\
&&
\int \mbox{d} \KK'    \, \frac{\delta \left( |K|'-|K| \right)}{4\pi K^2} 
\exp \left(    - 2 {K_r'}^2\sigma_r^2 - 2(K_z'-Q)^2\sigma_z^2 \right),
\end{eqnarray*}
where we approximated $|K'|  \simeq Q$ and performed the temporal integral.
Taking $\mbox{d} \KK ' = {K'}^2 \mbox{d}K' \mbox{d} \cos \theta \mbox{d}\phi$ and integrating over $K'$ and $\phi$ we arrive at
\begin{eqnarray*}
&& G^{(2)}(\KK,\DK) \simeq  \frac{ (N a  \sigma_z\sigma_r)^2}{2\pi^3} 
\exp \left( -\frac{\Delta K_r^2 \sigma_r^2 + \Delta K_z^2 \sigma_z^2}{2}  \right)
\\
&& \int_{-1}^1 \mbox{d} z \,    
\exp \left(    - 2 K^2 \sigma_r^2 (1-z^2) - 2(K z -Q)^2\sigma_z^2 \right),
\end{eqnarray*}
where $z=\cos \theta$. We now introduce $y=1-z$ and $\delta K = K-Q$. In these new variables, the integrand takes the form
$ \exp \left( - 2K^2\sigma_r^2(2y-y^2) + 2 (\delta K(1-y) -Qy )^2 \sigma_z^2 \right) $.
As shown in Section~\ref{Sbb} the value of $K$ is close to $Q$, so the term $ - 4 K^2 \sigma_r^2 y$
 gives the characteristic width in $y$ being approximately equal
to $\frac{1}{4Q^2 \sigma_r^2}$. Due to the condition from Eq.~(\ref{Cond1}), and the fact that $\sigma_r \geqslant a_{hor}$,
given by Eq.~(\ref{srahor2}), this width is much smaller than unity, and thus the term $ 2 K^2\sigma_r^2y^2$ can be neglected.
The last term reads $-2 (\delta K(1-y) \sigma_z - Qy\sigma_z)^2$.
We have  $Qy\sigma_z \approx  \frac{\sigma_z}{4Q \sigma_r^2} \leq \frac{\sigma_z \sigma_r^2}{Q a_{hor}^4} \ll 1$,
where we used $\sigma_r \geq a_{hor}$ and the fast collision condition given in Eq.~(\ref{szybkieC}). 
Additionally, as $y \ll 1$ we can approximate $\delta K(1-y) \sigma_z \simeq \delta K \sigma_z$. 
Therefore, we obtain 
\begin{eqnarray*}
&& \int_{-1}^1 \mbox{d} z \, \exp \left(    - 2 K^2 \sigma_r^2 (1-z^2) - 2(K z -Q)^2\sigma_z^2 \right)  
\\
&& \simeq
\int_0^\infty \mbox{d} y \,  \exp \left(    - 4 Q^2 \sigma_r^2 y  - 2\delta K\sigma_z^2 \right). 
\end{eqnarray*} 
As a result of all the approximations, we finally obtain
\begin{eqnarray*}
 G^{(2)}(\KK,\DK) &\simeq&   \frac{ (Na\sigma_z)^2}{(2\pi)^3Q^2} 
\exp \left( -\frac{\Delta K_r^2 \sigma_r^2 + \Delta K_z^2 \sigma_z^2}{2}  \right)\times
\\
& & \times\exp \left( - 2\delta K\sigma_z^2 \right).
\end{eqnarray*}

The semiclassical expression from Eq.~(\ref{G2cl1}) in the strong confinement case takes the form
\begin{eqnarray*}
&& G^{(2)}(\KK,\DK) \simeq Q (Na a_{hor}^2)^2 \sigma_z \frac{2^{3/2}}{\pi^{7/2}} \int_0^\infty \mbox{d} \tau \times
\\
&&
\times \int \!\!\mbox{d} \KK'    \, \frac{\delta \left( |K|'-|K| \right)}{4\pi K^2} 
e^{  - 2\alpha^2 \tau^2  - 2{K_r'}^2 a_{hor}^2(1+\tau^2)  }\times
\\
\\
&&
\times \exp \left(  -\frac{ \Delta K_r^2 a_{hor}^2+ \Delta K_z^2 \sigma_z^2}{2  } - 2(K_z'-Q)^2\sigma_z^2  \right),
\end{eqnarray*}
where we put $K' \simeq Q$. We now perform the temporal integral and, as before, partly perform 
the integral over $\KK'$ obtaining:
\begin{eqnarray*}
&& G^{(2)}(\KK,\DK) \simeq Q a_{hor}^4 \sigma_z \frac{(Na)^2}{2\pi^3}
\exp \left( -\frac{\Delta K_r^2a_{hor}^2}{2} \right) \times
\\
&& \times\exp \left( - \frac{ \Delta K_z^2 \sigma_z^2 }{2}  \right)
 \int_{-1}^1 \mbox{d} z  \,    \sqrt{\frac{1}{\alpha^2 +K^2 a_{hor}^2(1-z^2)   }  }\times
\\
&&
\times\exp \left( - 2K^2a_{hor}^2 (1-z^2)   - 2(Kz-Q)^2\sigma_z^2\right).
\end{eqnarray*}
We again introduce $y = 1-z$ and $\delta K = K - Q $, which can be used to rewrite the term $\exp \left( - 2K^2a_{hor}^2 (1-z^2) - 2(Kz-Q)^2\sigma_z^2\right) =\exp \left( -
2K^2a_{hor}^2 (2y-y^2) - 2(\delta K (1-y)-Q y)^2\sigma_z^2\right) $.  As before, the term $-4 K^2 a_{hor}^2 y $ implies that the width of $y$ is maximally
${1}/{4Q^2a_{hor}^2} \ll 1$.  Consequently, we neglect the term $2K^2a_{hor}^2 y^2 $ and approximate $ - 2(\delta K (1-y)-Q y)^2\sigma_z^2 \simeq - 2(\delta K
-Q y)^2\sigma_z^2 $. As $K \simeq Q$, we also approximate the term $-4 K^2 a_{hor}^2 y \simeq -4 Q^2 a_{hor}^2 y $.  Furthermore, we approximate the term present in the
denominator $K^2 a_{hor}^2(1-z^2) \simeq 2 Q^2 a_{hor}^2 y $ and extend the limits of integration from $2$ to $\infty$.  As a result we obtain
\begin{eqnarray*}
&& G^{(2)}(\KK,\DK) \simeq  \frac{(Na a_{hor}^2)^2}{4 \pi^3} 
e^{ -\frac{\Delta K_r^2a_{hor}^2 + \Delta K_z^2 \sigma_z^2 }{2}  }\times
\\
\\
&&
\times \frac{1}{\alpha } \int_0^\infty \mbox{d} z  \,    \frac{1}{\sqrt{\alpha^2 + z }} 
\exp \left( - 2 z   - \frac{1}{2\alpha^2}(2\delta K Q a_{hor}^2 -z)^2\right),
\end{eqnarray*}
where we changed the variables to  $z = 2Q^2 a_{hor}^2 y $.

Below, we derive expression for the back to back part of the two particle correlation function
$ |M(\KK,\DK)|^2$ averaged over $\KK$. 
From Eq.~(\ref{Mtdrugie}) we have
\begin{widetext}
\begin{equation} \label{Mav}
\int \!\!\mbox{d} \KK \, |M(\KK,\DK)|^2 = \int\!\! \frac{\mbox{d} \KK \mbox{d} \x_1 \mbox{d} \x_2}{\hbar^2(2\pi)^6}
\int_{0}^\infty\!\! \mbox{d} t_1 \mbox{d} t_2 \,  
e^{ - i \DK (\x_1-\x_2) + i \frac{\hbar}{m} \left[K^2 + \frac{\Delta K^2}{4} - Q^2 \right] (t_1-t_2) }  \tilde B(\KK,\x_1,t_1) \tilde B^*(\KK,\x_2,t_2) 
\end{equation}
\end{widetext}
where $ \tilde B(\KK,\x,t) = B(\KK,\x,t) \exp \left(i  \frac{\hbar Q^2}{m} t\right) $,
$\DK = \K_1+\K_2$, and we used ${(k_1^2+k_2^2)}/{2} = K^2 + {\Delta K^2}/{4} $. 
As shown in Section~\ref{Sbb} the value of $K$ is close to $Q$, so we substitute $K = Q + \delta K$.
Then, the  Eq.~(\ref{Mav}) takes the following form
\begin{widetext}
\begin{eqnarray} \nonumber
\int \mbox{d} \KK \, |M(\KK,\DK)|^2 &=&  \frac{1}{ \hbar^2(2\pi)^6}  \int \mbox{d} \Omega_\KK 
\int_{0}^\infty \mbox{d} t_1 \mbox{d} t_2 \int \mbox{d} \x_1 \mbox{d} \x_2  \,  
e^{  - i \DK (\x_1-\x_2) + i \frac{\hbar}{m} \left(K^2 + \frac{\Delta K^2}{4} - Q^2 \right) (t_1-t_2) }\times
\\ \nonumber
\\ \label{Mav2}
& & \times \tilde B(\KK,\x_1,t_1) \tilde B^*(\KK,\x_2,t_2) \int_0^\infty K^2 \mbox{d} K \, \exp \left(  i \frac{\hbar}{m} (2Q \delta K + \delta K^2)(t_1-t_2) \right),
\end{eqnarray}
\end{widetext}
where we decomposed the integral over $\KK$; note that $B(\KK,\x,t)$ does not depend on $K$ but only on the direction ${\bf e}_\KK$.  
Due to the fact that $\delta K \ll Q$, we approximate 
$\int_0^\infty K^2 \mbox{d} K \simeq Q^2 \int_{-\delta Q}^{\delta Q}  \mbox{d} \delta K $,
where $\delta Q$ is much smaller than $Q$ but still bigger than the width in $\delta K$. 
Then, the last integral in Eq.~(\ref{Mav2}) takes the form
\begin{eqnarray*}
&& \int_{-\Delta Q}^{\Delta Q}  \mbox{d} \delta K \, \exp \left( i \frac{\hbar}{m} (2 Q \delta K + \delta K^2) (t_1-t_2) \right)
\\
&& = Q \int_{-x_0}^{x_0}  \mbox{d} x \,  \exp \left( i \frac{\hbar Q^2 (t_1-t_2)}{m}(2x+x^2)  \right),
\end{eqnarray*}
where  $x = \delta K/Q$ and $x_0 = \delta Q /Q$. If we take $x_0 = 0.1$ the above integral gives us a 
peaked function in $t_1-t_2$ with a width
of $(t_1-t_2) = {5m}/{\hbar Q^2}$. For larger $(t_1-t_2)$ the function oscillates rapidly. 
The condition in Eq.~(\ref{Cond1}) together with the fact that $Q \sigma_z \gg 1$ 
imply that $\frac{5m}{\hbar Q^2} \ll \tau_c, \tau_r $, which shows that during
the time $(t_1-t_2) = {5m}/{\hbar Q^2} $ the change of the wavefunctions $\tilde \psi_{\pm Q}(\x,t)$ are negligible.
Thus, the integration over $\delta K$ results in the Dirac delta function 
$\int_{-\Delta Q}^{\Delta Q}  \mbox{d} \delta K \, \exp \left( i \frac{\hbar}{m} (2 Q \delta K + \delta K^2) (t_1-t_2) \right)  \simeq 
 \frac{\pi m}{\hbar Q} \delta(t_1-t_2)$. As a result, we obtain
\begin{eqnarray} \nonumber
&& \int \mbox{d} \KK \, |M(\KK,\DK)|^2 \simeq  \frac{ \pi m Q }{ (2\pi)^6\hbar^3} \times
\\ \label{Mavera}
&& 
\times\int_{0}^\infty\!\! \mbox{d} t \int\!\! \mbox{d} \Omega_\KK \left| \int\!\! \mbox{d} \x \,  
 e^{ - i \DK \cdot \x }  B(\KK,\x,t) \right|^2.
\end{eqnarray}

The classical expression in Eq.~(\ref{G2classical}) integrated over $\KK$ takes the form 
\begin{eqnarray}\nonumber
&& \int\!\! \mbox{d} \KK \, G^{(2)}_{bb}(\KK,\DK) =  \frac{2\hbar \sigma_{tot} }{m}\int\!\! \mbox{d} \KK' \int_0^\infty\!\! \mbox{d} t  \int\!\! \mbox{d} \x \, |2K'|\times
\\ 
&&
\times W_{+Q}\!\left(\x,\KK'+\frac{\DK}{2},t \right) W_{-Q}\! \left(\x,-\KK'+ \frac{\DK}{2},t \right).     
\end{eqnarray}
Approximating now  $K'\simeq Q $, substituting Wigner functions given by
Eq.~(\ref{Wigner}) and performing the integral over $\KK',$ which results in the Dirac delta function, one
obtains exactly the same formula as in Eq.~(\ref{Mavera}) with $B(\KK,\x,t) = B(\x,t)$ given by Eq.~(\ref{Bp}).

Now,  we calculate the expression in Eq.~(\ref{Mavera}) subject to the condition presented in Eq.~(\ref{conditionNn2}),  when 
$B(\KK,\x,t) \simeq B(\x,t)$, in the case of gaussian ansatz, see Eqs.~(\ref{vatnowe}) and (\ref{parwar}). 
Using Eq.~(\ref{Bp}) and performing gaussian integrals we obtain
\begin{eqnarray} 
&& \int \mbox{d} \KK \, |M(\KK,\DK)|^2 \simeq v_0 \frac{(Na)^2}{8\pi^2 \sigma_z^2}
 \int_{0}^\infty \mbox{d} t \,e^{ - 2\frac{v_0^2t^2}{\sigma_z^2}} \times\quad
\nonumber \\ 
&& \times
 \frac{ 1}{ \sigma_r^4(t)|a_r(t)|^2 |a_z(t)|   } 
 e^{ - \frac{1}{8} \left( \frac{\Delta K_r^2}{|a_r(t)|^2\sigma_r^2(t)} +
\frac{\Delta K_z^2 }{|a_z(t)|^2\sigma_z^2} \right) }.
\label{srednieM}
\end{eqnarray}
This expression can be compared to the momentum density of $\psi_{\pm Q}$, given by Eq.~(\ref{vatnowe}), which takes the form
\begin{eqnarray*}
 |\psi_{\pm Q}(\K,t)|^2 &=&   \frac{N \pi^{3/2}}{2 \sigma_r^2(t) \sigma_z |a_r(t)|^2 |a_z(t)|  } \times
\\
& &
\times e^{  - \frac{1}{4} \left( \frac{k_r^2}{|a_r(t)|^2\sigma_r^2(t)} +
\frac{k_z^2 }{|a_z(t)|^2\sigma_z^2} \right)  },
\end{eqnarray*}
where $\psi(\K,t) = \int \mbox{d} \x \, e^{-i\K\x} \psi(\x,t)$, and we used Eq.~(\ref{parwar}).
By comparing the two expressions above we find that 
\begin{eqnarray*}
\int \mbox{d} \KK \, |M(\KK,\DK)|^2 &\propto& \int_0^\infty \mbox{d} t \, 
\frac{1}{\sigma_r^2(t)\sigma_z}   e^{ - 2\frac{v_0^2t^2}{\sigma_z^2} }\times
\\
& & \times\left|\psi_{\pm Q} \left(\frac{\DK}{\sqrt{2}},t \right)\right|^2.
\end{eqnarray*}
The expression in Eq.~(\ref{srednieM}) integrated over $\Delta K_z$ takes the form: 
\begin{eqnarray} \nonumber
&& \int \mbox{d} \Delta K_z \int \mbox{d} \KK \, |M(\KK,\DK)|^2 \simeq C_b 
 \int_{0}^\infty \frac{\mbox{d} \tau}{1+\tau^2}  \times
\\ \nonumber
&& 
 \times \exp\left(- 2\frac{\alpha^2}{\beta^2} \tau^2 -  \frac{\Delta K_r^2(1+\tau^2/\beta^2)}{2(1+\tau^2)}  \right),
\end{eqnarray}
where $C_b = \frac{4Q (Na)^2}{(2\pi)^{3/2} \sigma_z^2\sigma_r^2 }$ and  $\tau = \beta \omega_r t$. In deriving this last expression we used Eq.~(\ref{parwar}).

\section{Metastable helium experiment parameters} \label{metahelium}

In the metastable helium experiment the excited states of atoms are used for which the scattering length is $a=7.51$ nm.
The parameters of the Palaiseau experiment are \cite{paryz0} 
\begin{eqnarray*}
 N= 10^5, \ \frac{\omega_r}{2\pi} = 1150 \ \mathrm{Hz}, \ \frac{\omega_r}{2\pi} = 47 \  \mathrm{Hz},
 \ v_0 = 9.81 \frac{\mathrm{cm}}{\mathrm{s}},
\end{eqnarray*}
while the parameters of Vienna experiment are \cite{Wieden} 
\begin{eqnarray*}
 N= 2\times 10^6, \ \frac{\omega_r}{2\pi} = 800 \ \mathrm{Hz}, \ \frac{\omega_r}{2\pi} = 47 \  \mathrm{Hz},
 \ v_0 = 9.81 \frac{\mathrm{cm}}{\mathrm{s}}.
\end{eqnarray*}
With these parameters we solve numerically Eq.~(\ref{sz0}) to obtain $\sigma_z$, and substitute it to
Eq.~(\ref{sr0}) to obtain $\sigma_r$. With these results, we calculate $\alpha$ and $\beta$, which in the 
case of Palaiseau experiment take the values $\alpha \simeq 0.22$, $\beta \simeq 3.3$, while for 
Vienna experiment $\alpha \simeq 0.2$, $\beta \simeq 11$.

\section{Formulas needed in Section \ref{lc} } \label{ApLoc}

\subsection{Derivation of  Eq.~(\ref{G1formula}) }\label{derf}

From the definition of the Wigner function we have
\begin{eqnarray*}
G^{(1)}\left( \x + \frac{\Delta \x}{2}, \x - \frac{\Delta \x}{2}, T\right) = \int \mbox{d} \K \, e^{-i\K \Delta \x} W(\x,\K;T)
\end{eqnarray*}
Inserting this expression into Eq.~(\ref{G1def}), we obtain
\begin{eqnarray} \nonumber
&& G^{(1)}\left(\K +\frac{\Delta \K}{2},\K -\frac{\Delta \K}{2} \right) = \lim_{T \rightarrow \infty }
\exp \left( i \frac{\hbar \K \Delta \K}{m}T \right)
\\ \nonumber
&& \left( \frac{\hbar T}{m}  \right)^3\int \mbox{d} \K' \, \exp \left(-i \frac{\hbar \K' \Delta \K}{m} T \right) W \left(\frac{\hbar}{m} \K T,\K';T \right),
\end{eqnarray}
where we introduced $\K = \frac{\K_1+\K_2}{2}$ and $\Delta \K = \K_1-\K_2$.
Introducing new variable $\x' = \frac{\hbar (\K-\K')T}{m}$ and using Eq.~(\ref{Wc}), we obtain
\begin{eqnarray} \nonumber
&& G^{(1)}\left(\K +\frac{\Delta \K}{2},\K -\frac{\Delta \K}{2} \right) = \lim_{T \rightarrow \infty } \int \mbox{d} \x' \, e^{ i \Delta \K \x' } \times
\\ \nonumber
&& \times\int_0^T\!\! \mbox{d} t \, f\left(\frac{\hbar}{m} \K t + \x' \left( 1 - \frac{t}{T} \right),\K, t  \right).
\end{eqnarray}
Introducing now $\x = \x'+ \frac{\hbar}{m} \K t$, we finally obtain
\begin{eqnarray} \nonumber
&& G^{(1)}\left(\K +\frac{\Delta \K}{2},\K -\frac{\Delta \K}{2} \right) = 
\\ \nonumber
&&  \int \mbox{d} \x \int_0^\infty \mbox{d} t \,  \exp \left(i \Delta \K \left( \x -\frac{\hbar}{m} \K t  \right) \right)  f\left(\x,\K, t  \right).
\end{eqnarray}

\subsection{Derivation of the source function $f(\x,\K,t)$} \label{wypf}

In this Appendix we derive Eq.~(\ref{sourcef7n}), valid as long as the condition in Eq.~(\ref{conditionNn2}) is satisfied.
The condition yields $B(\K,\x,t) \simeq B(\x,t)$, and the anomalous density is then given by Eq.~(\ref{Mkw}).
Using Eqs.~(\ref{Mkw}) and (\ref{Bp})   together with Eqs.~(\ref{Gkn}) and (\ref{G1formula}), we find that
\begin{eqnarray} \nonumber
 f(\x , \K, t ) &=&   \frac{2}{(2\pi)^3\hbar^2} \int_{-t}^{t} \mbox{d} \Delta t  \, \exp \left( - i \frac{2\hbar k^2}{m} \Delta t \right)\times
\\ \label{sourcef}
& & \times B_p^*(\x,\K,t, -\Delta t) B_p(\x,\K,t, \Delta t),
\end{eqnarray}
where
\begin{eqnarray}\nonumber
B_p(\x,\K,t,\Delta t) &=& \int \mbox{d} \Delta \x \, K_f(\Delta \x,\Delta t) \times
\\ \label{Bpwzor}
& & \times B\left(\x + \Delta \x + \frac{\hbar \K \Delta t}{m}, t -\Delta t\right),
\end{eqnarray}
and $K_f$ denotes the free propagator.

The formulas given by Eqs.~(\ref{sourcef}) and (\ref{Bpwzor}) are the basis for understanding the physics of the processes.
We first show, that only for $|\Delta t| < \Delta t_0 = 2 C  {m \sigma_r}/{\hbar Q}$ 
the integrand in these equations is nonzero.
Here, $C$ is of the order of unity;  it means that $C$ can be equal to few but not to few tens.
This estimation stems from considering the integration over $\Delta x$. 
As the system has cylindrical symmetry, we take $\K = (k_x,0,k_z)$ without the lost of generality.
The part of the integrand with  $\Delta x$ dependence consist of free propagator $ K_f \left( \Delta x, \Delta t  \right)
\propto \exp \left( i \frac{ m \Delta t^2}{2 \hbar \Delta t } \right)$ multiplied by the source function $B$.
We introduce the phase of $B$ as $B=|B|e^{i \phi}$ and write 
$ K_f \left( \Delta x, \Delta t  \right)B \propto |B|\exp \left( i \frac{ m \Delta x^2}{2 \hbar \Delta t } + i \phi \right) $.
The function $B$ has the width in the radial direction  approximately equal to $\sigma_r$, since it
is proportional $\psi_{+Q}\psi_{-Q}$. 
This means that the function $|B|$ is nonzero only for
\begin{equation} \label{wa111}
\left|x + \Delta x + \frac{\hbar k_x \Delta t}{m} \right| < C \sigma_r,
\end{equation}
where $C$ is of the order of a unity.
Let us now note that in Eq.~(\ref{sourcef}) we deal with the product of two $B_p$ functions.
The analysis of the integral of $B_p^*$ results in analogous condition 
\begin{equation} \label{wa222}
\left|x + \Delta x' - \frac{\hbar k_x \Delta t}{m} \right| < C \sigma_r,
\end{equation}
where $\Delta \x'$ denotes the integration variable.  
The above conditions, Eqs.~(\ref{wa111}) and (\ref{wa222}), can be rewritten as
\begin{eqnarray*}
\left|x \!+\! \frac{\Delta x \!+\! \Delta x' }{2} \right| < C \sigma_r \  \mathrm{and} \
\left|\frac{\hbar k_x \Delta t}{m} \!+\!\frac{\Delta x \!-\! \Delta x' }{2}  \right| < C \sigma_r.
\end{eqnarray*}
As shown in Appendix \ref{appendixAD}, the phase $\phi$ is maximally equal to $\epsilon Q x$ where $\epsilon \ll 1$.
In such  case the function $\exp \left( i \frac{ m \Delta x^2}{2 \hbar \Delta t } + i \phi \right) $ 
is an  oscillating function with decreasing period of oscillation when moving from the central point 
$\Delta x_0 = \epsilon \frac{\hbar Q \Delta t}{m}$. The characteristic width  of $\Delta x$ equals 
$ 2\sqrt{  \frac{\hbar |\Delta t|}{m}}$. 
As a result, the second of the above inequality takes the form
\begin{eqnarray*}
\left|\frac{\hbar k_x \Delta t}{m} \left( 1 + \frac{Q}{k_x}\frac{\epsilon - \epsilon'}{2}  \right)
\pm 2\sqrt{  \frac{\hbar|\Delta t|}{m}}  \right| < C \sigma_r.
\end{eqnarray*}
According to the assumption stated in Section \ref{model}, we restrict our analysis to the situation for which $k_x > \frac{Q}{2}$.  Thus, $\left|\frac{Q}{k_x}\frac{\epsilon -
  \epsilon'}{2} \right| < |\epsilon - \epsilon' | $ and can be neglected.  Let us notice that for $\Delta t$ equal to $C \frac{m \sigma_r}{\hbar k_x}$, we have
$\frac{\hbar k_x \Delta t}{m} = C \sigma_r $, and
\begin{eqnarray*}
 2\sqrt{  \frac{\hbar|\Delta t|}{m}} = 
2 \sqrt{C} \frac{\sigma_r}{\sqrt{k_x \sigma_r}} < 2\sqrt{2C} \frac{\sigma_r}{\sqrt{Q\sigma_r}} = \Delta_0 \ll \sigma_r,
\end{eqnarray*}
because $Q\sigma_r \gg 1$.
The above inequality implies that $ \Delta t <  C  \frac{m \sigma_r}{\hbar k_x}$.
Using  $k_x > \frac{Q}{2}$, we obtain
\begin{eqnarray*}
\Delta t < \Delta t_0 = 2 C  \frac{m \sigma_r}{\hbar Q}.
\end{eqnarray*}
We have shown that the timescale of $\Delta t$ is ${m \sigma_r}/{\hbar Q} $, a time
that is needed for the scattered particle to leave the cloud. 

Basing on the results presented above, we now continue with approximating the expression in Eq.~(\ref{Bpwzor}).
First, we notice that due to condition in Eq.~(\ref{conditionNn2}) and the fact that $\sigma_z \gg \sigma_r$,
we have $\Delta t_0 \ll \tau_c, \tau_r$. This means that the changes of the wavefunctions $\psi_{\pm Q}$
can be neglected during the time $\Delta t_0$. Thus, according to the Eq.~(\ref{Bp}), we have
\begin{eqnarray} \nonumber
 B\left(\tilde \x + \Delta \x, t-\Delta t \right) &\simeq& 2 g \psi_{+Q}\psi_{-Q}\left(\tilde \x + \Delta \x, t\right)\times
\\  \label{cos11}
& & \times \exp \left( i \frac{\hbar Q^2}{m} (t-\Delta t) \right),
\end{eqnarray}
where $\tilde \x = \x + {\hbar \K \Delta t}/{m}$.
We found above that the effective width in $\Delta x$
given by the propagator $K_f(\Delta \x,\Delta t)$ is $\Delta_0 \ll \sigma_r$.
As the propagator is a function of $\Delta \x$ thus the width in $\Delta y$ and $\Delta z$ is also $\Delta_0$.
On such a distance the change of the function $| \psi_{+Q} \psi_{-Q}|$ can be neglected.
As a result, we obtain
\begin{equation}\label{cos12}
|\psi_{+Q}\psi_{-Q}\left(\tilde \x + \Delta \x , t \right)| \simeq |\psi_{+Q}\psi_{-Q}\left(\tilde \x, t \right)|.
\end{equation}
Finally, we analyze the phase of $\psi_{+Q}\psi_{-Q}(\tilde \x + \Delta \x,t)$.  Using gaussian ansatz given by Eq.~(\ref{vatnowe}), together with Eq.~(\ref{parwar}), we
find that the $x$ dependent phase of the integrand in Eq.~(\ref{Bpwzor}) equals
\begin{eqnarray*}
   \beta \omega_rt \frac{ (\tilde x + \Delta x)^2 }{\sigma_r^2(t)}  +\frac{m\Delta x^2}{2\hbar \Delta t}.
\end{eqnarray*}
It can be rewritten as 
\begin{eqnarray*}
\frac{m}{2\hbar \Delta t} (1+\epsilon)
\left( \Delta x + \tilde x \frac{\epsilon}{1+\epsilon} \right)^2 +  
\beta \omega_rt \frac{ \tilde x^2 }{\sigma_r^2(t)} (1+\epsilon),
\end{eqnarray*}
where $\epsilon ={ 2 \hbar \Delta t \beta \omega_r t}/{m \sigma_r^2(t)}  $,  and its maximal value 
can be estimated to be  $ \frac{2\hbar \Delta t_0}{m a_{hor}^2}  = 4 C \frac{\sigma_r}{Q a_{hor}^2} $.
According to the condition (\ref{Cond1}), this is much smaller than unity, and we obtain $1+\epsilon \simeq 1$. 
Thus the phase is equal to 
\begin{equation}\label{cos13}
\frac{m}{2\hbar \Delta t} \left( \Delta x + \epsilon \tilde x  \right)^2 +  
\beta \omega_rt \frac{ \tilde x^2 }{\sigma_r^2(t)}. 
\end{equation}
The same reasoning can be repeated in the case of $y$ and $z$ coordinates.
Thus, the Eq.~(\ref{cos12}), together with Eq.~(\ref{cos13}), gives
\begin{eqnarray*}
&&K_f(\Delta \x,\Delta t) \psi_{+Q}\psi_{-Q}(\tilde \x + \Delta \x,t) 
\\
&&
\simeq \psi_{+Q}\psi_{-Q}(\tilde \x,t) 
K_f \left( \Delta x + \epsilon_x \tilde x, \Delta y + \epsilon_y \tilde y,  \Delta z + \epsilon_z \tilde z, \Delta t\right).
\end{eqnarray*}
Using the above together with  Eq.~(\ref{cos11}) make Eq.~(\ref{Bpwzor}) to take the form
\begin{eqnarray}\nonumber
&& B_p(\x,\K,t,\Delta t) = \int \mbox{d} \Delta \x \, K_f(\Delta \x,\Delta t) \times
\\ \nonumber
&& \times B\left(\x + \Delta \x + \frac{\hbar \K \Delta t}{m}, t -\Delta t\right),
\\ \nonumber
&& \simeq 2g \psi_{+Q}\psi_{-Q}\left(\tilde \x,t\right) 
\exp \left( i \frac{\hbar Q^2}{m} (t-\Delta t) \right) \times
\\ \nonumber
&& \times \int \mbox{d} \Delta \x \, 
K_f \left( \Delta x + \epsilon_x \tilde x, \Delta y + \epsilon_y \tilde y,  \Delta z + \epsilon_z \tilde z, \Delta t\right)
\\
&& = 2g \psi_{+Q}\psi_{-Q}\left(\tilde \x,t\right) 
\exp \left( i \frac{\hbar Q^2}{m} (t-\Delta t) \right)
\end{eqnarray}
In order to perform the above approximation we assumed the quadratic form of the phase 
of the wavefunctions $\psi_{\pm Q}$ with position independent coefficients as given by the gaussian ansatz.
However we note that it is enough that the phase would be well approximated by quadratic function on a distance
$\Delta_0 \ll \sigma_r$. As this is generally true thus the above reasoning apllies for 
true (not only variational) solution of the GP equation.
Using the above equation the expression in Eq.~(\ref{sourcef}) for the source function simplifies to
\begin{eqnarray} \nonumber
&& f(\x , \K, t ) \simeq   \frac{8g^2}{(2\pi)^3\hbar^2} \int_{-t}^{t} \mbox{d} \Delta t  \, 
\exp \left( - i \frac{2\hbar (k^2-Q^2)}{m} \Delta t \right)
\\ \nonumber
&& \psi_{+Q}^*\psi_{-Q}^*\left(\x-\frac{\hbar \K \Delta t}{m},t\right)\times
\psi_{+Q}\psi_{-Q}\left(\x+\frac{\hbar \K \Delta t}{m},t\right).
\end{eqnarray}

Furthermore, introducing new variable $\delta k = k-Q$ we obtain $k^2 - Q^2 = 2 Q \delta k \left(1 + \frac{\delta k}{2Q} \right)$.
As $|\delta k| \ll Q$ we can approximate $k^2-Q^2 \simeq 2 Q\delta k$ and, additionally,
$\frac{\hbar \K \Delta t}{m} \simeq \frac{\hbar Q \Delta t}{m} {\bf e}_\K $, where ${\bf e}_\K = \frac{\K}{k}$.
The timescale of $t$ is  equal or larger to one of the characteristic times $\tau_c$ or $\tau_r$.
As we noticed before, $\Delta t_0 \ll \tau_c,\tau_r$, thus, most of the collision takes place
at times $t$ much larger than $\Delta t_0$. Therefore, we  extend the limit of integration
in the above integral to infinity,$\int_{-t}^t \simeq \int_{-\infty}^\infty$.
With these approximations, we arrive at 
\begin{eqnarray} \nonumber
&& f(\x , \K, t ) =   \frac{2 \hbar \sigma_{tot}}{\pi^2 m Q} \int_{-\infty}^{\infty} \mbox{d} \delta r  \, 
\exp \left( - i 4 \delta k \delta r \right)\times
\\ \label{sourcef4}
&& \times\psi_{+Q}^*\psi_{-Q}^*\left(\x - \delta r {\bf e}_\K, t \right)  
\psi_{+Q}\psi_{-Q}
\left(\x + \delta r {\bf e}_\K, t \right),
\end{eqnarray}
where we changed the variables $\delta r = {\hbar Q \Delta t}/{m} $.
Sill, this formula can be further simplified basing on appropriate approximations.
Specifically, from Eq.~(\ref{parwar}) we obtain that the maximal characteristic change of the function $\psi_{\pm Q}$ in the $z$ direction
on the distance  $\sigma_r$ is approximately equal to
\begin{eqnarray*}
&& a_z(\infty) \left((\sigma_z + \sigma_r)^2 - \sigma_z^2 \right)
\simeq a_z(\infty) 2 \sigma_z \sigma_r  
\\
&& = \frac{\sigma_r}{\sigma_z} \left( 1 - i \left(\beta - \frac{1}{\beta} \right) \frac{\pi}{2} \right).
\end{eqnarray*}
As $\beta = \sigma_r^2/a_{hor}^2$, this quantity is much smaller than unity, as long as the condition given by Eq.~(\ref{condclas}) is satisfied.
Then, we can neglect the dependence on ${\bf e}_z$ in term $\delta r {\bf e}_\K  $, arriving at
\begin{eqnarray} \nonumber
&& f(\x , \K, t ) =   \frac{2 \hbar \sigma_{tot}}{\pi^2 m Q} \int_{-\infty}^{\infty} \mbox{d} \delta r  \, 
\exp \left( - i 4 \delta k \delta r \right)\times
\\ \label{sourcef5}
&& \times\psi_{+Q}^*\psi_{-Q}^*\left(\x - \delta r {\bf e}_{\K,r}, t \right)
\psi_{+Q}\psi_{-Q}
\left(\x + \delta r {\bf e}_{\K,r}, t \right),
\end{eqnarray}
where ${\bf e}_{\K,r} = \frac{k_x {\bf e}_x + k_y{\bf e}_y}{k}$.

\subsection{Source function: semiclassical model} \label{wypfclas}

We show here that the expression from Eq.~(\ref{sourcef7n}) can be obtained from a semiclassical model introduced in Section~\ref{Sbb}.
The formula for a source function in such a model takes the form:
\begin{eqnarray} \label{fcl}
 f_{cl}(\x,\K,t)  =  \frac{\hbar \sigma_{tot}}{\pi m} \int \mbox{d} \K'\, \frac{1}{K}  h \left( K ,\DK, \x,t \right),
\end{eqnarray}
where the function
\begin{eqnarray}\label{h}
&& h(K,\DK,\x,t) = \int \mbox{d} \KK' \, \delta \left( K'-K \right) \times
\\ \nonumber
&& 
\times W_{+Q} \left(\x,\KK'+\frac{\DK}{2},t \right) W_{-Q}\left(\x,-\KK'+\frac{\DK}{2},t \right),
\end{eqnarray}
with $\KK = \frac{\K-\K'}{2}$ and $\DK = \K + \K'$.

We first analyze the function $h$ given by Eq.~(\ref{h}) in the case of a gaussian ansatz, see Eq.~(\ref{vatnowe}).
Using the formula for the Wigner function from Eq.~(\ref{Wigner}) we arrive at
\begin{equation}\label{hgauss}
h(K,\DK,\x,t) = h_1(\x,t)h_2(\DK,\x,t) h_3(K,t),
\end{equation}
where:
\begin{eqnarray} \nonumber
&& h_1(\x,t) =  \left(\frac{4 N}{(2\pi)^3} \right)^2 \exp \left( - 2\frac{x^2+y^2}{\sigma_r^2(t)}  - 2\frac{z^2 + v_0^2t^2}{\sigma_z^2} \right),
\\
&& h_2 (\DK,\x,t) = \exp \left( - \frac{1}{2} (\Delta K_z - \delta_z)^2\sigma_z^2  \right) \times
\\ \label{h2}
&& 
\quad\quad\quad\quad\quad\quad \times e^{  -\frac{1}{2}  \left((\Delta K_x - \delta_x)^2 +(\Delta K_x - \delta_y)^2\right) \sigma_r^2(t) },
\\ \nonumber
&& h_3(K,t) = \int \mbox{d} \KK' \, \delta \left( K'-K \right) \times
\\ \label{h3}
&&  \quad\quad\quad\quad\quad\quad \times e^{  -  2 {K'}_r^2 \sigma_r(t)^2- 2  ({K'}_z-Q - \delta_{Kz})^2 \sigma_z^2   },
\end{eqnarray}
and
\begin{eqnarray*}
 \delta_x &=& 2 \beta \frac{\omega_rt x}{\sigma_r^2(t)}, \  \  \ \ \delta_y = 2 \beta \frac{\omega_rt y}{\sigma_r^2(t)},
\\
\delta_z &=&  2 \left( \beta - \frac{1}{\beta} \right) \arctan(\omega_r t) \frac{z }{\sigma_z^2},
\\
\delta_{Kz} &=& -\left( \beta - \frac{1}{\beta} \right)\arctan(\omega_r t) \frac{ v_0 t}{\sigma_z^2}.
\end{eqnarray*}

We now analyze the above functions discussing  the values of $K_{max}$ and $\DK_{max}$ 
for which the function $h(K_{max},\DK_{max},\x,t)$ has maximum, together with the widths in $K$ and  $\Delta K_{r,z}$ around the maximum.
From the form  of $h_3$ one can deduce that the width in $K$ is approximately equal to 
$\Delta_K = \frac{1}{\sigma_z} + \frac{1}{2Q\sigma_r^2}$ with the maximum located between $Q - \delta_K$,
where $\delta_K < \frac{\beta}{\sigma_z} = \frac{\sigma_r^2}{a_{hor}^2 \sigma_z}$, and $Q - \delta_K + \Delta_K$. 
Additionally, it can be seen from the $h_2$ that 
 $|\Delta K_{max;x,y}| <  \frac{2\beta}{\sigma_r} = \frac{2\sigma_r}{a_{hor}^2}$
and $|\Delta K_{max;z}| < \frac{2\beta}{\sigma_z} $, with the width in $\Delta K_{r,z}$ around the maximum
equal to $\frac{1}{\sigma_r}$ and $\frac{1}{\sigma_z} $, respectively.
The width in $K$ is much smaller than the width in $\Delta K_{x,y}$.
We make use of this fact below in performing an approximation.

To this end, we first find $\KK_0$ and $\DK_0$ satisfying $\K = \KK_0 + \frac{\DK_0}{2}$, for which the function $h(K_0,\DK_0,\x,t)$ has maximum.  Note, that without the
restriction $\K = \KK_0 + \frac{\DK_0}{2}$, we would have $|\KK_0| = K_{max}$ and $\DK_0 = \DK_{max}$, but here it does not need to be the case.  Therefore, we introduce
$\delta \K' = \K' + \KK_0 - \frac{\DK_0}{2} $.  We obtain
\begin{eqnarray*} 
 K \simeq K_0  - \frac{1}{2} {\bf e}_0 \cdot  \delta \K' \ \ \mathrm{and} \ \ \ \DK =\DK_0 +  \delta \K',
\end{eqnarray*}
where ${\bf e}_0 = \frac{\KK_0}{K_0}$.  Next, we change the variables of integration in Eq.~(\ref{fcl}) from $\K'$ to $\DK$, and further on to $\delta \K'$.  Thus, the
integral over $k_x'$ takes the following form
\begin{eqnarray*}
\int \mbox{d} \delta k'_x \, 
 \frac{1}{K}   h \left(K_0 - \frac{1}{2} {\bf e}_0 \cdot  \delta \K' ,\DK_0 +  \delta \K' ,\x,t \right). 
\end{eqnarray*}
In the considerations presented above, we found that the width in $K$ is much smaller than the width in $\Delta K_x$.
The integration over $\delta k'_x$ can be effectively changed into integration over $K$ 
($\int \mbox{d}\delta k'_x  \rightarrow  \frac{2}{e_{0,x}} \int \mbox{d} K   $)
keeping $\Delta K_x$ constant, and equal to $\Delta K_{0,x}$. This is true, if 
$e_{0,x} =  {\bf e}_0 \cdot {\bf e}_x$ is of the order of unity.
We take $\K = (k_x,0,k_z)=k(\sin \theta,0,\cos \theta)$ without the lost of generality.
As we shall see below, for such a choice $e_{0,x} > \frac{1}{2}$.
Thus, we arrive at
\begin{equation} \label{fcl1}
 f_{cl}(\x,\K,t)  \simeq  \frac{2\hbar \sigma_{tot}}{\pi m Q e_{k,x}} \int \mbox{d} \Delta K_y \mbox{d} \Delta K_z \mbox{d} K\,  
 h \left( K ,\DK, \x,t \right),
\end{equation}
where we also approximated $1/K \simeq 1/Q$ and $e_{0,x} \simeq e_{k,x}$.  
Note, that the dependence on $\K$ is hidden in $\Delta K_{0,x}$.  
In what follows, we explicitly find this dependence.

To this end, we start from the equality
\begin{eqnarray*}
\K = K_0 {\bf e}_0 + \frac{\DK_0}{2}.
\end{eqnarray*}
Taking ${\bf e}_0  = (\sin \theta_0 \cos \phi_0, \sin \theta_0 \sin \phi_0, \cos \theta_0 )$
each of the components of $\K$ takes the form:
\begin{eqnarray}\label{phirow1}
k \sin \theta &=&  K_0 \sin \theta_0 \cos \phi_0 + \frac{\Delta K_{0,x}}{2},
\\ \label{phirow2}
0 &=& K_0\sin \theta_0 \sin \phi_0 +  \frac{\Delta K_{0,y}}{2},
\\ \label{phirow3}
k \cos \theta &=& K_0 \cos \theta_0 + \frac{\Delta K_{0,z}}{2}. 
\end{eqnarray}
Using the fact that $|\Delta K_{0,z}| \ll Q$, $k \simeq K_0 \simeq Q$, and introducing $\theta_0 = \theta + \delta \theta$, the Eq.~(\ref{phirow3}) takes the approximate
form
\begin{eqnarray}\nonumber
\delta \theta &\simeq&  - \frac{\delta k}{Q}  \cot \theta +  \frac{K_0-Q}{Q} \cot \theta + \frac{\Delta K_{0,z} }{2 Q \sin \theta} 
\\ \label{dtheta}
&=& - \frac{\delta k}{Q}  \cot \theta + r_1,
\end{eqnarray}
where $\delta k = k-Q$.
Consistently with the approximations, that were undertaken in order to obtain this formula, we have  $|\delta \theta| \ll 1 $.
We then have $ |\Delta K_{0,z}| \simeq |\Delta K_{max;z}| < \frac{2\beta}{\sigma_z}  $, which together with 
$|\sin \theta| > \frac{1}{2}$ results in  $  \left|\frac{\Delta K_{0,z} }{2 Q \sin \theta} \right| \leq \frac{2 \sigma_r^2}{Q a_{hor}^2 \sigma_z} $.
Estimating now $|K_0-Q| \leq  \delta_K + \Delta_K$, we obtain
$ \left|\frac{K_0-Q}{Q} \cot \theta \right| \leq \frac{\sigma_r^2}{Q a_{hor}^2 \sigma_z} + \frac{1}{2Q^2 \sigma_r^2} + \frac{1}{Q\sigma_z}$,
where we used $|\cot \theta| < 1$. Thus, we obtain
\begin{equation}\label{rrow}
|r_1| \leqslant 3 \frac{\sigma_r^2}{Q a_{hor}^2 \sigma_z} +\frac{1}{2Q^2 \sigma_r^2} + \frac{1}{Q\sigma_z}.
\end{equation}
As $ |\Delta K_{0,y} | \ll K_0 \simeq Q$, Eq.~(\ref{phirow2}) can be solved approximately, leading to
\begin{equation}\label{phi0rozw}
\phi_0 \simeq - \frac{\Delta K_{0,y}}{2 Q \sin \theta}.
\end{equation}
Thus, we have $|\delta \theta| \ll 1$ and $|\phi| \ll 1$, which results in $e_{0,x} = e_{k,x} = {\bf e}_\K \cdot {\bf e}_x$.
According to the assumption stated in Section \ref{model}, we have $e_{0,x} > \frac{1}{2} $, and so the same applies to ${\bf e}_{0,x}$.
Using $|\delta \theta| \ll 1$ and $|\phi| \ll 1$,  Eq.~(\ref{phirow1}) can be approximately rewritten as
\begin{eqnarray*}
\frac{\Delta K_{0,x}}{2} \simeq \delta k \sin \theta - Q \cos \theta \delta \theta - (K_0-Q) \sin \theta + Q \sin \theta \frac{\phi_0^2}{2}.
\end{eqnarray*}
With the help of Eqs.~(\ref{dtheta}) and (\ref{phi0rozw}), we rewrite the above equation as
\begin{eqnarray*}
\frac{\Delta K_{0,x}}{2} &\simeq& \frac{\delta k}{ \sin \theta} - Q r_1\cos \theta  - (K_0-Q) \sin \theta +\frac{(\Delta K_{0,y})^2}{8 Q \sin \theta}
\\ 
& &   = \frac{\delta k}{ \sin \theta} + r_2.
\end{eqnarray*}
Proceeding in the same way as in the case of $r_1$, additionally invoking Eq.~(\ref{rrow}),
we obtain that
\begin{eqnarray*}
|r_2| \leqslant 4 \frac{\sigma_r^2}{a_{hor}^2 \sigma_z} + \frac{1}{Q\sigma_r^2} + \frac{2}{\sigma_z} +  \frac{\sigma_r^2}{Q a_{hor}^4 }.
\end{eqnarray*}
If the righthand side of the above is  smaller than the width od $\frac{\Delta K_{0,x}}{2}$, equal to $\frac{1}{2\sigma_r}$, which gives
\begin{equation}\label{war10}
8 \frac{\sigma_r^3}{a_{hor}^2 \sigma_z} + \frac{2}{Q\sigma_r} + \frac{4 \sigma_r}{\sigma_z} +  \frac{2\sigma_r^3}{Q a_{hor}^4 } \ll 1,
\end{equation}
then we obtain
\begin{equation}\label{dkx00}
\Delta K_{0,x} \simeq \frac{2 \delta k}{ \sin \theta}. 
\end{equation}
Let us now note that, as $\sigma_r \geqslant a_{hor}$, $\sigma_z \gg \sigma_r$, and from the condition given in Eq.~(\ref{conditionNn2}) we obtain
\begin{eqnarray*}
\frac{2}{Q\sigma_r} + \frac{4 \sigma_r}{\sigma_z} +  \frac{2\sigma_r^3}{Q a_{hor}^4 } \ll 1.
\end{eqnarray*}
Using this results, the condition in Eq.~(\ref{war10}) can be brought to the following form
\begin{eqnarray*}
8 \frac{\sigma_r^3}{a_{hor}^2 \sigma_z}  \ll 1.
\end{eqnarray*}
Inserting Eq.~(\ref{dkx00}) into Eq.~(\ref{fcl1}), using Eqs.~(\ref{h}) and (\ref{Wigner}), and performing the integrals, we arrive at the semiclassical form of the
source function,
\begin{eqnarray*} 
&& f_{cl}(\x,\K,t)  \simeq  \frac{\hbar \sigma_{tot}}{\pi^2 m Q e_{k,x}} \int \mbox{d} \Delta x  \, \exp \left( i \frac{2 \delta k}{e_{k,x}} \Delta x  \right)\times
\\
&&
\times\psi_{+Q}^*\psi_{-Q}^*\left(\x + \frac{\Delta x {\bf e}_x}{2},t \right) 
\psi_{+Q}\psi_{-Q}\left(\x - \frac{\Delta x {\bf e}_x}{2},t \right).
\end{eqnarray*}
In deriving this results, we additionally used Eq.~(\ref{barpsi}).
If we change the variable $\delta r e_{k,x} = - \frac{\Delta x}{2} $ in this equation, 
we arrive at the quantum formula given in Eq.~(\ref{sourcef5}).

\section{Validity of the perturbative approach}\label{VP}

In this Appendix we inspect the validity of the perturbative approach.
The assumption presented in Eq.~(\ref{conditionNn2}) implies that we can effectivelly neglect the mean-field potential $2g|\psi(\x,t)|^2$
in the Hamiltonian $H_0$, given by Eq.~(\ref{H0}). As a result, Eq.~(\ref{glowne}) takes the form
\begin{eqnarray*}
i \hbar \partial_t \hat \delta (\x ,t) =  - \frac{\hbar^2}{2m} \triangle \hat \delta (\x ,t) +  B (\x ,t) \hat \delta^\dagger (\x ,t).
\end{eqnarray*}
Substituting here
\begin{eqnarray*}
\hat \delta(\x,t) = (2\pi)^{-3/2} \int \mbox{d} \K \, \exp \left( i \K\x - i\frac{\hbar k^2}{2m} t \right) \hat \delta (\K,t),
\end{eqnarray*}
we obtain
\begin{eqnarray}\label{glowne3}
i \hbar \partial_t \hat \delta (\K ,t) =  \int \mbox{d} \K' \, A(\K,\K',t) \hat \delta^\dagger (\K ,t),
\end{eqnarray}
where 
\begin{eqnarray} \nonumber
 A(\K,\K',t) &=&  \int \frac{\mbox{d} \x }{(2\pi)^3}  \exp \left( - i (\K+\K')\x \right) \times
\\ \label{rowA}
& & \times\exp \left( i\frac{\hbar(k^2+{k'}^2)}{2m} t \right) B(\x,t).
\end{eqnarray}
With the definitions $M(\K_1,\K_2,t) = \langle \hat \delta(\K_1,t)\hat \delta(\K_2,t) \rangle  $ and 
 $G^{(1)}(\K_1,\K_2,t) = \langle \hat \delta^\dagger(\K_1,t)\hat \delta(\K_2,t) \rangle  $, from  Eq.~(\ref{glowne3}) we obtain that
 \begin{eqnarray*}
&& i \hbar \partial_t M(\K_1,\K_2,t) =  A(\K_1,\K_2,t) + \int \mbox{d} \K' \, 
\\
&& 
 \left(  A(\K_1,\K',t) G^{(1)}(\K',\K_2,t) +  A(\K_2,\K',t) G^{(1)}(\K',\K_1,t) \right).
 \end{eqnarray*}
It can be proved that  the anomalous density $ M(\K_1,\K_2) = i M(\K_1,\K_2,\infty) $. 
Therefore, after integrating the equation above we obtain
\begin{eqnarray} \nonumber
&&   M(\K_1,\K_2) =  \frac{1}{\hbar} \int_0^\infty  \mbox{d} t \,   A(\K_1,\K_2,t)  + \frac{1}{\hbar}\int_0^\infty  
\mbox{d} t \int \mbox{d} \K' 
\\ \nonumber
&& 
 (  A(\K_1,\K',t) G^{(1)}(\K',\K_2,t) +  A(\K_2,\K',t) G^{(1)}(\K',\K_1,t) ).
\end{eqnarray}
Now we apply the first order perturbation, which amounts to neglecting the second line of this equation. As a result, we obtain 
\begin{equation} \label{m1234}
M(\K_1,\K_2) = \frac{1}{\hbar}
\int_0^\infty \mbox{d} t \, A(\K_1,\K_2,t),
\end{equation}
 which is exactly formula given in Eq.~(\ref{Mkw}).  
The perturbation theory, expressed in Eqs.~(\ref{Mkw}) and (\ref{Gkn}),
used in the paper are valid as long as 
$ \frac{1}{\hbar}\int_0^\infty  \mbox{d} t \int \mbox{d} \K' (  A(\K_1,\K',t) G^{(1)}(\K',\K_2,t) +  A(\K_2,\K',t) G^{(1)}(\K',\K_1,t) )$
 is much smaller than 
$\frac{1}{\hbar} \int_0^\infty \mbox{d} t \, A(\K_1,\K_2,t) $.  

We now estimate the value of
\begin{equation}\label{cosest}
\frac{1}{\hbar} \int_0^\infty  \mbox{d} t \int \mbox{d} \delta \K' \,  A(\K_1,\K_2 + \delta \K' ,t) G^{(1)}(\K_2 + \delta \K',\K_2,t),
\end{equation}
where we used $\K' = \K_2 + \delta \K'$. 
For simplicity of the calculation, we take $ \K_1 = -\K_2 = \K = Q {\bf e}_\K $.
Using that, together with Eqs.~(\ref{Bp}) and (\ref{rowA}), we obtain
\begin{eqnarray} \nonumber
 A(\K, -\K +\delta \K',t) &=& \int \frac{\mbox{d} \x }{(2\pi)^3} \exp \left( - i \delta \K'\x - i v_0 {\bf e}_\K \delta \K' t\right) 
\\ \nonumber
& &\exp\left( i \frac{\hbar}{2m} (\delta \K')^2 t  \right)
 2 g \psi_{+Q}\psi_{-Q}(\x,t).
\end{eqnarray}
In rough approximation $\psi_{+Q}\psi_{-Q}(\x,t)$ decomposes into $\x$ and $t$ dependent parts.
Additionally as $|\delta \K'| \ll Q$ we have in most directions of $\K$, 
$\left|\frac{\hbar}{2m} (\delta \K')^2 \right| \ll |v_0 {\bf e}_\K \delta \K'|$ which makes us to neglect the term
$\exp\left( i \frac{\hbar}{2m} (\delta \K')^2 t  \right)$.
As a result the approximate form of $A$ reads
\begin{equation}\label{A0}
 A(\K, -\K +\delta \K',t) = \hbar M_0(\K)  A_0(\delta \K') e^{ - i v_0 {\bf e}_\K \delta \K' t}  A_1(t).
\end{equation}
Inserting the above into Eq.~(\ref{m1234}) we obtain
\begin{eqnarray*}
M(\K,-\K+\delta \K') &=& \frac{1}{\hbar} \int_0^\infty \mbox{d}t \, A(\K, -\K +\delta \K',t)
\\
&=& M_0(\K)  A_0(\delta \K') \int_0^\infty \mbox{d}t \,  e^{ - i v_0 {\bf e}_\K \delta \K' t}  A_1(t)
\end{eqnarray*}
As $\DK = \delta \K'$, the function $A_0(\delta \K') $  has the same width as the anomalous density, equal to
${1}/{\sigma_{r,z}}$ in the respective directions.
The fact that the width in  $K = \left| \K - \frac{\delta \K'}{2} \right| \simeq Q  - \frac{1}{2} {\bf e}_\K \delta \K'$ 
of the anomalous density is equal to $\Delta_K$, implies that the width in $t$ of the function $A_1(t)$ is equal to 
${1}/{2 v_0 \Delta_K}$.
We normalized $A_0(0) = 1$ and $ \int_0^\infty \mbox{d}t\, A_1(t) =1 $, which results in that $M_0(\K)$
is equal to the anomalous density $M(\K,-\K)$ in the first order perturbation theory.

We now focus our attention  on $ G^{(1)}(\K_1,\K_2,t)$. Using derivation analogous to the ones presented in Appendix~\ref{ApLoc}, one can show that
\begin{eqnarray*}
 G^{(1)}\left(\K,\Delta \K,t\right) =   \int \mbox{d} \x \int_0^t \mbox{d} t' \,  
e^{i \Delta \K \left( \x   -  v_0 t' {\bf e}_\K \right) } f(\x,\K, t).
\end{eqnarray*}
In a rough approximation
\begin{eqnarray} \nonumber
&& G^{(1)}(-\K + \delta \K',-\K,t) \approx  \rho(-\K) G_2( \delta \K')  \times
\\ \label{cosw}
&&
\times \int_0^t \mbox{d} t' \,  
\exp \left( - i v_0 t' {\bf e}_{-\K} \delta \K' \right)   f_0(t'),
\end{eqnarray}
where $\rho$ is the density and $G_2$ is given by Eq.~(\ref{G2wzor}). 
The width in ${\bf e}_{-\K} \delta \K' $ of   $G_1$  is equal to $\Delta_k$,
which implies that the width in $t'$ of the function $f_0(t')$ is $ {1}/{v_0 \Delta_k}$. 
The fact that $ G^{(1)}(-\K,-\K,\infty) = \rho(-\K)$ implies that $\int_0^\infty \mbox{d} t' \, f_0(t') =1 $.
Inserting now Eqs.~(\ref{A0}) and (\ref{cosw}) into Eq.~(\ref{cosest}) we obtain
\begin{eqnarray*}
&& M_0(\K) \rho(-\K) \int \mbox{d} \delta \K' \, A_0(\delta \K')G_2( \delta \K')\times
\\
&&
\times\int_0^\infty  \!\!\mbox{d} t   \int_0^t \mbox{d} t' \,
 \exp \left( - i v_0 {\bf e}_\K \delta \K' (t-t') \right) A_1(t) f_0(t').
\end{eqnarray*}
Due to the fact that  $\Delta_K \approx \Delta_k$,  
the widths of both functions $A_1$ and $f_0$ are similar.
Therefore, the temporal integrals  give the width in ${\bf e}_\K \delta \K'$ approximately equal to $\Delta_k$.
Furthermore, the widths in $\delta \K'$ of the functions $A_0$ and $G_2$ is similar.
As a result, we obtain
\begin{eqnarray*}
&& \int \mbox{d} \delta \K' \, A_0(\delta \K')G_2( \delta \K')\int_0^\infty  \mbox{d} t   \int_0^t \mbox{d} t' \, e^{ - i v_0 {\bf e}_\K \delta \K' (t-t') } \times
\\
&&
\times A_1(t) f_0(t')  \approx V_c,
\end{eqnarray*}
where $V_c$ is the correlation volume.
Note that back to back and local correlation volumes are similar thus in rough approximation we 
take them to be the same and denote by $V_c$.

We conclude that the considered term is approximately equal to 
\begin{eqnarray*}
  M_0(\K) \rho(-\K) V_c,
\end{eqnarray*}
where the first order perturbation term is equal to $M_0(\K)$.
The term $ \frac{1}{\hbar}\int_0^\infty  \mbox{d} t \int \mbox{d} \K'   A(\K_2,\K',t) G^{(1)}(\K',\K_1,t) $
shall take similar value as written above. Thus 
the condition for the validity of the perturbation approach is that $\rho(-\K) V_c \ll 1$, i.e.,
the mean number of particles scattered into the correlation volume has to be much smaller than unity.


\begin{thebibliography}{99}
\bibitem{dopiska} We neglect the quantum depletion of the condensates since we are interested in the scattered particles.
\end{thebibliography}
\end{document}